\documentclass[11pt,letterpaper]{article}
\usepackage[margin=0.8in]{geometry}
\usepackage{ifpdf}
\ifpdf
    \usepackage[pdftex]{graphicx}
    \usepackage[update]{epstopdf}
\else
	\usepackage{graphicx}
\fi
\usepackage{hyperref}
\usepackage{wrapfig,bbm}
\usepackage{enumitem,color}
\usepackage[title]{appendix}
\usepackage{booktabs} 
\usepackage{array}
\usepackage{amsmath,amsthm,tabu}
\usepackage{amssymb}
\usepackage{amstext}
\usepackage{cite,setspace}
\usepackage[linesnumbered,ruled]{algorithm2e}
\usepackage{pdflscape}

\newtheorem{theorem}{Theorem}[section]
\newtheorem{lemma}[theorem]{Lemma}

\newtheorem{prop}[theorem]{\textbf{Proposition}}
\newtheorem{hypo}[theorem]{\textbf{Hypothesis}}
\newtheorem{remark}[theorem]{Remark}
\newtheorem{fact}[theorem]{Fact}
\newtheorem{definition}[theorem]{\textbf{Definition}}

\newenvironment{IEEEproof}{\proof}{\endproof}

\begin{document}

\title{Symmetry, Outer Bounds, and Code Constructions: A Computer-Aided Investigation on the Fundamental Limits of Caching}
\author{Chao Tian}
\maketitle

\begin{abstract}
We illustrate how computer-aided methods can be used to investigate the fundamental limits of the caching systems, which~are significantly different from the conventional analytical approach usually seen in the information theory literature.
The linear programming (LP) outer bound of the entropy space serves as the starting point of this approach; however, our effort goes significantly beyond using it to prove information inequalities. We first identify and formalize the symmetry structure in the problem, which~enables us to show the existence of optimal symmetric solutions. A symmetry-reduced linear program is then used to identify the boundary of the memory-transmission-rate tradeoff for several small cases, for which we obtain a set of tight outer bounds. General hypotheses on the optimal tradeoff region are formed from these computed data, which~are then analytically proven. This~leads to a complete characterization of the optimal tradeoff for systems with only two users, and~certain partial characterization for systems with only two files. Next, we show that by carefully analyzing the joint entropy structure of the outer bounds for certain cases, a novel code construction can be reverse-engineered, which~eventually leads to a general class of codes. Finally, we show that outer bounds can be computed through strategically relaxing the LP in different ways, which~can be used to explore the problem computationally. This~allows us firstly to deduce generic characteristic of the converse proof, and~secondly to compute outer bounds for larger problem cases, despite the seemingly impossible computation scale.
\end{abstract}

\section{Introduction}

We illustrate how computer-aided methods can be used to investigate the fundamental limits of the caching systems, which~is in clear contrast to the conventional analytical approach usually seen in the information theory literature. The~theoretical foundation of this approach can be traced back to the linear programming (LP) outer bound of the entropy space~\cite{Yeung:97}. The~computer-aided approach has been previously applied in~\cite{Tian:JSAC13,TianLiu:15,Tian:15-2,li2017multilevel} on distributed data storage systems to derive various outer bounds, which in many cases are tight. In~this work, we first show that the same general methodology can be tailored to the caching problem effectively to produce outer bounds in several cases, but more importantly, we show that data obtained through computation can be used in several different manners to deduce meaningful structural understanding of the fundamental limits and optimal code constructions. 

The computer-aided investigation and exploration methods we propose are quite general; however, we tackle the caching problem in this work. Caching systems have attracted much research attention recently. In~a nutshell, caching is a data management technique that can alleviate the communication burden during peak traffic time or data demand time, by prefetching and prestoring certain useful content at the users' local caches. Maddah-Ali and Niesen~\cite{MaddahAliNiesen:14} recently considered the problem in an information theoretical framework, where the fundamental question is the optimal tradeoff between local cache memory capacity and the content delivery transmission rate. It was shown in~\cite{MaddahAliNiesen:14} that coding can be very beneficial in this setting, while uncoded solutions suffer a significant loss. Subsequent works extended it to decentralized caching placements~\cite{MaddahAliNiesen:14Networking}, caching~with nonuniform demands~\cite{niesen2017coded}, online caching placements~\cite{pedarsani2016online}, hierarchical caching~\cite{karamchandani2016hierarchical}, caching with random demands~\cite{ji2017order}, among other things. There have been significant research activities recently~\cite{ghasemi2017improved,sengupta2017improved,ajaykrishnan2015critical,chen2016fundamental,Sahraei:15,Amiri:17,Wan:16,yu2018exact, tian2018caching,gomez2018fundamental} in both refining the outer bounds and finding stronger codes for caching. Despite these efforts, the~fundamental tradeoff had not been fully characterized except for the case with only two users and two files~\cite{MaddahAliNiesen:14} before our work. This~is partly due to the fact that the main focus of the initial investigations~\cite{MaddahAliNiesen:14,MaddahAliNiesen:14Networking,niesen2017coded,pedarsani2016online} was on systems operating in the regime where the number of files and the number of users are both large, for which the coded solutions can provide the largest gain over the uncoded counterpart. However, in many applications, the~number of simultaneous data requests can be small, or the collection of users or files need to be divided into subgroups in order to account for various service and request inhomogeneities; see, {e.g.,}~\cite{niesen2017coded}. More importantly, precise and conclusive results on such cases with small numbers of users or files can provide significant insights into more general cases, as we shall show in this work. 

In order to utilize the computational tool in this setting, the~symmetry structure in the problem needs be understood and used to reduce the problem to a manageable scale. The~symmetry-reduced LP is then used to identify the boundary of the memory-transmission-rate tradeoff for several cases. General hypotheses on the optimal tradeoff region are formed from these data, which~are then analytically proven. This~leads to a complete characterization of the optimal tradeoff for systems with two users, and~certain partial characterization for systems with two files. Next, we show that by carefully analyzing the joint entropy structure of the outer bounds, a novel code construction can be reverse-engineered, which~eventually leads to a general class of codes. Moreover, data can also be used to show that a certain tradeoff pair is not achievable using linear codes. Finally, we show that outer bounds can be computed through strategically relaxing the LP in different ways, which~can be used to explore the problem computationally. This~allows us firstly to deduce generic characteristic of the converse proof, and~secondly to compute outer bounds for larger problem cases, despite the seemingly impossible computation scale.

Although some of the tightest and most conclusive results on the optimal memory-transmission-rate tradeoff in caching systems are presented in this work, our main focus is in fact to present the generic computer-aided methods that can be used to facilitate information theoretic investigations in a practically-important research problem setting. For this purpose, we~will provide the necessary details on the development and the rationale of the proposed techniques in a semi-tutorial (and thus less concise) manner. The~most important contribution of this work is three methods for the investigation of fundamental limits of information systems: (1) computational and data-driven converse hypothesis, (2) reverse-engineering optimal codes, and~(3) computer-aided exploration. We believe that these methods are sufficiently general, such that they can be applied to other coding and communication problems, particularly those related to data storage and management.

The rest of the paper is organized as follows. In~Section \ref{sec:pre}, existing results on the caching problem and some background information on the entropy LP framework are reviewed. The~symmetry structure of the caching problem is explored in detail in Section \ref{sec:symmetry}. In~Section \ref{sec:hypo}, we show how the data obtained through computation can be used to form hypotheses, and~then analytically prove them. In~Section \ref{sec:reverseEng}, we show that the computed data can also be used to facilitate reverse-engineering new codes, and~also to prove that a certain memory-transmission-rate pair is not achievable using linear codes. In~Section \ref{sec:generalization}, we provide a method to explore the structure of the outer bounds computationally, to obtain insights into the problem and derive outer bounds for large problem cases. A few concluding remarks are given in Section \ref{sec:conclusion}, and~technical proofs and some computer-produced proof tables are relegated to the~Appendix. 

\section{Preliminaries}
\label{sec:pre}

\begin{figure}[tb]
\centering
\includegraphics[width=9cm]{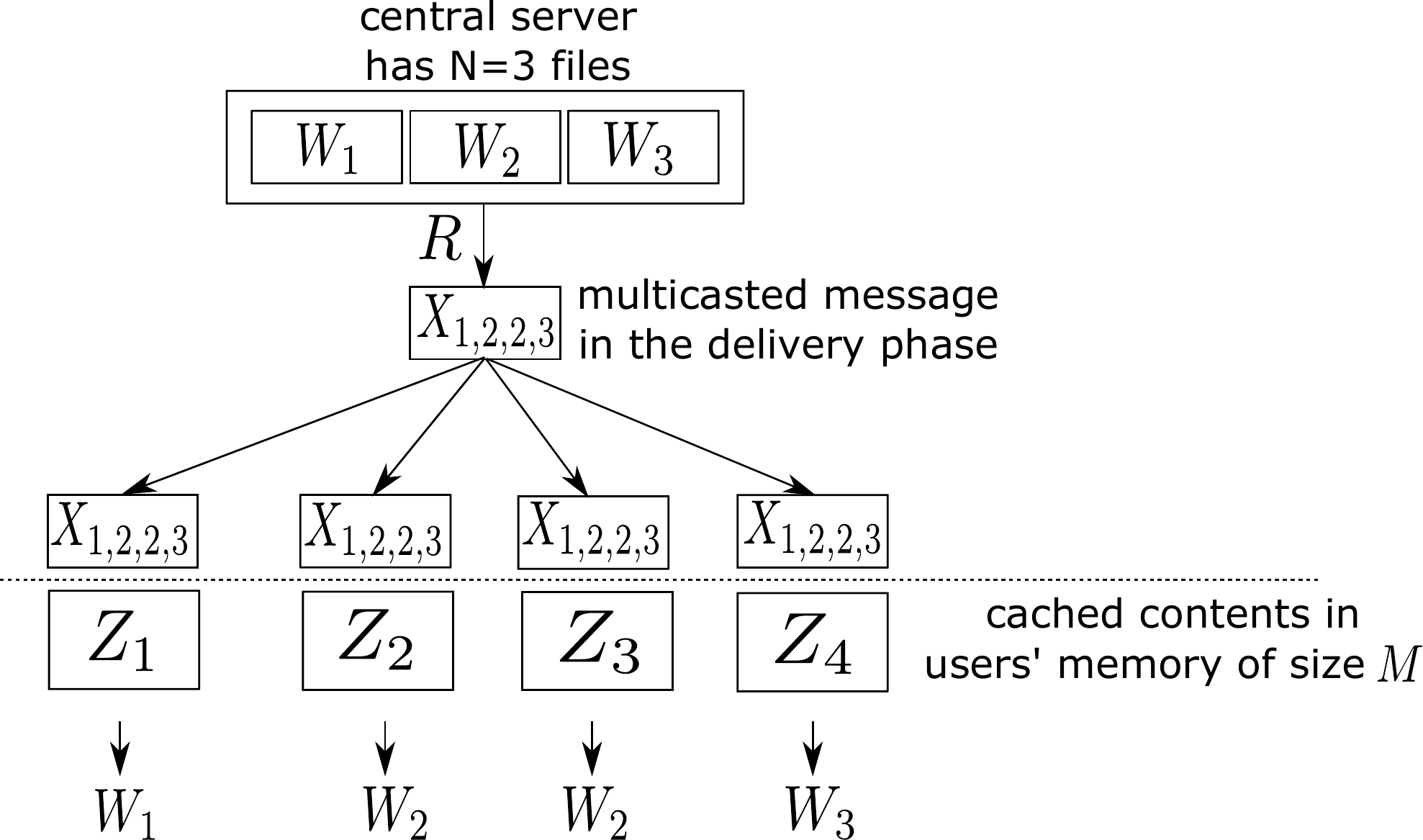}
\caption{An example caching system, where there are $N=3$ files and $K=4$ users. In this case the users request files $(1,2,2,3)$, respectively, and thus the multicast common information is written as $X_{1,2,2,3}$.\label{fig:system}}
\end{figure}

\subsection{The Caching System Model}

There are a total of  $N$ mutually independent files of equal size and $K$ users in the system. 
The overall system operates in two phases: in the placement phase, each user stores in his local cache some content from these files; in the delivery phase, each user will request one file, and the central server transmits (multicasts) certain common content to all the users to accommodate their requests. Each user has a local cache memory of capacity $M$, and the contents stored in the placement phase are determined without knowing a priori the precise requests in the delivery phase. The system should minimize the amount of multicast information which has rate $R$ for all possible combinations of user requests, under the memory cache constraint $M$, both of which are measured as multiples of the file size $F$. The primary interest of this work is the optimal tradeoff between $M$ and $R$. In the rest of the paper, we shall refer to a specific combination of the file requests of all users together as a {\em demand}, or a {\em demand pattern}, and reserve the word ``request'' as the particular file a user needs. Fig. \ref{fig:system} provides an illustration of the overall system.

Since we are investigating the fundamental limits of the caching systems in this work, the~notation for the various quantities in the systems needs to be specified. The~$N$ files in the system are denoted as $\mathcal{W}\triangleq\{W_1,W_2,\ldots,W_N\}$; the~cached contents at the $K$ users are denoted as $\mathcal{Z}\triangleq\{Z_1,Z_2,\ldots,Z_K\}$; and~the transmissions to satisfy a given demand are denoted as $X_{d_1,d_2,\ldots,d_K}$, {i.e.}, the~transmitted information $X_{d_1,d_2,\ldots,d_K}$ when user $k$ requests file $W_{d_k}$, $k=1,2,\ldots,K$. For simplicity, we shall write $(W_1,W_2,\ldots,W_n)$ simply as $W_{[1:n]}$, and $(d_1,d_2,\ldots,d_K)$ as $d_{[1:K]}$; when there are only two users in the system, we write $(X_{i,1},X_{i,2},\ldots,X_{i,j})$ as $X_{i,[1:j]}$. There are other simplifications of the notation for certain special cases of the problem, which~will be introduced as they become necessary.

The cache content at the $k$-th user is produced directly from the files through the encoding function $f_k$, and~the transmission content from the files through the encoding function $g_{d_{[1:K]}}$, {i.e.,}
$$Z_k=f_k(W_{[1:N]}),\quad X_{d_{[1:K]}}=g_{d_{[1:K]}}(W_{[1:N]}),$$
the second of which depends on the particular demands $d_{[1:K]}$. Since the cached contents and transmitted information are both deterministic functions of the files, we have:
\begin{align}
H(Z_kW_1,W_2,\ldots,W_N)&=0,\quad k=1,2,\ldots,K,\\
H(X_{d_1,d_2,\ldots,d_K}W_1,W_2,\ldots,W_N)&=0,\quad d_k\in \{1,2,\ldots,N\}.\label{eqn:encoding}
\end{align}

It is also clear that: 
\begin{align}
H(W_{d_k}Z_k, X_{d_1,d_2,\ldots,d_K})=0, \label{eqn:reconstruction}
\end{align} 
{i.e.,} the file $W_{d_k}$ is a function of the cached content $Z_k$ at user $k$ and the transmitted information when user $k$ requests $W_{d_k}$. 
The memory satisfies the constraint:
\begin{align}
M\geq H(Z_i),\quad i\in \{1,2,\ldots,K\}, \label{eqn:memoryconstraint}
\end{align}
and the transmission rate satisfies:
\begin{align}
R\geq H(X_{d_1,d_2,\ldots,d_K}),\quad d_k\in \{1,2,\ldots,N\}.\label{eqn:transmissionconstraint}
\end{align}

Any valid caching code must satisfy the specific set of conditions in (\ref{eqn:encoding})--(\ref{eqn:transmissionconstraint}). 
A slight variant of the problem definition allows vanishing probability of error, {i.e.,} the probability of error asymptotically approaches zero as $F$ goes to infinity; all the outer bounds derived in this work remain valid for this variant with appropriate applications of Fano's inequality~\cite{CoverThomas}.

\subsection{Known Results on Caching Systems}

The first achievability result on this problem was given in~\cite{MaddahAliNiesen:14}, which~is directly quoted below. 

\begin{theorem}[Maddah-Ali and Niesen~\cite{MaddahAliNiesen:14}] For $N$ files and $K$ users each with a cache size $M\in \{0,N/K,2N/K,\ldots,N\}$, \label{theorem:AliNiesen}
\begin{align}
R=K(1-M/N)\cdot \min\left\{\frac{1}{1+KM/N},\frac{N}{K}\right\}
\end{align}
is achievable. For general $0\leq M\leq N$, the~lower convex envelope of these $(M,R)$ points is achievable. 
\end{theorem}

The first term in the minimization is achieved by the scheme of uncoded placement together with coded transmission~\cite{MaddahAliNiesen:14}, while the latter term is by simple uncoded placement and uncoded transmission. More recently, Yu {et al.}~\cite{yu2018exact} provided the optimal solution when the placement is restricted to be uncoded. Chen {et al.}~\cite{chen2016fundamental} extended a special scheme for the case $N=K=2$ discussed in~\cite{MaddahAliNiesen:14} to the general case $N\leq K$, and~showed that the tradeoff pair $\left(\frac{1}{K},\frac{N(K-1)}{K}\right)$ is achievable. There~were also several other notable efforts in attempting to find better binary codes~\cite{Sahraei:15,Amiri:17,Wan:16, gomez2018fundamental}. Tian and Chen~\cite{tian2018caching} proposed a class of codes for $N\leq K$, the~origin of which will be discussed in more details in Section \ref{sec:reverseEng}. G{\'o}mez-Vilardeb{\'o}~\cite{gomez2018fundamental} also proposed a new code, which can provide further improvement in the small cache memory regime. Tradeoff points achieved by the codes in~\cite{tian2018caching} can indeed be optimal in some cases. 
It is worth noting that while all the schemes~\cite{MaddahAliNiesen:14,chen2016fundamental,Sahraei:15, Amiri:17, Wan:16,yu2018exact,gomez2018fundamental} are binary codes, the~codes in~\cite{tian2018caching} use a more general finite field.

A cut-set outer bound was also given in~\cite{MaddahAliNiesen:14}, which~is again directly quoted below. 
\begin{theorem}[Maddah-Ali and Niesen~\cite{MaddahAliNiesen:14}] For $N$ files and $K$ users each with a cache size $0\leq M\leq N$, \label{theorem:AliNiesenOuter}
\begin{align}
R\geq \max_{s\in\{1,2,\ldots,\min\{N,K\}\}}\left(s-\frac{sM}{\lfloor N/s\rfloor}\right).
\end{align}
\end{theorem}

Several efforts to improve this outer bound have also been reported, which~have led to more accurate approximation characterizations of the optimal tradeoff~\cite{sengupta2017improved,ghasemi2017improved, ajaykrishnan2015critical}. However, as mentioned earlier, even for the simplest cases beyond $(N,K)=(2,2)$, complete characterizations was not available before our work (firstly reported in~\cite{Tian:16symmetry}). In~this work, we specifically treat such small problem cases, and~attempt to deduce more generic properties and outer bounds from these cases. Some of the most recent work~\cite{yu2017characterizing,wang2017improved} that were obtained after the publication of our results~\cite{Tian:16symmetry} provide even more accurate approximations, the~best of which at this point of time is roughly a factor of $2$~\cite{yu2017characterizing}.

\subsection{The Basic Linear Programming Framework}

The basic linear programing bound on the entropy space was introduced by Yeung~\cite{Yeung:97}, which~can be understood as follows. Consider a total of $n$ discrete random variables $(X_1,X_2,\ldots,X_n)$ with a given joint distribution. There are a total of $2^n-1$ joint entropies, each associated with a non-empty subset of these random variables. It is known that the entropy function is monotone and submodular, and~thus, any valid $(2^n-1)$ dimensional entropy vector must have the properties associated with such monotonicity and submodularity, which~can be written as a set of inequalities. Yeung showed (see,~{e.g.},~\cite{Yeung:book}) that the minimal sufficient set of such inequalities is the so-called elemental inequalities:
\begin{align}
&H(X_i\{X_k, k\neq i\})\geq 0,\quad i\in \{1,2,\ldots,n\}\label{eqn:Shannontype1}\\
&I(X_i;X_j\{X_k, k\in \Phi\})\geq 0, \, \mbox{where } \Phi\subseteq \{1,2,\ldots,n\}\setminus\{i,j\},\, i\neq j.\label{eqn:Shannontype2}
\end{align}

The $2^n-1$ joint entropy terms can be viewed as the variables in a linear programming (LP) problem, and~there is a total of $n+{n \choose 2}2^{n-2}$ constraints in (\ref{eqn:Shannontype1}) and (\ref{eqn:Shannontype2}). In~addition to this generic set of constraints, each specific coding problem will place additional constraints on the joint entropy values. These can be viewed as a constraint set of the given problem, although the problem might also induce constraints that are not in this form or even not possible to write in terms of joint entropies. For example, in the caching problem, the~set of random variables are $\{W_i,i=1,2,\ldots,N\}\cup\{Z_i,i=1,2,\ldots,K\}\cup\{X_{d_1,d_2,\ldots,d_K}: d_k\in \{1,2,\ldots,N\}\}$, and~there is a total of $2^{N+K+N^K}-1$ variables in this LP; the~problem-specific constraints are those in (\ref{eqn:encoding})--(\ref{eqn:transmissionconstraint}), and~there are $N+K+N^K+{N+K+N^K\choose 2}2^{N+K+N^K-2}$ elemental entropy constraints, which~is in fact doubly exponential in the number of users $K$. 

\subsection{A Computed-Aided Approach to Find Outer Bounds}

In principle, with the aforedescribed constraint set, one can simply convert the outer bounding problem into an LP (with an objective function $R$ for each fixed $M$ in the caching problem, or more generally a linear combination of $M$ and $R$), and~use a generic LP solver to compute it. Unfortunately, despite the effectiveness of modern LP solvers, directly applying this approach on an engineering problem is usually not possible, since the scale of the LP is often very large even for simple settings. For example, for the caching problem, when $N=2, K=4$, there are overall 200 million elemental inequalities. The~key observation used in~\cite{Tian:JSAC13} to make the problem tractable is that the LP can usually be significantly reduced, by taking into account the symmetry and the implication relations in the~problem. 

The details of the reductions can be found in~\cite{Tian:JSAC13}, and~here, we only provide two examples in the context of the caching problem to illustrate the basic idea behind these reductions:
\begin{itemize}[topsep=3pt,itemsep=2pt]
\item Assuming the optimal codes are symmetric, which~will be defined more precisely later, the~joint entropy $H(W_2,Z_3,X_{2,3,3})$ should be equal to the joint entropy $H(W_1,Z_2,X_{1,2,2})$. This~implies that in the LP, we can represent both quantities using a single variable.
\item Because of the relation (\ref{eqn:reconstruction}), the~joint entropy $H(W_2,Z_3,X_{2,3,3})$ should be equal to the joint entropy $H(W_2,W_3,Z_3,X_{2,3,3})$. This~again implies that in the LP, we can represent both quantities using a single variable. 
\end{itemize}

The reduced primal LP problem is usually significantly smaller, which~allows us to find a lower bound for the tradeoff region for a specific instance with fixed file sizes. Moreover, after identifying the region of interest using these computed boundary points, a human-readable proof can also be produced computationally by invoking the dual of the LP given above. Note a feasible and bounded LP always has a rational optimal solution when all the coefficients are rational, and~thus, the bound will have rational coefficients. More details can again be found in~\cite{Tian:JSAC13}; however, this procedure can be intuitively viewed as follows. Suppose a valid outer bound in the constraint set has the form of: 
\begin{align}
\sum_{\Phi\subseteq \{1,2,\ldots,n\}}\alpha_{\Phi}H(X_k,k\in \Phi)\geq 0, \label{eqn:conjectureIneq}
\end{align}
then it must be a linear combination of the known inequalities, {i.e.}, (\ref{eqn:Shannontype1}) and (\ref{eqn:Shannontype2}), and~the problem-specific constraints, {e.g.}, (\ref{eqn:encoding})--(\ref{eqn:transmissionconstraint}) for the caching problem. To find a human-readable proof is essentially to find a valid linear combination of these inequalities, and~for the conciseness of the proof, the~sparsest linear combination is preferred. By utilizing the LP dual with an additional linear objective, we can find within all valid combinations a sparse (but not necessarily the sparsest) one, which~can yield a concise proof of the inequality (\ref{eqn:conjectureIneq}). 

It~should be noted that in \cite{Tian:JSAC13}, the region of interest was obtained by first finding a set of fine-spaced points on the boundary of the outer bound using the reduced LP, and then manually identifying the effective bounding segments using these boundary points. This~task can however be accomplished more efficiently using an approach proposed by Lassez and Lassez~\cite{Lassez:92}, as pointed out in \cite{Apte:15}. This prompted the author to implement this part of the computer program using this more efficient approach. For completeness, the~specialization of the Lassez algorithm to the caching problem, which~is much simplified in this setting, is provided in Appendix \ref{appendix:Lassez}.

The proof found through this approach can be conveniently written in a matrix to list all the linear combination coefficients, and~one can easily produce a chain of inequalities using such a table to obtain a more conventional human-readable proof. This~approach of generating human-readable proofs has subsequently been adopted by other researchers~\cite{li2017multilevel,Ho:14}. Though we shall present several results thus obtained in this current work in the tabulation form, our main goal is to use these results to present the computer-aided approach, and~show the effectiveness of our approach.

\section{Symmetry in the Caching Problem}
\label{sec:symmetry}

The computer-aided approach to derive outer bounds mentioned earlier relies critically on the reduction of the basic entropy LP using symmetry and other problem structures. In~this section, we consider the symmetry in the caching problem. Intuitively, if we place the cached contents in a permuted manner at the users, it will lead to a new code that is equivalent to the original one. Similarly, if we reorder the files and apply the same encoding function, the~transmissions can also be changed accordingly to accommodate the requests, which~is again an equivalent code. The~two types of symmetries can be combined, and~they induce a permutation group on the joint entropies of the subsets of the random variables $\mathcal{W}\cup\mathcal{Z}\cup\mathcal{X}$. 

For concreteness, we may specialize to the case $(N,K)=(3,4)$ in the discussion, and~for this case:
\begin{align}
&\mathcal{W}=\{W_1,W_2,W_3\},\quad \mathcal{Z}= \{Z_1,Z_2,Z_3,Z_4\},\quad \mathcal{X}=\{X_{d_1,d_2,d_3,d_4}:d_k\in \{1,2,3\}\}.
\end{align}

\subsection{Symmetry in User Indexing}

Let a permutation function be defined as $\bar{\pi}(\cdot)$ on the user index set of $\{1,2,\ldots,K\}$, which~reflects a permuted placement of cached contents $\mathcal{Z}$. Let the inverse of $\bar{\pi}(\cdot)$ be denoted as $\bar{\pi}^{-1}(\cdot)$, and~define the permutation on a collection of elements as the collection of the elements after permuting each element individually. The~aforementioned permuted placement of cached contents can be rigorously defined through a set of new encoding functions and decoding functions. Given the original encoding functions $f_k$ and $g_{d_{[1:K]}}$, the~new functions $f^{\bar{\pi}}_k$ and $g^{\bar{\pi}}_{d_{[1:K]}}$ associated with a permutation $\bar{\pi}$ can be defined~as:
\begin{align}
\bar{Z}_k&\triangleq f^{\bar{\pi}}_k(W_{[1:N]})\triangleq f_{\bar{\pi}(k)}(W_{[1:N]})=Z_{\bar{\pi}(k)},\nonumber\\
\bar{X}_{d_{[1:K]}}&\triangleq g^{\bar{\pi}}_{d_{[1:K]}}(W_{[1:N]})\triangleq g_{d_{\bar{\pi}^{-1}([1:K])}}(W_{[1:N]})=X_{d_{\bar{\pi}^{-1}([1:K])}}.\label{eqn:barfunction}
\end{align}

To see that with these new functions, any demand $d_{([1:K])}$ can be correctly fulfilled as long as the original functions can fulfill the corresponding reconstruction task, consider the pair $(f^{\bar{\pi}}_k(W_{[1:N]}),g^{\bar{\pi}}_{d_{[1:K]}}(W_{[1:N]}))$, which~should reconstruct $W_{d_k}$. This~pair is equivalent to the pair $(f_{\bar{\pi}(k)}(W_{[1:N]}),g_{d_{\bar{\pi}^{-1}([1:K])}}(W_{[1:N]}))$, and~in the demand vector $d_{\bar{\pi}^{-1}([1:K])}$, the~$\bar{\pi}(k)$ position is in fact $d_{\bar{\pi}^{-1}(\bar{\pi}(k))}=d_k$, implying that the new coding functions are indeed valid.

We can alternatively view $\bar{\pi}(\cdot)$ as directly inducing a permutation on $\mathcal{Z}$ as $\bar{\pi}(Z_k)=Z_{\bar{\pi}(k)}$, and~a permutation on $\mathcal{X}$ as: 
\begin{align}
\bar{\pi}(X_{d_1,d_2,\ldots,d_K})=X_{d_{\bar{\pi}^{-1}(1)},d_{\bar{\pi}^{-1}(2)},\ldots,d_{\bar{\pi}^{-1}(K)}}.
\end{align}

For example, the~permutation function $\bar{\pi}(1)=2,\bar{\pi}(2)=3,\bar{\pi}(3)=1,\bar{\pi}(4)=4$ will induce:
\begin{align}
(d_1,d_2,d_3,d_4)\rightarrow (\bar{d}_1,\bar{d}_2,\bar{d}_3,\bar{d}_4)=(d_3,d_1,d_2,d_4).
\end{align}

Thus, it will map $Z_1$ to $\bar{\pi}(Z_1)=Z_2$, but map $X_{1,2,3,2}$ to $X_{3,1,2,2}$, $X_{3,2,1,3}$ to $X_{1,3,2,3}$, and~$X_{1,1,2,2}$ to~$X_{2,1,1,2}$. 

With the new coding functions and the permuted random variables defined above, we have the following relation: 
\begin{align}
(\mathcal{W}^{\bar{\pi}},\mathcal{Z}^{\bar{\pi}},\mathcal{X}^{\bar{\pi}}){=}(\mathcal{W},\bar{\pi}(\mathcal{Z}),\bar{\pi}(\mathcal{X})), \label{eqn:equal}
\end{align}
where the superscript $\bar{\pi}$ indicates the random variables induced by the new encoding functions.

We call a caching code user-index-symmetric, if for any subsets $\mathcal{W}_o\subseteq \mathcal{W},\mathcal{Z}_o\subseteq\mathcal{Z},\mathcal{X}_o\subseteq\mathcal{X}$, and~any permutation $\bar{\pi}$, the~following relation holds:
\begin{align}
H(\mathcal{W}_o,\mathcal{Z}_o,\mathcal{X}_o)=H(\mathcal{W}_o,\bar{\pi}(\mathcal{Z}_o),\bar{\pi}(\mathcal{X}_o)).
\end{align} 

For example, for such a symmetric code, the~entropy $H(W_2,Z_2,X_{1,2,3,2})$ under the aforementioned permutation is equal to $H(W_2,Z_3,X_{3,1,2,2})$; note that $W_2$ is a function of $(Z_2,X_{1,2,3,2})$, and~after the mapping, it is a function of $(Z_3,X_{3,1,2,2})$. 

\subsection{Symmetry in File Indexing}

Let a permutation function be defined as $\hat{\pi}(\cdot)$ on the file index set of $\{1,2,\ldots,N\}$, which~reflects a renaming of the files $\mathcal{W}$. This~file-renaming operation can be rigorously defined as a permutation of the input arguments to the functions $f_k$ and $g_{d_{[1:K]}}$. Given the original encoding functions $f_k$ and $g_{d_{[1:K]}}$, the~new functions $f^{\hat{\pi}}_k$ and $g^{\hat{\pi}}_{d_{[1:K]}}$ associated with a permutation ${\hat{\pi}}$ can be defined as:
\begin{align}
\hat{Z}_k&\triangleq f^{\hat{\pi}}_k(W_{[1:N]})\triangleq f_k(W_{\hat{\pi}^{-1}([1:N])}),\nonumber\\
\hat{X}_{d_{[1:K]}}&\triangleq g^{\hat{\pi}}_{d_{[1:K]}}(W_{[1:N]})\triangleq g_{\hat{\pi}(d_{[1:K]})}(W_{\hat{\pi}^{-1}([1:N])}).\label{eqn:hatfunction}
\end{align}

We first show that the pair $(f^{\hat{\pi}}_k(W_{[1:N]}),g^{\hat{\pi}}_{d_{[1:K]}}(W_{[1:N]}))$ can provide reconstruction of $W_{d_k}$. This~pair by definition is equivalent to $(f_k(W_{\hat{\pi}^{-1}([1:N])}),g_{\hat{\pi}(d_{[1:K]})}(W_{\hat{\pi}^{-1}([1:N])}))$, where the $k$-th position of the demand vector $\hat{\pi}(d_{[1:K]})$ is in fact $\hat{\pi}(d_k)$. However, because of the permutation in the input arguments, this implies that the $\hat{\pi}(d_k)$-th file in the sequence $W_{\hat{\pi}^{-1}([1:N])})$ can be reconstructed, which~is indeed~$W_{d_k}$.

Alternatively, we can view $\hat{\pi}(\cdot)$ as directly inducing a permutation on $\hat{\pi}(W_k)=W_{\hat{\pi}(k)}$, and~it also induces a permutation on $\mathcal{X}$ as:
\begin{align}
\hat{\pi}(X_{d_1,d_2,\ldots,d_K})= X_{\hat{\pi}(d_1),\hat{\pi}(d_2),\ldots,\hat{\pi}(d_K)}.
\end{align}

For example, the permutation function $\hat{\pi}(1)=2,\hat{\pi}(2)=3,\hat{\pi}(3)=1$ maps $W_2$ to $\hat{\pi}(W_2)=W_3$, but maps $X_{1,2,3,2}$ to $X_{2,3,1,3}$, $X_{3,2,1,3}$ to $X_{1,3,2,1}$, and~$X_{1,1,2,2}$ to $X_{2,2,3,3}$.

With the new coding functions and the permuted random variables defined above, we have the following equivalence relation:
\begin{align}
(\mathcal{W}^{\hat{\pi}},\mathcal{Z}^{\hat{\pi}},\mathcal{X}^{\hat{\pi}})=&\left(W_{([1:N])},f_{[1:k]}(W_{\hat{\pi}^{-1}([1:N])}),\left\{g_{\hat{\pi}(d_{[1:K]})}(W_{\hat{\pi}^{-1}([1:N])}): d_{[1:K]}\in \mathcal{N}^K\right\}\right)\nonumber\\
\stackrel{d}{=}&\left(W_{\hat{\pi}([1:N])},f_{[1:k]}(W_{[1:N]}),\left\{g_{\hat{\pi}(d_{[1:K]})}(W_{[1:N]}): d_{[1:K]}\in \mathcal{N}^K\right\}\right)\nonumber\\
=&\left(\hat{\pi}(\mathcal{W}),\mathcal{Z},\hat{\pi}(\mathcal{X})\right),\label{eqn:equalindistribution}
\end{align}
where $\stackrel{d}{=}$ indicates equal in distribution, which~is due to the the random variables in $\mathcal{W}$ being independently and identically distributed, thus exchangeable.

We call a caching code file-index-symmetric, if for any subsets $\mathcal{W}_o\subseteq \mathcal{W},\mathcal{Z}_o\subseteq\mathcal{Z},\mathcal{X}_o\subseteq\mathcal{X}$, and~any permutation $\hat{\pi}$, the~following relation holds:
\begin{align}
H(\mathcal{W}_o,\mathcal{Z}_o,\mathcal{X}_o)=H(\hat{\pi}(\mathcal{W}_o),\mathcal{Z}_o,\hat{\pi}(\mathcal{X}_o)).
\end{align} 

For example, for such a symmetric code, $H(W_3,Z_3,X_{1,2,3,2})$ under the aforementioned permutation is equal to $H(W_1,Z_3,X_{2,3,1,3})$; note that $W_3$ is a function of $(Z_3,X_{1,2,3,2})$, and~after the mapping, $W_1$ is a function of $(Z_3,X_{2,3,1,3})$.

\subsection{Existence of Optimal Symmetric Codes}

With the symmetry structure elucidated above, we can now state our first auxiliary result.

\begin{prop}
For any caching code, there is a code with the same or smaller caching memory and transmission rate, which~is both user-index-symmetric and file-index-symmetric. 
\label{prop:symmetry}
\end{prop}

We call a code that is both user-index-symmetric and file-index-symmetric a {symmetric code}. This~proposition implies that there is no loss of generality to consider only symmetric codes. The~proof of this proposition relies on a simple space-sharing argument, where a set of base encoding functions and base decoding function are used to construct a new code. In~this new code, each file is partitioned into a total of $N!K!$ segments, each having the same size as suitable in the base coding functions. The~coding functions obtained as in (\ref{eqn:barfunction}) and (\ref{eqn:hatfunction}) from the base coding functions using permutations $\bar{\pi}$ and $\hat{\pi}$ are used on the $i$-th segments of all the files to produce random variables $\mathcal{W}^{\bar{\pi}\cdot\hat{\pi}}\cup\mathcal{Z}^{\bar{\pi}\cdot\hat{\pi}}\cup\mathcal{X}^{\bar{\pi}\cdot\hat{\pi}}$. Consider a set of random variables $(\mathcal{W}_o\cup\mathcal{Z}_o\cup\mathcal{X}_o)$ in the original code, and~denote the same set of random variables in the new code as $(\mathcal{W}'_o\cup\mathcal{Z}'_o\cup\mathcal{X}'_o)$. We have:
\begin{align}
H(\mathcal{W}'_o\cup\mathcal{Z}'_o\cup\mathcal{X}'_o)=\sum_{\bar{\pi},\hat{\pi}}H(\mathcal{W}^{\bar{\pi}\cdot\hat{\pi}}_o\cup \mathcal{Z}^{\bar{\pi}\cdot\hat{\pi}}_o\cup\mathcal{X}^{\bar{\pi}\cdot\hat{\pi}}_o)=\sum_{\bar{\pi},\hat{\pi}}H(\hat{\pi}(\mathcal{W}_o)\cup\bar{\pi}(\mathcal{Z}_o)\cup\bar{\pi}\cdot\hat{\pi}(\mathcal{X}_o)),
\end{align}
 because of (\ref{eqn:equal}) and (\ref{eqn:equalindistribution}).
Similarly, for another pair of permutations $(\bar{\pi}',\hat{\pi}')$, the~random variables $\hat{\pi}'(\mathcal{W}'_{o})\cup\bar{\pi}'(\mathcal{Z}'_{o})\cup\bar{\pi}'\cdot\hat{\pi}'(\mathcal{X}'_{o})$ in the new code will have exactly the same joint entropy value. It is now clear that the resultant code by space sharing is indeed symmetric, and~it has (normalized) memory sizes and a transmission rate no worse than the original one. A similar argument was used in~\cite{Tian:JSAC13} to show, with a more detailed proof, the existence of optimal symmetric solution in regenerating codes. In~a separate work~\cite{ZhangTian:17TCOM}, we investigated the properties of the induced permutation $\bar{\pi}\cdot\hat{\pi}$, and~particularly, showed that it is isomorphic to the power group~\cite{harary1953graph}; readers are referred to~\cite{ZhangTian:17TCOM} for more details. 

\subsection{Demand Types}

Even for symmetric codes, the~transmissions to satisfy different types of demands may use different rates. For example in the setting $N,K=(3,4)$, $H(X_{1,2,2,2})$ may not be equal to $H(X_{1,1,2,2})$, and~$H(X_{1,2,3,2})$ may not be equal to $H(X_{3,2,3,2})$. The~transmission rate $R$ is then chosen to be the maximum among all cases. This~motivates the notion of demand types. 

\begin{definition}
In an $(N,K)$ caching system, for a specific demand, let the number of users requesting file $n$ be denoted as $m_n$, $n=1,2,\ldots,N$. We call the vector obtained by sorting the values $\{m_1,m_2,\ldots,m_N\}$ in a decreasing order as the demand type, denoted as $\mathcal{T}$.
\end{definition}

Proposition \ref{prop:symmetry} implies that for optimal symmetric solutions, demands of the same type can always be satisfied with transmissions of the same rate; however, demands of different types may still require different rates. This~observation is also important in setting up the linear program in the computer-aided approach outlined in the previous section. Because we are interested in the worst case transmission rate among all types of demands, in the symmetry-reduced LP, an additional variable needs to be introduced to constrain the transmission rates of all possible types. 

For an $(N,K)$ system, determining the number of demand types is closely related to the integer partition problem, which~is the number of possible ways to write an integer $K$ as the sum of positive integers. There is no explicit formula, but one can use a generator polynomial to compute it~\cite{Andrews:book}. For~several small $(N,K)$ pairs, we list the demand types in Table \ref{tab:types}.

\begin{table}
\caption{Demand types for small $(N,K)$ pairs.}
\label{tab:types}

\centering

\begin{tabular}{cc}
\toprule
\textbf{(N,K)}&\textbf{Demand Types}\\\midrule
(2,3)&(3,0),(2,1)\\
(2,4)&(4,0),(3,1),(2,2)\\
(3,2)&(2,0,0),(1,1,0)\\
(3,3)&(3,0,0),(2,1,0),(1,1,1)\\
(3,4)&(4,0,0),(3,1,0),(2,2,0),(2,1,1)\\
(4,2)&(2,0,0,0),(1,1,0,0)\\
(4,3)&(3,0,0,0),(2,1,0,0),(1,1,1,0)\\
\bottomrule
\end{tabular}

\end{table}

It can be seen that when $N\leq K$, increasing $N$ induces more demand types, but this stops when $N>K$; however, increasing $K$ always induces more demand types. This~suggests it might be easier to find solutions for a collection of cases with a fixed $K$ and arbitrary $N$ values, but more difficult for that of a fixed $N$ and arbitrary $K$ values. This~intuition is partially confirmed with our results presented~next.

\section{Computational and Data-Driven Converse Hypotheses}
\label{sec:hypo}

Extending the computational approach developed in~\cite{Tian:JSAC13} and the problem symmetry, in this section, we first establish complete characterizations for the optimal memory-transmission-rate tradeoff for $(N,K)=(3,2)$ and $(N,K)=(4,2)$. Based on these results and the known result for $(N,K)=(2,2)$, we~are able to form a hypothesis regarding the optimal tradeoff for the case of $K=2$. An analytical proof is then provided, which~gives the complete characterization of the optimal tradeoff for the case of $(N,2)$ caching systems. We then present a characterization of the optimal tradeoff for $(N,K)=(2,3)$ and an outer bound for $(N,K)=(2,4)$. These results also motivate a hypothesis on the optimal tradeoff for $N=2$, which~is subsequently proven analytically to yield a partial characterization. Note that since both $M$ and $R$ must be nonnegative, we do not explicitly state their non-negativity from here~on. 

\subsection{The Optimal Tradeoff for $K=2$}

The optimal tradeoff for $(N,K)=(2,2)$ was found in \cite{MaddahAliNiesen:14}, which we restated below.
\begin{prop}[Maddah Ali and Niesen \cite{MaddahAliNiesen:14}]
Any memory-transmission-rate tradeoff pair for the $(N,K)=(2,2)$ caching problem must satisfy 
\begin{align}
2M+R\geq 2,\quad 2M+2R\geq 3,\quad M+2R\geq 2. \label{eqn:bounds2_2}
\end{align}
Conversely, there exist codes for any nonnegative $(M,R)$ pair satisfying (\ref{eqn:bounds2_2}).
\end{prop}

Our investigation thus starts with identifying the previously unknown optimal tradeoff for $(N,K)=(3,2)$ and $(N,K)=(4,2)$ using the computation approach outline in Section \ref{sec:pre}, the results of which are first summarized below as two propositions. 

\begin{prop}
\label{prop:NK32}
Any memory-transmission-rate tradeoff pair for the $(N,K)=(3,2)$ caching problem must satisfy 
\begin{align}
M+R\geq 2, \quad M+3R\geq 3. \label{eqn:bounds3_2}
\end{align}
Conversely, there exist codes for any nonnegative $(M,R)$ pair satisfying (\ref{eqn:bounds3_2}).
\end{prop}

\begin{prop}
\label{prop:NK42}
Any memory-transmission-rate tradeoff pair for the $(N,K)=(4,2)$ caching problem must satisfy 
\begin{align}
3M+4R\geq 8, \quad M+4R\geq 4. \label{eqn:bounds4_2}
\end{align}
Conversely, there exist codes for any nonnegative $(M,R)$ pair satisfying (\ref{eqn:bounds4_2}).
\end{prop}

\begin{figure}[tb]
\centering
\includegraphics[width=17cm]{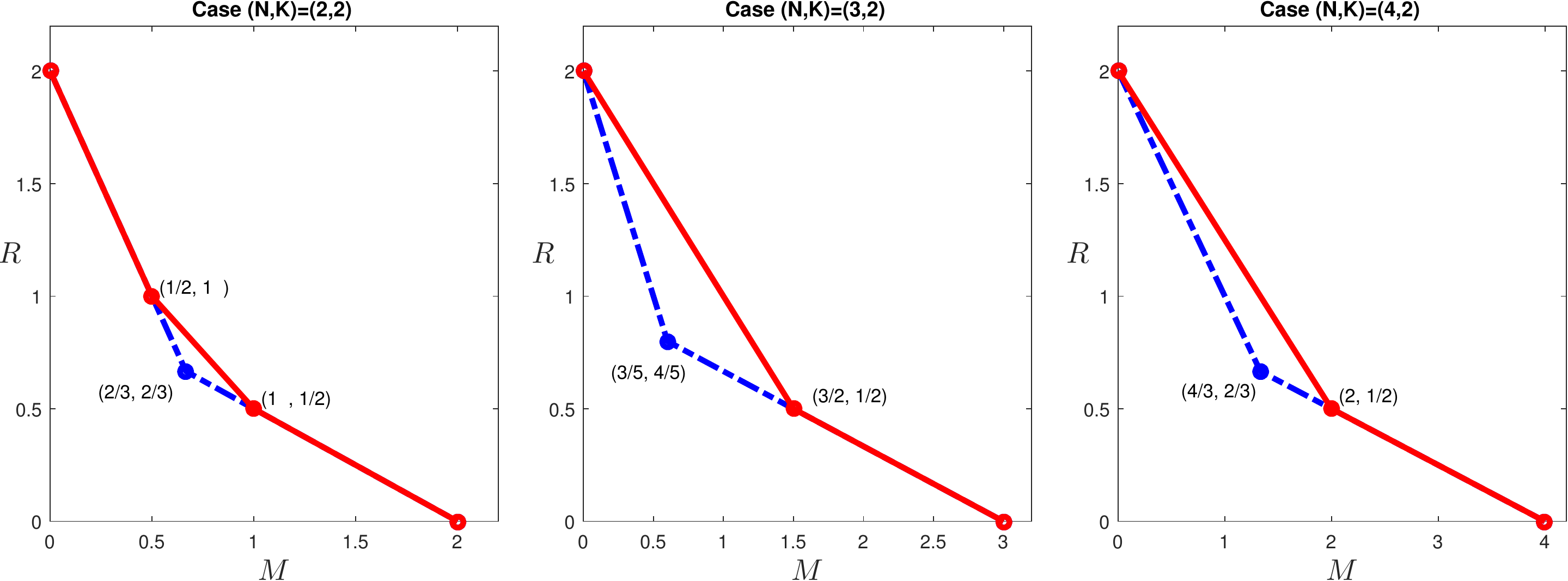}
\caption{The optimal tradeoffs for $(N,K)=(2,2)$, $(N,K)=(3,2)$ and $(N,K)=(4,2)$ caching systems. \label{fig:cachingN_2} The red solid lines give the optimal tradeoffs, while the blue dashed-dot lines are the cut-set outer bounds, included here for reference. }
\end{figure}

The proofs for Proposition \ref{prop:NK32} and Proposition \ref{prop:NK42} can be found in Appendix \ref{appendix:propNK32_42}, which are given in the tabulation format mentioned earlier. Strictly speaking, these two results are specialization of Theorem \ref{theorem:NK_N_2}, and there is no need to provide the proofs separately, however we provide them to illustrate the computer-aided approach.

The optimal tradeoff for these cases are given in Fig. \ref{fig:cachingN_2}. A few immediate observations are as follows
\begin{itemize}[topsep=3pt,itemsep=2pt]
\item For $(N,K)=(3,2)$ and $(N,K)=(4,2)$, there is only one non-trivial corner point on the optimal tradeoff, but for $(N,K)=(2,2)$ there are in fact two non-trivial conner points. 
\item The cut-set bound is tight at the high memory regime in all the cases.
\item The single non-trivial corner point for $(N,K)=(3,2)$ and $(N,K)=(4,2)$ is achieved by the scheme proposed in \cite{MaddahAliNiesen:14}. For the $(N,K)=(2,2)$ case, one of the corner point is achieved also by this scheme, but the other corner point requires a different code. 
\end{itemize}

Given the above observations, a natural hypothesis is as follows.
\begin{hypo}
There is only one non-trivial corner point on the optimal tradeoff for $(N,K)=(N,2)$ caching systems when $N\geq 3$, and it is $(M,R)=(N/2,1/2)$, or equivalently the two facets of the optimal tradeoff should be 
\begin{align}
3M+NR\geq 2N, \quad M+NR\geq N.
\end{align}
\end{hypo}
We are indeed able to analytically confirm this hypothesis, as stated formally in the following theorem.

\begin{theorem}
\label{theorem:NK_N_2}
For any integer $N\geq3$, any memory-transmission-rate tradeoff pair for the $(N,K)=(N,2)$ caching problem must satisfy 
\begin{align}
3M+NR\geq 2N, \quad M+NR\geq N. \label{eqn:boundsN_2}
\end{align}
Conversely, for any integer $N\geq3$, there exist codes for any nonnegative $(M,R)$ pair satisfying (\ref{eqn:boundsN_2}). For $(N,K)=(2,2)$, the memory-transmission-rate tradeoff must satisfy
\begin{align}
2M+R\geq 2,\quad 2M+2R\geq 3,\quad M+2R\geq 2. \label{eqn:bounds2_2b}
\end{align}
Conversely, for $(N,K)=(2,2)$, there exist codes for any nonnegative $(M,R)$ pair satisfying (\ref{eqn:bounds2_2b}).
\end{theorem}

Since the solution for the special case $(N,K)=(2,2)$ was established in \cite{MaddahAliNiesen:14}, we only need to consider the cases for $N\geq 3$. Moreover, for the converse direction, only the bound $3M+NR\geq 2N$ needs to be proved, since the other one can be obtained using the cut-set bound in \cite{MaddahAliNiesen:14}.
To prove the remaining inequality, the following auxiliary lemma is needed.

\begin{lemma}\label{lemma:peeling}
For any symmetric $(N,2)$ caching code where $N\geq 3$, and any integer $n=\{1,2,\ldots, N-2\}$,
\begin{align}
&(N-n)H(Z_1,W_{[1:n]},X_{n,n+1})\geq (N-n-2)H(Z_1,W_{[1:n]})+(N+n).\label{eqn:lemma}
\end{align}
\end{lemma}

Using Lemma \ref{lemma:peeling}, we can prove the converse part of Theorem \ref{theorem:NK_N_2} through an induction; the proofs of Theorem \ref{theorem:NK_N_2} and Lemma \ref{lemma:peeling} can be found in Appendix \ref{appendix:theoremN2}, both of which heavily rely on the symmetry specified in the previous section. Although some clues can be found in the proof tables for the cases $(N,K)=(3,2)$ and $(N,K)=(4,2)$, such as the effective joint entropy terms in the converse proof each having only a small number of $X$ random variables, finding the proof of Theorem \ref{theorem:NK_N_2} still requires considerable human effort, and was not completed directly through a computer program. One key observation simplifying the proof in this case is that as the hypothesis states, the optimal corner point is achieved by the scheme given in \cite{MaddahAliNiesen:14}, which is known only thanks to the computed bounds. In this specific case, the scheme reduces to splitting each file into half, and placing one half at the first user, and the other half at the second user; the corresponding delivery strategy is also extremely simple. We  combined this special structure and the clues from the proof tables to find the outer bounding steps.

\begin{remark} 
The result in~\cite{ghasemi2017improved} can be used to establish the bound $3M+NR\geq 2N$ when $K=2$, however only for the cases when $N$ is an integer multiple of three. For $N=4$, the~bounds developed in~\cite{sengupta2017improved,ajaykrishnan2015critical,ghasemi2017improved} give $M+2R\geq 3$, instead of $3M+4R\geq 8$, and~thus, they are loose in this case. After this bound was initially reported in~\cite{Tian:16symmetry}, Yu~{et al.}~\cite{yu2017characterizing} discovered an alternative proof.
\end{remark} 

\subsection{A Partial Characterization for $N=2$}
We first summarize the characterizations of the optimal tradeoff for $(N,K)=(2,3)$, and the computed outer bound for $(N,K)=(2,4)$, in two propositions. 
\begin{prop}
\label{prop:NK23}
The memory-transmission-rate tradeoff for the $(N,K)=(2,3)$ caching problem must satisfy: 
\begin{align}
2M+R\geq 2, \quad 3M+3R\geq 5,\quad M+2R\geq2. \label{eqn:bounds2_3}
\end{align}
Conversely, there exist codes for any nonnegative $(M,R)$ pair satisfying (\ref{eqn:bounds2_3}). 
\end{prop}

\begin{prop}
\label{prop:NK24}
The memory-transmission-rate tradeoff for the $(N,K)=(2,4)$ caching problem must satisfy: 
\begin{align}
2M+R\geq 2, \quad 14M+11R\geq 20,\quad  9M+8R\geq14, \quad 3M+3R\geq 5, \quad 5M+6R\geq 9, \quad M+2R\geq2. \label{eqn:bounds2_4}
\end{align} 
\end{prop}

\begin{figure}[tb]
\centering
\includegraphics[width=17cm]{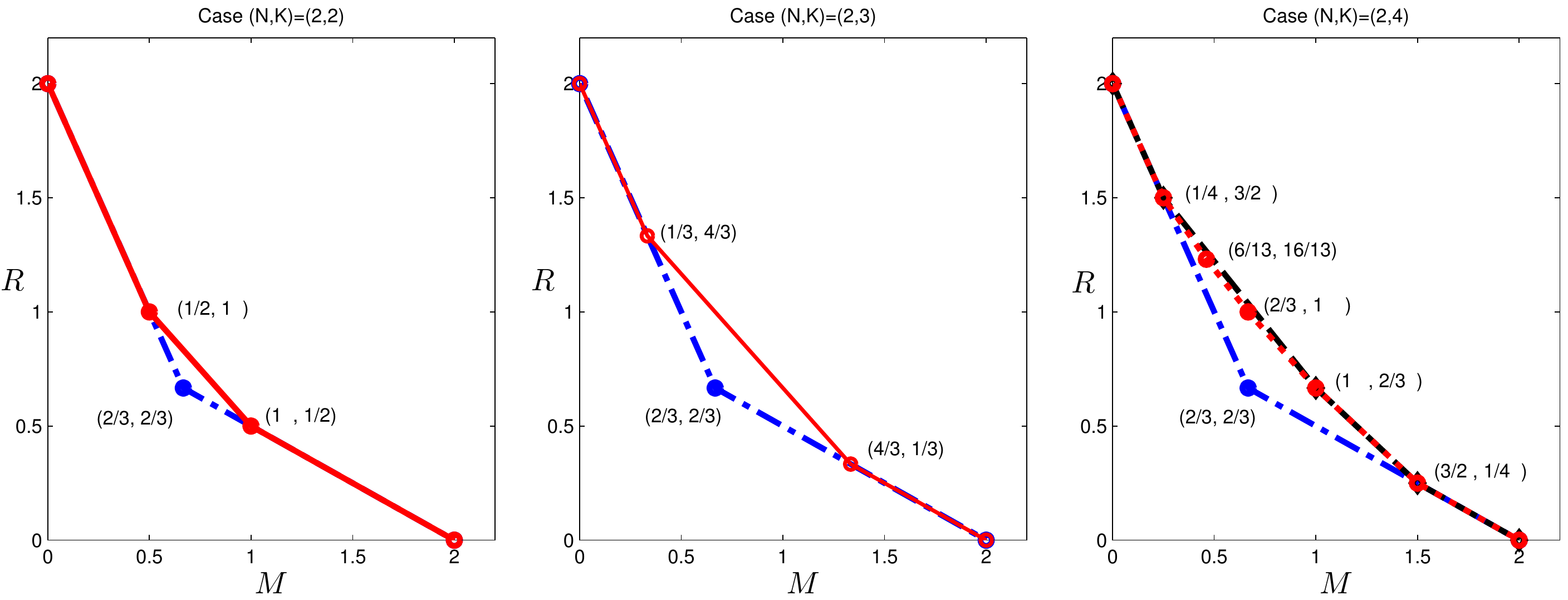}
\caption{The optimal tradeoffs for $(N,K)=(2,2)$, $(N,K)=(2,3)$ and computed outer bound $(N,K)=(2,4)$ caching systems. \label{fig:caching2_N} The red solid lines give the optimal tradeoffs for the first two case, and the red dotted line gives the computed outer bound $(N,K)=(2,4)$; The blue dashed-dot lines are the cut-set outer bounds, and the black dashed line is the inner bound using the scheme in \cite{MaddahAliNiesen:14} and \cite{chen2016fundamental}.}
\end{figure}

For Proposition \ref{prop:NK23}, the only new bound $3M+3R\geq 5$ is a special case of the more general result of Theorem \ref{theorem:2K} and we thus do not provide this proof separately. For Proposition \ref{prop:NK24}, only the second and the third inequalities need to be proved, since the fourth coincides with a bound in the $(2,3)$ case, the fifth is a special case of Theorem \ref{theorem:2K}, and the others can be produced from the cut-set bounds. The proofs for these two inequalities given in Appendix \ref{appendix:Prop4_9}. The optimal tradeoff for $(N,K)=(2,2),(2,3)$ and the outer bound for $(2,4)$ are depicted in Fig. \ref{fig:caching2_N}. A few immediate observations and comments are as follows:
\begin{itemize}[topsep=3pt,itemsep=2pt]
\item There are two non-trivial corner points on the outer bounds for $(N,K)=(2,2)$ and $(N,K)=(2,3)$, and there are five non-trivial corner points for $(N,K)=(2,4)$. 
\item The outer bounds coincide with known inner bounds for $(N,K)=(2,2)$ and $(N,K)=(2,3)$, but not $(N,K)=(2,4)$. The corner points at $R=1/K$  (and the corner point $(1,2/3)$ for $(N,K)=(2,4)$) are achieved by the scheme given in \cite{MaddahAliNiesen:14}, while the corner points at $M=1/K$ are achieved by the scheme given in \cite{chen2016fundamental}. For $(N,K)=(2,4)$, two corner points at the intermediate memory regime cannot be achieved by either the scheme in \cite{MaddahAliNiesen:14} or that in \cite{chen2016fundamental}. 
\item The cut-set outer bounds \cite{MaddahAliNiesen:14} are tight at the highest and lowest memory segments; a new bound for the second highest memory segment produced by the computer based method is also tight.  
\end{itemize}

\begin{remark}
The bounds developed in~\cite{sengupta2017improved,ajaykrishnan2015critical,ghasemi2017improved} give $2(M+R)\geq 3$ for $(N,K)=(2,3)$ and $(N,K)=(2,4)$, instead of $3M+3R\geq 5$, and~thus, they are loose in this case. When specializing the bounds in~\cite{yu2017characterizing}, it matches Proposition \ref{prop:NK23} for $(N,K)=(2,3)$, but it is weaker than Proposition \ref{prop:NK24} for $(N,K)=(2,4)$.
\end{remark}

From the above observations, we can hypothesize that for $N=2$, the number of corner points will continue to increase as $K$ increases above $4$, and at the high memory regime, the scheme \cite{MaddahAliNiesen:14} is optimal. More precisely, the following hypothesis appear to be a natural first step.
\begin{hypo}
\label{hypo:2K}
The first two non-trivial $(M,R)$ corner points of the optimal tradeoff for $(N,K)=(2,K)$ at the high memory regime when $K\geq 4$ are 
\begin{align}
\left(\frac{2(K-1)}{K},\frac{1}{K}\right)\quad \mbox{and}\quad\left(\frac{2(K-2)}{K},\frac{2}{K-1}\right). \label{eqn:twopoints}
\end{align}
Conversely,  when $K\geq 4$ and $N=2$, any $(M,R)$ pair must satisfy
\begin{align}
K(K+1)M+2(K-1)KR\geq 2(K-1)(K+2),
\end{align}
which is the line passing through the two corner points in (\ref{eqn:twopoints}).
\end{hypo}

This following theorem confirms that the hypothesis is indeed true.
\begin{theorem}\label{theorem:2K}
When $K\geq 3$ and $N=2$, any $(M,R)$ pair must satisfy 
\begin{align}
K(K+1)M+2(K-1)KR\geq 2(K-1)(K+2).\label{eqn:2K}
\end{align}
As a consequence, the uncoded-placement-coded-transmission scheme in \cite{MaddahAliNiesen:14} (with space-sharing) is optimal when $M\geq \frac{2(K-2)}{K}$, for the cases with $K\geq 4$ and $N=2$.
\end{theorem}

The first line segment at the high memory regime is $M+2R\geq 2$, which is given by the cut-set bound; its intersection with  (\ref{eqn:2K}) is indeed the first point in (\ref{eqn:twopoints}). The proof of this theorem now boils down to the proof of the bound (\ref{eqn:2K}). This requires a sophisticated induction, the digest of which is summarized in the following lemma. The symmetry of the problem is again heavily utilized throughout of the proof of this lemma. For notational simplicity, we use $X_{\rightarrow j}$ to denote $X_{1,1,\ldots,1,2,1,\ldots,1}$, {\em i.e.}, when the $j$-t user requests the second file, and all the other users request the first file; we also write a collection of such variables $(X_{\rightarrow j}, X_{\rightarrow j+1},\ldots,X_{\rightarrow k})$ as $X_{\rightarrow [j:k]}$. 
\begin{lemma}
\label{lemma:2K}
For $N=2$ and $K\geq 3$, the following inequality holds for $k\in \{2,3,\ldots,K-1\}$
\begin{align}
&(K-k+1)(K-k+2)H(Z_1,W_1,X_{\rightarrow [2:k]})\nonumber\\
&\,\geq [{(K-k)(K-k+1)}-2]H(Z_1,W_1,X_{\rightarrow [2:k-1]})+2H(W_1,X_{\rightarrow [2:k-1]})+2(K-k+1)H(W_1,W_2),\label{eqn:lemma2K}
\end{align}
where we have taken the convention $H(Z_1,W_1,X_{\rightarrow [2:1]})=H(Z_1,W_1)$
\end{lemma}
The proof of Lemma \ref{lemma:2K} is given in Appendix \ref{appendix:lemma2K}. Theorem \ref{theorem:2K} can now be proved straightforwardly. 

\begin{IEEEproof}[Proof of Theorem \ref{theorem:2K}]
We first write the following simple inequalities
\begin{align}
H(Z_1)+H(X_{\rightarrow 2})\geq H(Z_1,X_{\rightarrow 2})=H(Z_1,W_1,X_{\rightarrow 2}).
\end{align}
Now applying Lemma \ref{lemma:2K} with $k=2$ gives
\begin{align}
&(K-1)K[H(Z_1)+H(X_{\rightarrow 2})]\nonumber\\
&\quad\geq [K^2-3K]H(Z_1,W_1)+2H(W_1)+2(K-1)H(W_1,W_2).
\end{align}
Observe that
\begin{align}
H(Z_1,W_1)=H(W_1|Z_1)+H(Z_1)\geq \frac{1}{2}H(W_1,W_2|Z_1)+H(Z_1)=\frac{1}{2}H(W_1,W_2)+\frac{1}{2}H(Z_1),
\end{align}
where in the first inequality the file index symmetry $H(W_1|Z_1)=H(W_2|Z_1)$ has been used.
We can now continue to write
\begin{align}
&(K-1)K[H(Z_1)+H(X_{\rightarrow 2})]\nonumber\\
&\quad\geq \frac{K^2-3K}{2}[H(W_1,W_2)+H(Z_1)]+2H(W_1)+2(K-1)H(W_1,W_2),
\end{align}
which has some a common term $H(Z_1)$ on both sizes with different coefficients. Removing the common term and multiplying both sides by two lead to
\begin{align}
&K(K+1)H(Z_1)+2(K-1)KH(X_{\rightarrow 2})\nonumber\\
&\geq [{(K-2)(K-1)}-2+4(K-1)]H(W_1,W_2)+4H(W_1)\nonumber\\
&=2K^2+2K-4,
\end{align}
where the equality relies on the assumption that $W_1$ and $W_2$ are independent files of unit size. Taking into consideration the memory and transmission rate constraints (\ref{eqn:memoryconstraint}) and (\ref{eqn:transmissionconstraint}) now completes the proof.
\end{IEEEproof}

Lemma \ref{lemma:2K}  provides a way to reduce the number of $X$ variables in $H(Z_1,X_{\rightarrow [2:k]})$, and thus is the core of the proof. 
Even with the hypothesis regarding the scheme in \cite{MaddahAliNiesen:14} being optimal, deriving the outer bound (particularly the coefficients in the lemma above) directly using this insight is far from being straightforward. Some of the guidance in finding our derivation was in fact obtained through a strategic computational exploration of the outer bounds. This information is helpful because the computer-generated proofs are not unique, and some of these solutions can appear quite arbitrary, however, to deduce general rules in the proof requires a more structured proof instead. In Section \ref{sec:generalization}, we present in more details this new exploration method, and discuss how insights can be actively identified in this particular case.


\section{Reverse-Engineering Code Constructions}
\label{sec:reverseEng}

In the previous section, outer bounds of the optimal tradeoff were presented for the case $(N,K)=(2,4)$, which is given in Fig. \ref{fig:caching2_N}. Observe that the corner points
\begin{align}
\left(\frac{2}{3},1\right) \quad \mbox{and} \quad\left(\frac{6}{13},\frac{16}{13}\right),
\end{align}
cannot be achieved by existing codes in the literature. The former point can indeed be achieved with a new code construction. This construction was first presented in \cite{tian2018caching}, where it was generalized more systematically to yield a new class of codes for any $N\leq K$, whose proof and analysis are more involved.
In this paper, we focus on how a specific code for this corner point was found through a reverse engineering approach, which should help dispel the mystery on this seemingly arbitrary code construction. 

\subsection{The Code to Achieve $\left(\frac{2}{3},1\right)$ for $(N,K)=(2,4)$}

\begin{table*}[tb!]
\begin{center}
\caption{Caching content for $(N,K)=(2,4)$\label{tab:newcorner1}}
\begin{tabular}{|c || c | c | c | c|}
\hline
User 1 &$A_1+B_1$ & $A_2+B_2$ & $A_3+B_3$ & $A_1+A_2+A_3+2(B_1+B_2+B_3)$\\\hline
User 2 &$A_1+B_1$ & $A_4+B_4$ & $A_5+B_5$ & $A_1+A_4+A_5+2(B_1+B_4+B_5)$\\\hline
User 3 &$A_2+B_2$ & $A_4+B_4$ & $A_6+B_6$ & $A_2+A_4+A_6+2(B_2+B_4+B_6)$\\\hline
User 4 &$A_3+B_3$ & $A_5+B_5$ & $A_6+B_6$ & $A_3+A_5+A_6+2(B_3+B_5+B_6)$\\
\hline
\end{tabular}
\end{center}
\end{table*}

The two files are denoted as $A$ and $B$, each of which is partitioned into $6$ segments of equal size, denoted as $A_i$ and $B_i$, respectively, $i=1,2,\ldots,6$. Since we count the memory and transmission in multiple of the file size, the corner point $\left(\frac{2}{3},1\right)$ means needs each user to store $4$ symbols, and the transmission will use $6$ symbols. The contents in the cache of each user are given in Table \ref{tab:newcorner1}. By the symmetry of the cached contents, we only need to consider the demand $(A,A,A,B)$, {\em i.e.}, the first three users requesting $A$ and user $4$ requesting $B$, and the demand $(A,A,B,B)$, {\em i.e.,} the first two users requesting $A$ and the other two requesting $B$. 
Assume the file segments are in $\mathbb{F}_{5}$ for concreteness.

\begin{itemize}[topsep=3pt,itemsep=2pt]
\item For  the demands $(A,A,A,B)$, the transmission is as follows,
\begin{align*}
\text{Step $1$:  }&B_1,B_2,B_4;\\
\text{Step $2$: }&A_3+2A_5+3A_6,A_3+3A_5+4A_6;\\
\text{Step $3$:  }&A_1+A_2+A_4.
\end{align*}
After step $1$, user 1 can recover $(A_1,A_2)$; furthermore, he has $(A_3+B_3,A_3+2B_3)$ by eliminating known symbols $(A_1,A_2,B_1,B_2)$, from which $A_3$ can be recovered. After step $2$, he can obtain $(2A_5+3A_6,3A_5+4A_6)$ to recover $(A_5,A_6)$. Using the transmission in step $3$, he can obtain $A_4$ since he has $(A_1,A_2)$. User 2 and user 3 can use a similar strategy to reconstruct all file segments in $A$. User 4 only needs $B_3,B_5,B_6$ after step $1$, which he already has in his cache, however they are contaminated by file segments from $A$. Nevertheless, he knows $A_3+A_5+A_6$ by recognizing 
\begin{align}
&(A_3+A_5+A_6)=2\sum_{i=3,5,6}(A_i+B_i)\nonumber\\
&\qquad\quad-[A_3+A_5+A_6+2(B_3+B_5+B_6)].
\end{align}
Together with the transmission in step $2$, user 4 has three linearly independent combinations of $(A_3,A_5,A_6)$. After recovering them, he can remove these interferences from the cached content for $(B_3,B_5,B_6)$.

\item 
For the demand $(A,A,B,B)$, we can send 
\begin{align*}
\text{Step $1$:  }& B_1,A_6;\\
\text{Step $2$: }&A_2+2A_4,A_3+2A_5,B_2+2B_3,B_4+2B_5.
\end{align*}
User 1 has $A_1,B_1,A_6$ after step $1$, and he can also form
\begin{align*}
B_2+B_3=&[A_2+A_3+2(B_2+B_3)]\\
&\qquad\qquad -(A_2+B_2) -(A_3+B_3),
\end{align*}
and together with $B_2+2B_3$ in the transmission of step $2$, he can recover $(B_2,B_3)$, and thus $A_2,A_3$. He still needs $(A_4,A_5)$, which can be recovered straightforwardly from the transmission $(A_2+2A_4,A_3+2A_5)$ since he already has $(A_2,A_3)$. Other users can use a similar strategy to decode their requested files.
\end{itemize}

\subsection{Extracting Information for Reverse Engineering}

It is clear at this point that for this case of $(N,K)=(2,4)$, the code to achieve this optimal corner point is not straightforward. Next we discuss a general approach to deduce the code structure from the LP solution, which leads to the discovery of the code in our work. The approach is based on the following assumptions: the outer bound is achievable ({\em i.e.}, tight), moreover, there is a (vector) linear code that can achieve this performance. 

Either of the two assumptions above may not be hold in general, and in such a case our attempt will not be successful. Nevertheless, though linear codes are known to be not sufficient for all network coding problem \cite{Dougherty:05}, existing results in the literature suggest that vector linear codes are surprisingly versatile and powerful. Similarly, though it is known that Shannon-type inequalities, which are the basis for the outer bounds computation, are not sufficient to characterize rate region for all coding problems \cite{Zhang:97,Zhang:98}, they are surprisingly powerful, particularly in coding problems with strong symmetry structures \cite{Yeung:99,Tian:11}.

There are essentially two types of information that we can extract from the primal LP and dual LP:
\begin{itemize}[topsep=3pt,itemsep=2pt]
\item From the effective information inequalities: since we can produce a readable proof using the dual LP, if a code can achieve this corner point, then the information inequalities in the proof must hold with equality for the joint entropy values induced by this code, which reveals a set of conditional independence relations among random variables induced by this code;
\item From the extremal joint entropy values at the corner points: although we are only interested in the tradeoff between the memory and transmission rate, the LP solution can provide the whole set of joint entropy values at an extreme point. These values can reveal a set of dependence relations among the random variables induced by any code that can achieve this point.
\end{itemize}

Though the first type of information is important, its translation to code constructions appears difficult. On the other hand, the second type of information appears to be more suitable for the purpose of code design, which we adopt next.

One issue that complicates our task is that the entropy values such extracted are not always unique, and sometimes have considerable slacks. For example, for different LP solutions at the same operating point of $(M,R)=\left(\frac{2}{3},1\right)$, the joint entropy $H(Z_1,Z_2)$ can vary between $1$ and $4/3$. We can identify such a slack in any joint entropy in the corner point solutions by considering a regularized primal LP: for a fixed rate value $R$ at the corner point in question as an upper bound, the objective function can be set as
\begin{align}
\mbox{minimize: }& H(Z_1)+\gamma H(Z_1,Z_2)
\end{align}
instead of 
\begin{align}
\mbox{minimize: }& H(Z_1),
\end{align}
subject to the same original symmetric LP constraints at the target $M$. By choosing a small positive $\gamma$ value, {\em e.g.}, $\gamma=0.0001$, we can find the minimum value for $H(Z_1,Z_2)$ at the same $(M,R)$ point; similarly, by choosing a small negative $\gamma$ value, we can find the maximum value for $H(Z_1,Z_2)$ at the same $(M,R)$ point. Such slacks in the solution add  uncertainty to the codes we seek to find, and may indeed imply the existence of multiple code constructions. For the purpose of reverse engineering the codes, we focus on the joint entropies that do not have any slacks, {\em i.e.}, the ``stable''  joint entropies in the solution. 

\subsection{Reverse-Engineering the Code for $(N,K)=(2,4)$}

\begin{table}
\caption{Stable joint entropy values at the corner point $\left(\frac{2}{3},1\right)$ for $(N,K)=(2,4)$. \label{tab:reversecode}}
\begin{center}
\begin{tabular}{|c|c|c|}
\hline
Joint entropy & Computed value\\\hline\hline
$H(Z_1|W_1)$&  3\\\hline
$H(Z_1,Z_2|W_1)$&  5\\\hline
$H(Z_1,Z_2,Z_3|W_1)$ &6\\\hline\hline
$H(X_{1,2,2,2}|W_1)$&3\\\hline
$H(Z_1,X_{1,2,2,2}|W_1)$&4\\\hline
$H(X_{1,1,1,2}|W_1)$&3\\\hline
$H(Z_1,X_{1,1,1,2}|W_1)$&4\\\hline
$H(Z_1,Z_2,X_{1,1,1,2}|W_1)$&5\\\hline
\hline
$H(X_{1,1,2,2}|W_1)$&3\\\hline
$H(Z_1,X_{1,1,2,2}|W_1)$&4\\\hline
$H(Z_1,Z_2,X_{1,1,2,2}|W_1)$&5\\\hline
\end{tabular}
\end{center}
\end{table}

With the method outlined above, we identify the following stable joint entropy values in the $(N,K)=(2,4)$ case for the operating point $\left(\frac{2}{3},1\right)$ listed in Table \ref{tab:reversecode}. The values are normalized by multiplying everything by 6. For simplicity, let us assume that each file has 6 units of information, written as $W_1=(A_1,A_2,\ldots,A_6)\triangleq A$ and $W_2=(B_1,B_2,\ldots,B_6)\triangleq B$, respectively. 
This is a rich set of data, but a few immediate observations are given next. 
\begin{itemize}[topsep=3pt,itemsep=2pt]
\item The quantities can be categorized into three groups: the first is without any transmission, the second is the quantities involving the transmission to fulfill the demand type $(3,1)$, and the last for demand type $(2,2)$.
\item The three quantities $H(Z_1|W_1),H(Z_1,Z_2|W_1)$ and $H(Z_1,Z_2,Z_3|W_1)$ provide the first important clue. The values indicate that for each of the two files, each user should have 3 units in its cache, and the combination of any two users should have 5 units in their cache, and the combination of any three users should have all 6 units in their cache. This strongly suggests placing each piece $A_i$ (and  $B_i$) at two users. Since each $Z_i$ has 4 units, but it needs to hold 3 units from each of the two files, coded placement  (cross files) is thus needed. At this point, we place the corresponding symbols in the caching, but keep the precise linear combination coefficients as undetermined. 
\item The next critical observation is that $H(X_{1,2,2,2}|W_1)=H(X_{1,1,1,2}|W_1)=H(X_{1,1,2,2}|W_1)=3$. This implies that the transmission has 3 units of information on each file alone. However, since the operating point dictates that $H(X_{1,2,2,2})=H(X_{1,1,1,2})=H(X_{1,1,2,2})=6$, it further implies that in each transmission, 3 units are for the linear combinations of $W_2$, and 3 units are for those of $W_1$; in other words, the linear combinations do not need to mix information from different files.
\item Since each transmission only has 3 units of information from each file, and each user has only 3 units of information from each file, they must be linearly independent of each other.  
\end{itemize}

The observation and deductions are only from the perspective of the joint entropies given in Table \ref{tab:reversecode}, without much consideration of the particular coding requirement. For example, in the last item discussed above, it is clear that when transmitting the 3 units of information regarding a file (say file $W_2$), they should be simultaneously useful to other users requesting this file, and to the users not requesting this file. This intuition then strongly suggests each transmitted linear combination of $W_2$ should be a subspace of the $W_2$ parts at some users not requesting it. Using these intuitions as guidance, finding the code becomes straightforward after a few trial-and-errors. In \cite{tian2018caching} we were able to further generalize this special code to a class of codes for any case when $N\leq K$; readers are referred to \cite{tian2018caching} and \cite{tian2017uncoded} for more details on these codes.

\subsection{Disproving Linear-Coding Achievability}

The reverse engineering approach may not always be successful, either because the structure revealed by the data is very difficult to construct explicitly, or because linear codes are not sufficient to achieve this operating point. 
In some other cases, the determination can be done explicitly. In the sequel we present an example for $(N,K)=(3,3)$, which belongs to the latter case. An outer bound for $(N,K)=(3,3)$ is presented in the next section, and among the corner points, the pair $(M,R)=(\frac{2}{3},\frac{4}{3})$ is the only one that cannot be achieved by existing schemes. Since the outer bound appears quite strong, we may conjecture this pair is also achievable and attempt to construct a code. Unfortunately, as we shall show next, there does not exist such a (vector) linear code. Before delving into the data provided by the LP, readers are encouraged to consider proving directly that this tradeoff point cannot be achieved by linear codes, which does not appear to be straightforward to the author. 

We shall assume each file has $3m$ symbols in certain finite field, where $m$ is a positive integer. The LP produces the joint entropy values (in terms of the number of finite field symbols, not in multiples of file size as in the other sections of the paper) in Table. \ref{tab:reversecode33} at this corner point, where only the conditional joint entropies relevant to our discussion next are listed. The main idea is to use these joint entropy values to deduce structures of the coding matrices, and then combining these structures with the coding requirements to reach a contradiction.

\begin{table}
\caption{Stable joint entropy values at the corner point $\left(\frac{2}{3},\frac{4}{3}\right)$ for $(N,K)=(3,3)$. \label{tab:reversecode33}}
\begin{center}
\begin{tabular}{|c|c|c|}
\hline
Joint entropy & Computed value\\\hline\hline
$H(Z_1|W_1)$& $2m$\\\hline
$H(Z_1|W_1,W_2)$& $m$\\\hline
$H(Z_1,Z_2|W_1,W_2)$& $2m$\\\hline
$H(Z_1,Z_2,Z_3|W_1,W_2)$& $3m$\\\hline\hline
$H(X_{1,2,3})$ & $4m$\\\hline
$H(X_{1,2,3}|W_1)$& $3m$\\\hline
$H(X_{1,2,3}|W_1,W_2)$& $2m$\\\hline
\end{tabular}
\end{center}
\end{table}

The first critical observation is that $H(Z_1|W_1,W_2)=m$, and the user-index-symmetry implies that  $H(Z_2|W_1,W_2)=H(Z_3|W_1,W_2)=m$. Moreover $H(Z_1,Z_2,Z_3|W_1,W_2)=3m$, from which we can conclude that excluding file $W_1$ and $W_2$, each user stores $m$ linearly independent combinations of the symbols of file $W_3$,  which are also linearly independent among the three users. Similar conclusions hold for files $W_1$ and $W_2$. Thus, without loss of generality, we can view the linear combinations of $W_i$ cached by the users, after excluding the symbols from the other two files, as the basis of file $W_i$. In other words, this implies that through a change of basis for each file, we can assume without loss of generality that user-$k$ stores $2m$ linear combinations in the following form
\begin{align}
V_k\cdot \left[\begin{array}{c}
W_{1,[(k-1)m+1:km]}\\W_{2,[(k-1)m+1:km]}\\W_{3,[(k-1)m+1:km]}
  \end{array}
  \right]
\end{align}
where $W_{n,j}$ is the $j$-th symbol of the $n$-th file, and $V_k$ is a matrix of dimension $2m\times 3m$; $V_k$ can be partitioned into submatrices of dimension $m\times m$, which are denoted as $V_{k;i,j}$, $i=1,2$ and $j=1,2,3$. Note that symbols at different users are orthogonal to each other without loss of generality.

Without loss of generality, assume the transmitted content $X_{1,2,3}$ is
\begin{align}
G\cdot  \left[\begin{array}{c}
  W_{1,[1:3m]}\\W_{2,[1:3m]}\\W_{3,[1:3m]}
  \end{array}
  \right]
\end{align}
where $G$ is a matrix of dimension $4m\times 9m$; we can partition it into blocks of $m\times m$, and each block is referred to as $G_{i,j}$, $i=1,2,\ldots,4$ and $j=1,2,\ldots,9$. Let us first consider user 1, which has the following symbols
\begin{align}
\left[
  \begin{array}{ccccccccc}
    V_{k;1,1}& 0 & 0 & V_{k;1,2} & 0  &0 &  V_{k;1,3} &  0    &0\\
    V_{k;2,1}& 0 & 0 & V_{k;2,2} & 0  &0 &  V_{k;2,3} &  0    &0\\\cmidrule(lr){1-9}
    G_{1,1}  & G_{1,2}& \multicolumn{6}{c}{...}                        & G_{1,9}\\
    \multicolumn{1}{c}{\vdots}&\multicolumn{1}{c}{\vdots}&\multicolumn{6}{c}{\vdots} &\multicolumn{1}{c}{\vdots}\\
    G_{4,1}  & G_{4,2}& \multicolumn{6}{c}{...}                        & G_{4,9}\\    
  \end{array} \right]\cdot
  \left[\begin{array}{c}
  W_{1,[1:3m]}\\W_{2,[1:3m]}\\W_{3,[1:3m]}
  \end{array}
  \right] \label{eqn:matrix}
\end{align}

The coding requirement states that $X_{1,2,3}$ and $Z_1$ together can be used to recover file $W_1$, and thus one can recover all the symbols of $W_1$ knowing (\ref{eqn:matrix}). Since $W_1$ can be recovered, its symbols can be eliminated in (\ref{eqn:matrix}), {\em i.e.,} 
\begin{align}
\left[
  \begin{array}{cccccc}
    V_{k;1,2} & 0  &0 &  V_{k;1,3} &  0    &0\\
    V_{k;2,2} & 0  &0 &  V_{k;2,3} &  0    &0\\\cmidrule(lr){1-6}
    G_{1,4}   & G_{1,5}& \multicolumn{3}{c}{...}                        & G_{1,9}\\
    \multicolumn{1}{c}{\vdots}&\multicolumn{1}{c}{\vdots}&\multicolumn{3}{c}{\vdots} &\multicolumn{1}{c}{\vdots}\\
    G_{4,4}  & G_{4,4}& \multicolumn{3}{c}{...}                        & G_{4,9}\\    
  \end{array} \right]\cdot
  \left[\begin{array}{c}
  W_{2,[1:3m]}\\W_{3,[1:3m]}
  \end{array}
  \right]\label{eqn:recoverW2W3}
\end{align}
in fact becomes known. Notice Table \ref{tab:reversecode33} specifies $H(Z_1|W_1)=2m$, and thus the matrix
\begin{align}
\left[
  \begin{array}{cc}
 V_{k;1,2} &  V_{k;1,3} \\
 V_{k;2,2} &  V_{k;2,3}
    \end{array}
    \right]
    \label{eqn:fullrankV}
\end{align}
is in fact full rank, and thus from the top part of (\ref{eqn:recoverW2W3}), $W_{2,[1:m]}$ and $W_{3,[1:m]}$ can be recovered. In summary, through elemental row operations and column permutations, the matrix in (\ref{eqn:matrix}) can be converted into the following form
\begin{align}
\left[
  \begin{array}{ccccccccc}
    U_{1,1}  & U_{1,2} & U_{1,3}& 0             & \multicolumn{4}{c}{...}&0\\
    U_{2,1}  & U_{2,2} & U_{2,3}& 0             & \multicolumn{4}{c}{...}&0\\
    U_{3,1}  & U_{3,2} & U_{3,3}& 0             & \multicolumn{4}{c}{...}&0\\
      0         &    0       &   0       &  U_{4,4}   & U_{5,7} & 0 & 0&0&0\\
      0         &    0       &   0       &  U_{4,4}   & U_{5,7} & 0 & 0&0&0\\
      0         &    0       &   0       &    0          &     0      & U_{6,5} &U_{6,6}&U_{6,8}& U_{6,9} 
  \end{array} \right]\cdot
  \left[\begin{array}{c}
  W_{1,[1:3m]}\\W_{2,[1:m]}\\W_{3,[1:m]}\\W_{2,[m+1:3m]} \\W_{3,[m+1:3m]}
  \end{array}
  \right], \label{eqn:matrix2}
\end{align}
where diagonal block square matrices are of full rank $3m$ and $2m$, respectively, and $U_{i,j}$'s are the resultant block matrices after the row operations and column permutations. This further implies that the matrix $[U_{6,5}, U_{6,6}, U_{6,8}, U_{6,9}]$ has maximum rank $m$,and it follows that the matrix
\begin{align}
\left[
  \begin{array}{cccc} 
 G_{1,5} &G_{1,6}&G_{1,8}& G_{1,9}\\
 \vdots   & \vdots &\vdots   & \vdots\\
 G_{4,5} &G_{4,6}&G_{4,8}& G_{4,9}
  \end{array} \right], \label{eqn:matrix3}
\end{align}
{\em i.e.}, the submatrix of $G$ by taking thick columns $(5,6,8,9)$, has only maximum rank $m$. However, due to the symmetry, we can also conclude that the submatrix of $G$ taking only thick columns $(1,3,7,9)$ and that taking only thick columns $(1,2,4,5)$ both have only maximum rank $m$. As a consequence the matrix $G$ has rank no larger than $3m$, but this contradicts  the condition that $H(X_{1,2,3})=4m$ in Table \ref{tab:reversecode33}. We can now conclude that this memory-transmission-rate pair is not achievable with any linear codes\footnote{Strictly speaking, our argument above holds under the assumption that the joint entropy values produced by LP are precise rational values, and the machine precision issue has thus been ignored. However, if the solution is accurate only up to machine precision, one can introduce a small slack value $\delta$ into the quantities, {\em e.g.,} replacing $3m$ with $(3\pm\delta)m$, and using a similar argument show that the same conclusion holds. This extended argument however becomes notationally rather lengthy, and we thus omitted it here for simplicity.}.

\section{Computational Exploration and Bounds for Larger Cases}
\label{sec:generalization}

In this section we explore the fundamental limits of the caching systems in more details using a computational approach. Due to the (doubly) exponential growth of the LP variables and constraints, directly applying the method outlined in Section \ref{sec:pre} becomes infeasible for larger problem cases. This is the initial motivation for us to investigate single-demand-type systems where only a single demand type is allowed. Any outer bound on the tradeoff of such a system is an outer bound for the original one, and the intersection of these outer bounds is thus also an outer bound. This investigation further reveals several hidden phenomena. For example, outer bounds for different single-demand-type systems are stronger in different regimes, and moreover, the LP bound for the original system is not simply the intersection of all outer bounds for single-demand-type systems, but in certain regimes they do match.

Given the observations above, we take the investigation one step further by choosing only a small subset of demands instead of the complete set in a single demand type. This allows us to obtain results for cases which initially appear impossible to compute. For example, even for $(N,K)=(2,5)$, there are a total of $2+5+2^5=39$ random variables, and the number of constraints in LP after symmetry reduction is more than $10^{11}$, which is significantly beyond current LP solver capability\footnote{The problem can be further reduced using problem specific implication structures as outlined in Section \ref{sec:pre}, but our experience suggests that even with such additional reduction the problem may still too large for a start-of-the-art LP solver without additional reduction.}. However, by strategically considering only a small subset of the demand patterns, we are indeed able to find meaningful outer bounds, and moreover, use the clues obtained in such computational exploration to complete the proof of Theorem \ref{theorem:2K}. We shall discuss the method we develop, and also present several example results for larger problem cases. 

\subsection{Single-Demand-Type Systems}
\label{sec:case33}

As mentioned above, in a single-demand-type caching systems, the demand must belong to a particular demand type. We first present results on two cases $(N,K)=(2,4)$ and $(N,K)=(3,3)$, and then discuss our observations using these results. 

\begin{prop}
\label{theorem:result2_4}
Any memory-transmission-rate tradeoff pair for the $(N,K)=(2,4)$ caching problem must satisfy the following conditions for \textbf{single demand type} $(4,0)$:  
\begin{align}
M+2R\geq 2\label{eqn:bounds2_4_type4_0},
\end{align}
and conversely any non-negative $(M,R)$ pair satisfying (\ref{eqn:bounds2_4_type4_0}) is achievable for single demand type $(4,0)$;
it must satisfy for \textbf{single demand type} $(3,1)$:
\begin{align}
2M+R\geq 2,\quad 8M+6R\geq 11, \quad 3M+3R\geq 5, \quad 5M+6R\geq 9,\quad M+2R\geq 2,\label{eqn:bounds2_4_type3_1}
\end{align}
and conversely any non-negative $(M,R)$ pair satisfying (\ref{eqn:bounds2_4_type3_1}) is achievable for single demand type $(3,1)$;
it must satisfy for \textbf{single demand type} $(2,2)$ 
\begin{align}
2M+R\geq 2,\quad 3M+3R\geq 5,\quad M+2R\geq 2, \label{eqn:bounds2_4_type2_2}
\end{align}
and conversely any non-negative $(M,R)$ pair satisfying (\ref{eqn:bounds2_4_type2_2}) is achievable for single demand type $(2,2)$.
\end{prop}

\begin{figure}
\centering
\includegraphics[width=9cm]{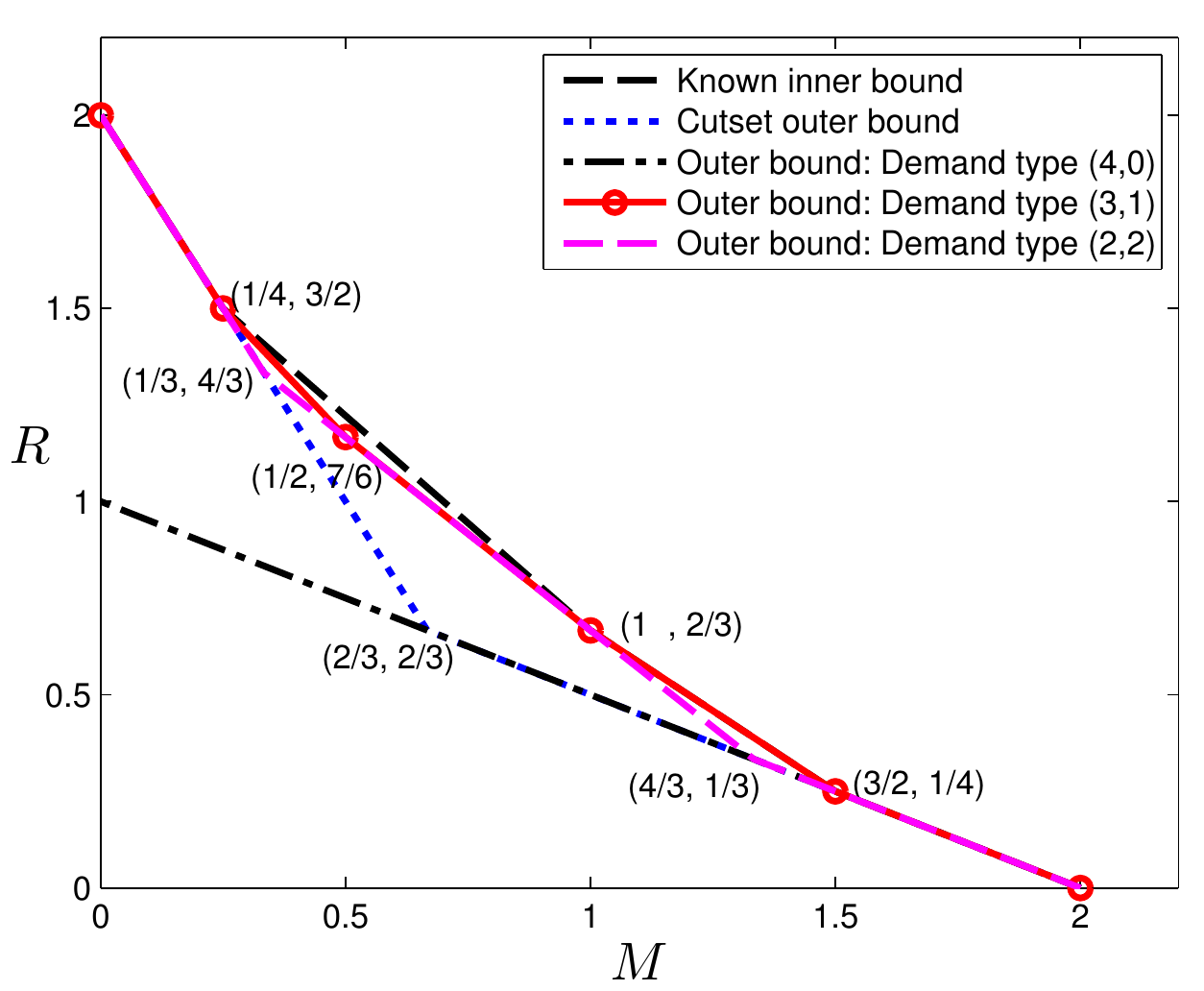}
\caption{Tradeoff outer bounds for $(N,K)=(2,4)$ caching systems. \label{fig:figure2_4}}
\end{figure}

The optimal $(M,R)$ tradeoffs are illustrated in Fig. \ref{fig:figure2_4} with the known inner bound, {\em i.e.}, those in \cite{MaddahAliNiesen:14,chen2016fundamental} and the one given in the last section, and the computed out bound of the original problem given in Section \ref{sec:hypo}. Here the demand type $(3,1)$ in fact provides the tightest outer bound which matches the known inner bound for $M\in[0,1/4]\cup[2/3,2]$. The converse proofs of (\ref{eqn:bounds2_4_type3_1}) and (\ref{eqn:bounds2_4_type2_2}) are obtained computationally, the details of which can be found in Appendix \ref{appendix6_1}. In fact only the middle three inequalities in (\ref{eqn:bounds2_4_type3_1}) and the second inequality in (\ref{eqn:bounds2_4_type2_2}) need to be proved, since the others are due to the cut-set bound. Although the original caching problem requires codes that can handle all types of demands, the optimal codes for single demand type systems turn out to be quite interesting by its own right, and thus we provide the forward proof of Theorem \ref{theorem:result2_4}  in Appendix \ref{appendix:theorem:result2_4}.

The computed outer bounds for single-demand-type systems for $(N,K)=(3,3)$ are summarized below; the proofs can be found in Appendix \ref{appendix33}. 

\begin{prop}
\label{theorem:firstresult3_3A}
Any memory-transmission-rate tradeoff pair for the $(N,K)=(3,3)$ caching problem must satisfy the following conditions for \textbf{single demand type $(3,0,0)$}:  
\begin{align}
 M+3R\geq 3, \label{eqn:3_3bounds3_0_0}
\end{align}
and conversely any non-negative $(M,R)$ pair satisfying (\ref{eqn:3_3bounds3_0_0}) is achievable for single demand type $(3,0,0)$;
it must satisfy for \textbf{single demand type $(2,1,0)$}:
\begin{align}
&M+R\geq2, \quad 2M+3R\geq 5,\quad M+3R\geq 3, \label{eqn:3_3bounds2_1_0}
\end{align}
and conversely any non-negative $(M,R)$ pair satisfying (\ref{eqn:3_3bounds2_1_0}) is achievable for single demand type $(2,1,0)$;
it must satisfy for \textbf{single demand type} $(1,1,1)$:
\begin{align}
&3M+R\geq 3,\quad 6M+3R\geq 8, \quad  M+R\geq2,\quad 
&12M+18R\geq 29,\quad 3M+6R\geq 8,\quad M+3R\geq 3. \label{eqn:3_3bounds1_1_1}
\end{align}
\end{prop}

\begin{figure}
\centering
\includegraphics[width=9cm]{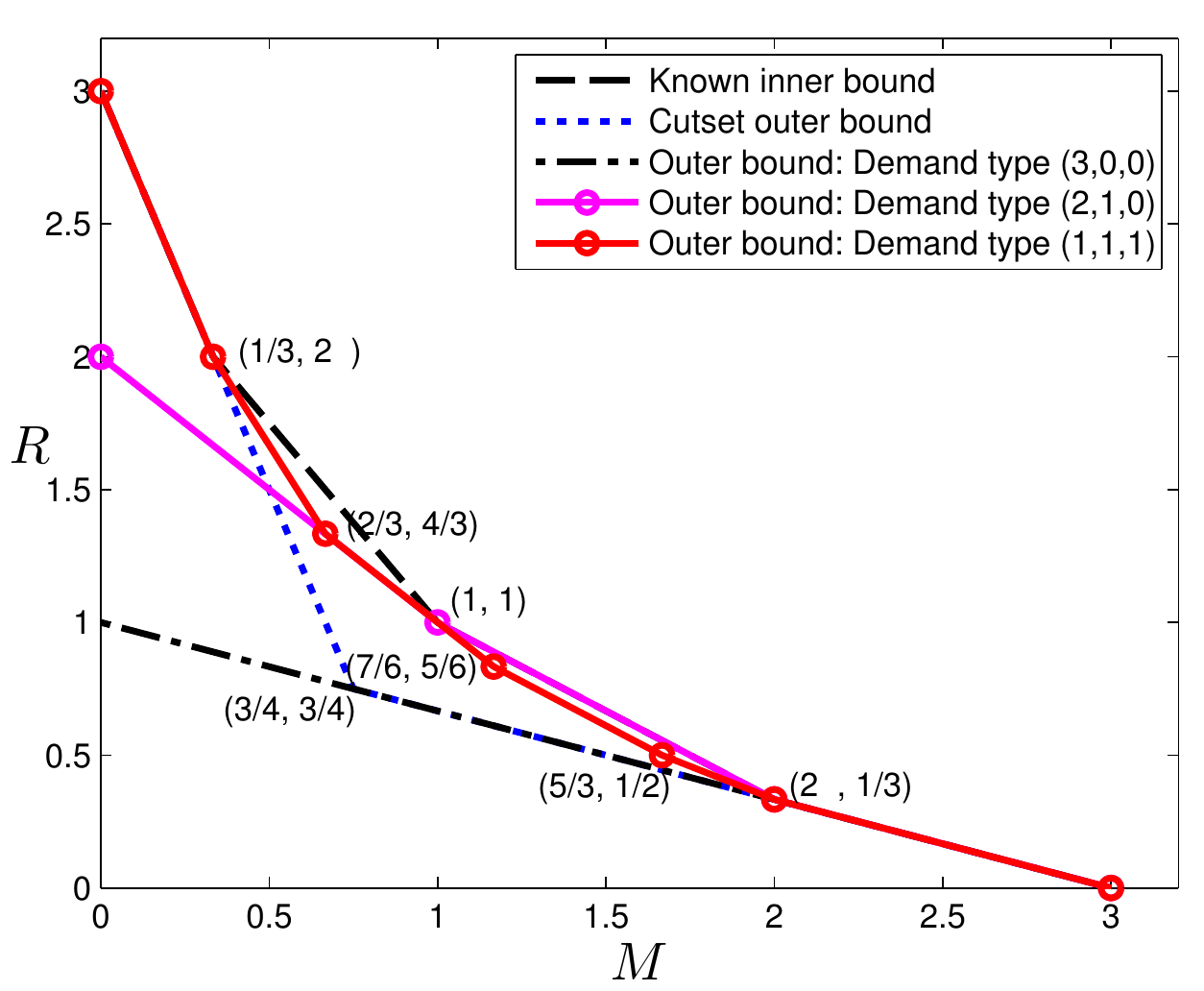}
\caption{Tradeoff outer bounds for $(N,K)=(3,3)$  caching\label{fig:figure3_3}}
\end{figure}

These outer bounds are illustrated in Fig. \ref{fig:figure3_3}, together with the best known inner bound by combining \cite{MaddahAliNiesen:14} and \cite{chen2016fundamental}, and the cut-set outer bound for reference. The bound is in fact tight for $M\in[0,1/3]\cup[1,3]$. Readers may notice that Proposition \ref{theorem:firstresult3_3A} provides complete characterizations for the first two demand types, but not the last demand type. As we have shown in Section \ref{sec:reverseEng}, the point $(\frac{2}{3},\frac{4}{3})$ in fact cannot be achieved using linear codes. 

\begin{remark}
The bound developed in~\cite{sengupta2017improved} gives $6M+3R\geq 8$ and $2M+4R\geq5$, and~that in~\cite{ajaykrishnan2015critical} gives $(M+R)\geq 2$ in addition to the cut-set bound.
\end{remark}

We can make the following observations immediately:
\begin{itemize}[topsep=3pt,itemsep=2pt]
\item The single-demand-type systems for few files usually produce tighter bounds at high memory regime, while those for more files usually produce tighter bounds at low memory regime. For example, the first high-memory segment of the bounds can be obtained by considering only demands that request a single file which coincidently is also the cut-set bound; for $(N,K)=(3,3)$, the bound obtained from the demand type $(2,1,0)$ is stronger than that from $(1,1,1)$ in the range $M\in [1,2]$.
\item Simply intersecting the single-demand-type outer bounds does not produce the same bound as that obtained from a system with the complete set of demands. This can be seen from the case $(N,K)=(2,4)$ in the range $M\in [1/4,2/3]$.
\item The outer bounds produced by single-demand-type systems in many cases match the bound when more comprehensive demands are considered. This is particularly evident in the case $(N,K)=(2,4)$ in the range $M\in[0,1/4]\cup[2/3,2]$. 
\end{itemize}

These observations provide further insights on the difficulty of the problem. For instance, for $(N,K)=(2,4)$, the demand type $(3,1)$ is the most demanding case, and code design for this demand type should be considered as the main challenge. More importantly, these observation suggests that it is possible to obtain very strong bounds by considering only a small subset of demands, instead of the complete set of demands. In the sequel we further explore this direction.

\subsection{Equivalent Bounds Using Subsets of Demands}

Based on the observations in the previous subsection, we conjecture that in some cases, equivalent bounds can be obtained by using only a smaller number of requests, and moreover, these demands do not need to form a complete demand type class, and next we show that this is indeed the case. 
To be more precise, we are relaxing the LP, by including only elemental inequality constraints that involve joint entropies of random variables within a subset of the random variables $\mathcal{W}\cup\mathcal{Z}\cup\mathcal{X}$, and other constraints are simply removed. However the symmetry structure specified in Section \ref{sec:symmetry} is still maintained to reduce the problem. 
This approach is not equivalent to forming the LP on a caching system where only those files, users and demands are present, since in this alternative setting, symmetric solutions may induce loss of optimality. 

There are many choices of subsets with which the outer bounds can be computed, and we only provide a few that are more relevant which confirm our conjecture:
\begin{fact}
\label{fact:toquote}
In terms of the computed outer bounds, the following facts were observed:
\begin{itemize}
\item For the $(N,K)=(2,4)$ case, the outer bound in Proposition \ref{prop:NK24} can be obtained by restricting to the subset of random variables $\mathcal{W}\cup\mathcal{Z}\cup\{X_{1,1,1,2},X_{1,1,2,2}\}$.
\item For the $(N,K)=(2,4)$ case, the outer bound in Proposition \ref{theorem:result2_4} in the range $M\in[1/3,2]$ for single demand type $(3,1)$ can be obtained by restricting to the subset of random variables $\mathcal{W}\cup\mathcal{Z}\cup\{X_{2,1,1,1},X_{1,2,1,1},X_{1,1,2,1},X_{1,1,1,2}\}$.
\item For the $(N,K)=(3,3)$ case, the intersection of the outer bounds in Proposition \ref{theorem:firstresult3_3A} can be obtained by restricting to the subset of random variables $\mathcal{W}\cup\mathcal{Z}\cup\{X_{2,1,1},X_{3,1,1},X_{3,2,1}\}$.
\item For the $(N,K)=(3,3)$ case, the outer bound in Proposition \ref{theorem:firstresult3_3A} in the range $M\in[2/3,3]$ for single demand type $(2,1)$ can be obtained by restricting to the subset of random variables $\mathcal{W}\cup\mathcal{Z}\cup\{X_{2,1,1},X_{3,1,1}\}$.
\end{itemize}
\end{fact}

These observations reveal that the subset of demands can be chosen rather small to produce strong bounds. For example, for the $(N,K)=(2,4)$ case, including only  joint entropies involving 8 random variables $\mathcal{W}\cup\mathcal{Z}\cup\{X_{1,1,1,2},X_{1,1,2,2}\}$ will produce the strongest bound as including all $22$ random variables. Moreover, for specific regimes, the same bound can be produced using an even smaller number of random variables (for the case $(N,K)=(3,3)$), or with a more specific set of random variables (for the case $(N,K)=(2,4)$ where in the range [1/3,2], including only some of the demand type $(3,1)$ is sufficient). Equipped with these insights, we can attempt to tackle larger problem cases, which would have appeared impossible to  computationally produce meaningful outer bounds for. In the sequel, this approach is applied for two purposes: (1) to identify generic structures in converse proofs, and (2) to produce outer bounds for large problem cases. 

\subsection{Identifying Generic Structures in Converse Proofs}

Recall our comment given after the proof of Theorem \ref{theorem:2K} that finding this proof is not straightforward. One critical clue was obtained when applying the exploration approach discussed above. When restricting the set of included random variables to a smaller set, the overall problem is being relaxed, however, if the outer bound thus obtained remains the same, it implies that the sought-after outer bound proof only needs to rely on the joint entropies within this restricted set. For the specific case of $(N,K)=(2,5)$, we have the following fact.
\begin{fact}
For $(N,K)=(2,5)$, the bound $15M+20R\geq 28$ in the range $M\in [6/5,8/5]$ can be obtained by restricting to the subset of random variables $\mathcal{W}\cup\mathcal{Z}\cup\{X_{2,1,1,1,1},X_{1,2,1,1,1},X_{1,1,2,1,1},X_{1,1,1,2,1},X_{1,1,1,1,2}\}$.
\end{fact}

Together with the second item in Fact \ref{fact:toquote}, we can naturally conjecture that in order to prove the hypothesized outer bound in Hypothesis \ref{hypo:2K}, only the dependence structure within the set of random variables $\mathcal{W}\cup\mathcal{Z}\cup X_{\rightarrow [1:K]}$ needs to be considered, and all the proof steps can be written using mutual information or joint entropies of them alone. Although this is still not a trivial task, the possibility is significantly reduced, {\em e.g., } for the $(N,K)=(2,5)$ case to only $12$ random variables, with a much simpler structure than that of the original problem with $39$ random variables. Perhaps more importantly, such a restriction makes it feasible to identify common route of derivation in the converse proof and then generalize it, from which we obtain the proof of Theorem \ref{theorem:2K}. 

\subsection{Computing Bounds for Larger Problem Cases}

We now present a few outer bounds for larger problem cases, and make comparison with other known bounds in the literature. This is not intended to be a complete list of results we obtain, and more results will be made online after they are computed. 

\begin{figure}[tb]
\centering
\includegraphics[width=16cm]{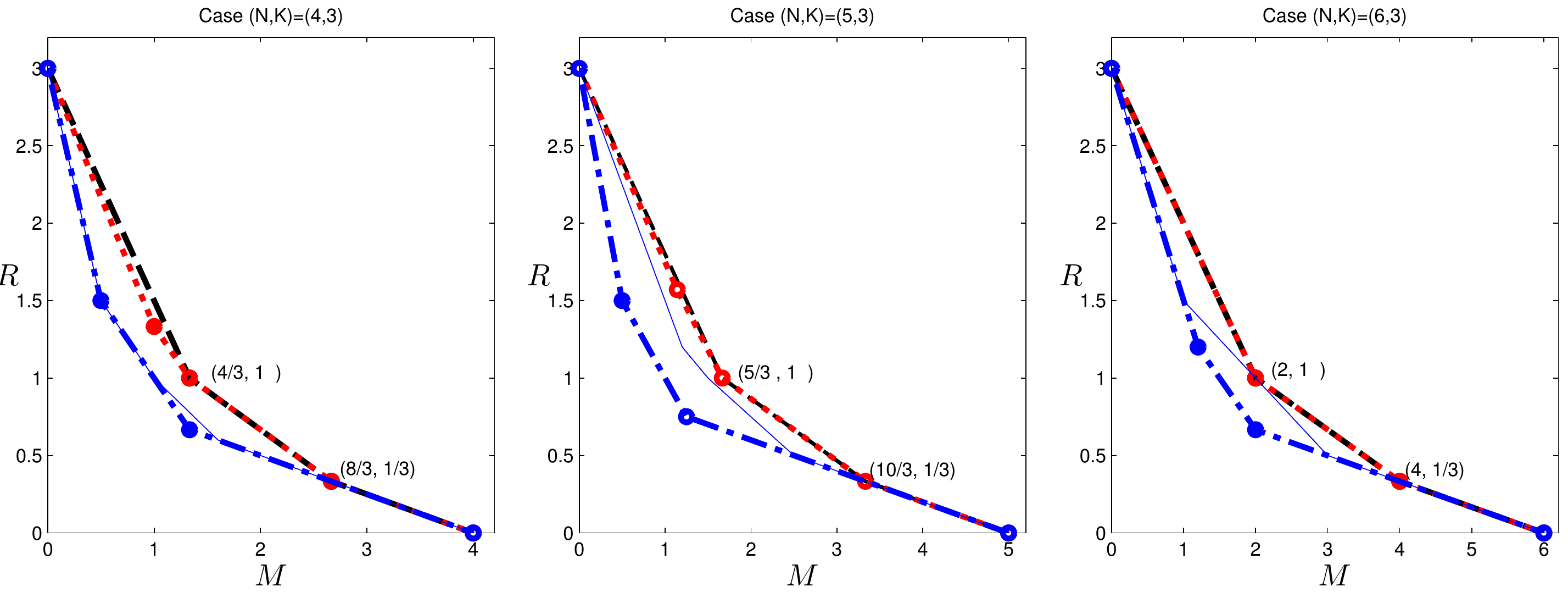}
\caption{The computed outer bounds for $(N,K)=(4,3)$, $(N,K)=(5,3)$ and $(N,K)=(6,3)$ caching systems. \label{fig:cachingN_3} The red dotted lines give the computed outer bounds, the blue dashed-dot lines are the cut-set outer bounds, the black dashed lines are the inner bound using the scheme in \cite{MaddahAliNiesen:14}, and the thin blue lines are the outer bounds given in \cite{ghasemi2017improved}. Only nontrivial outer bound corner points that match inner bounds are explicitly labeled. }
\end{figure}

In Fig. \ref{fig:cachingN_3}, we provide results for $(N,K)=(4,3)$, $(N,K)=(5,3)$ and $(N,K)=(6,3)$. Included are the computed outer bounds, the inner bound by  the scheme in \cite{MaddahAliNiesen:14}, the cut-set outer bounds, and for reference the outer bounds given in \cite{ghasemi2017improved}.  We omit the bounds in \cite{sengupta2017improved} and  \cite{ajaykrishnan2015critical} to avoid too much clutter in the plot, however they do not provide better bounds than that in \cite{ghasemi2017improved} for these cases. It can be seen that the computed bounds are in fact tight in the range $M\in[4/3,4]$ for  $(N,K)=(4,3)$, 
$M\in[5/3,5]$ for $(N,K)=(5,3)$, and tight in general for  $(N,K)=(6,3)$; in these ranges, the scheme given in \cite{MaddahAliNiesen:14} is in fact optimal. Unlike our computed bounds, the outer bound in \cite{ghasemi2017improved} does not provide additional tight results beyond those already determined using the cut-set bound, except the single point $(M,R)=(2,1)$ for $(N,K)=(6,3)$. 
\begin{figure}[tb]
\centering
\includegraphics[width=16cm]{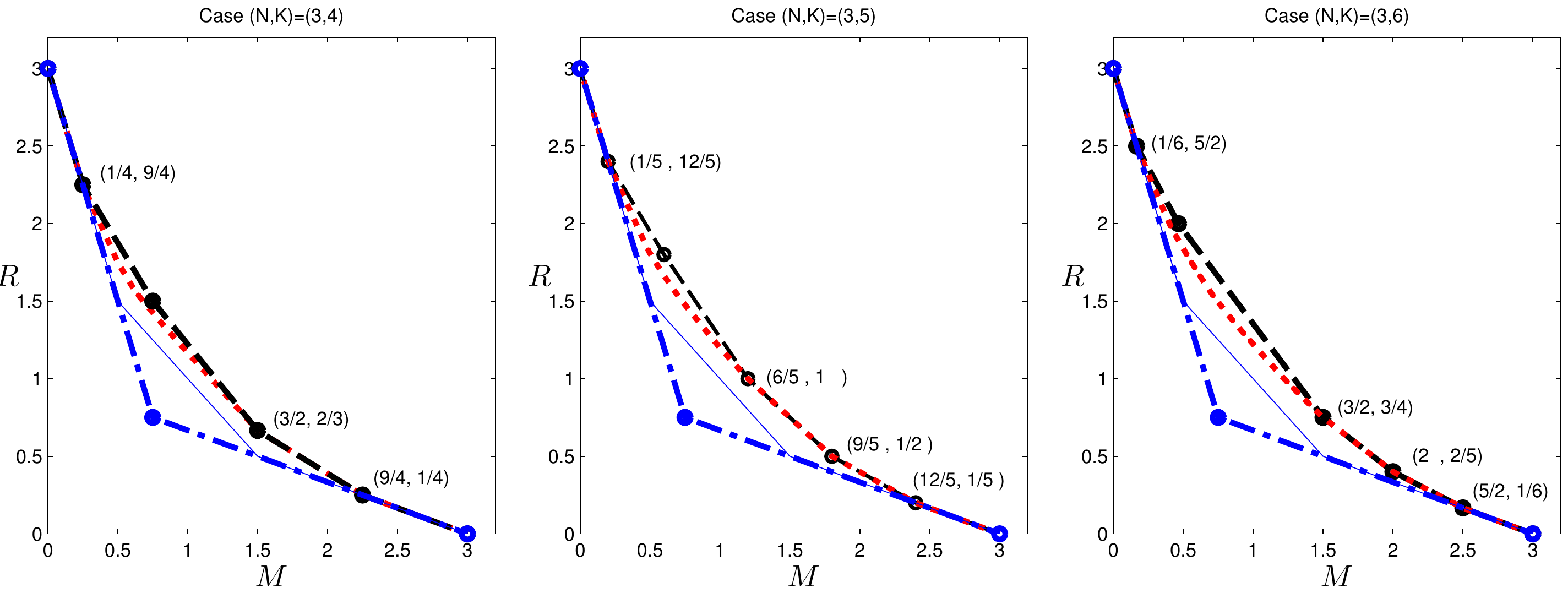}
\caption{The computed outer bounds for $(N,K)=(3,4)$, $(N,K)=(3,5)$ and $(N,K)=(3,6)$ caching systems. \label{fig:caching3_N} The red dotted lines give the computed outer bounds, the blue dashed-dot lines are the cut-set outer bounds, the black dashed lines are the inner bound using the scheme in \cite{MaddahAliNiesen:14} and \cite{tian2018caching}, and the thin blue lines are the outer bounds given in \cite{ghasemi2017improved}. Only nontrivial outer bound corner points that match inner bounds are explicitly labeled. }
\end{figure}

In Fig. \ref{fig:caching3_N}, we provide results for $(N,K)=(3,4)$, $(N,K)=(3,5)$ and $(N,K)=(3,6)$. Included are the computed outer bounds, the inner bound by the code in \cite{MaddahAliNiesen:14} and that in \cite{tian2018caching}, the cut-set outer bound, and for reference the outer bounds in \cite{ghasemi2017improved}.  The bounds in \cite{sengupta2017improved} and  \cite{ajaykrishnan2015critical} are again omitted. It can be seen that the computed bounds are in fact tight in the range $M\in[0,1/4]\cup[3/2,3]$ for  $(N,K)=(3,4)$, 
$M\in[0,1/5]\cup[6/5,3]$ for $(N,K)=(3,5)$, and $M\in[0,1/6]\cup[3/2,3]$ for  $(N,K)=(3,6)$. Generally, in the high memory regime, the scheme given in \cite{MaddahAliNiesen:14} is in fact optimal, and in the low memory regime, the schemes in \cite{chen2016fundamental,tian2018caching} are optimal. It can be see that the outer bound in \cite{ghasemi2017improved} does not provide additional tight results beyond those already determined using the cut-set bound. The bounds given above in fact provide grounds and directions for further investigation and hypotheses on the optimal tradeoff, which we are currently exploring. 

\section{Conclusion}
\label{sec:conclusion}

We presented a computer-aided investigation on the fundamental limit of the caching problem, including data-driven hypothesis forming which leads to several complete or partial characterizations of the memory-transmission-rate tradeoff, a new code construction reverse-engineered through the computed outer bounding data, and a computerized exploration approach that can reveal hidden structures in the problem and also enables us to find surprisingly strong outer bounds for larger problem cases.

It is our belief that this work provides strong evidence on the effectiveness of the computer-aided approach in the investigation of the fundamental limits of communication, data storage and data management systems. Although at the first sight, the exponential growth the LP problem would prevent any possibility of obtaining meaningful results on engineering problems of interest, our experience in \cite{Tian:JSAC13}\cite{TianLiu:15} and the current work suggest otherwise. By incorporating the structure of the problem, we develop more domain-specific tools in such investigations, and were able to obtain results that appear difficult for human experts to obtain directly. 

Our effort can be viewed as both data-driven and computational, and thus more advanced data analysis and machine learning technique may prove useful. Particularly, the computer-aided exploration approach is clearly a human-in-the-loop process, which can benefit from more automation based on reinforcement learning techniques. Moreover, the computed generated proofs may involve a large number of inequalities  and joint entropies, and more efficient classification or clustering of these inequalities  and joint entropies can reduce the human burden in the subsequent analysis. It is our hope that this work can serve as a starting point to introduce more machine intelligence and the corresponding computer-aided tools into information theory and communication research in the future.   

\begin{appendices}

\section{Finding Corner Points of the LP Outer Bounds}
\label{appendix:Lassez}

Since this is an LP problem, and also due to the problem setting, only the lower hull of the outer bound region between the two quantities $M$ and $R$ is of interest. The general algorithm in \cite{Lassez:92} is equivalent to the procedure given in Algorithm \ref{algorithm:lassez} in this specific setting. In this algorithm, the set $\mathcal{P}$ in the input is the initial extreme points of the tradeoff region, which are trivially known from the problem setting. The variables and constraints in the LP are given as outlined in Section \ref{sec:pre} for a fixed $(N,K)$ pair, which are populated and considered fixed. The output set $\mathcal{P}$ is the final computed extreme points of the outer bound. The algorithm can be intuitively explained as follows: starting with two known extreme points, if there are any other corner points, they must lie below the line segment connecting these two points, and thus an LP that minimizes the bounding plane alone the direction of this line segment must be able to find a lower value; if so, the new point is also an extreme point and we can repeat this procedure again. 

In the caching problem, the tradeoff is between two quantities $M$ and $R$. We note here if there are more than two quantities which need to be considered in the tradeoff, the algorithm is more involved, and we refer the readers to \cite{Lassez:92} and \cite{Apte:15} for more details on such settings. 

\begin{algorithm}
\label{algorithm:lassez}
    \SetKwInOut{Input}{Input}
    \SetKwInOut{Output}{Output}
    \Input{$N$, $K$, $\mathcal{P}=\{(N,0),(0,\min(N,K))\}$}
    \Output{$\mathcal{P}$}
    $n=2$; $i=1$;\\
   \While{$i<n$}{
   Compute the line segment connecting $i$-th and $(i+1)$-th $(M,R)$ pairs in $\mathcal{P}$, as $M+\alpha R=\beta$;\\
   Set the objective of the LP as $M+\alpha R$, and solve LP for solution $(M^*,R^*)$ and objective $\beta^*$;\\   
      \eIf{$\beta^*<\beta$}
      {
        Insert $(M^*,R^*)$ in $\mathcal{P}$ between the $i$-th and $(i+1)$-th $(M,R)$ pairs;\\
        $n=n+1$;        
      }      
      {
      	 $i=i+1$;
      }
   }
    \caption{An algorithm to identify the corner points of the LP outer bound}
\end{algorithm}

\section{Proofs of Proposition \ref{prop:NK32} and Proposition \ref{prop:NK42}}
\label{appendix:propNK32_42}

\begin{table}
\centering
\begin{tabular}{|c|c|}
\hline
$T_{ 1}$ & $F$ \\
$T_{ 2}$ & $R$ \\
$T_{ 3}$ & $H(X_{1,2})$ \\
$T_{ 4}$ & $H(W_{1})$ \\
$T_{ 5}$ & $H(W_{1},W_{2},W_{3})$ \\
$T_{ 6}$ & $H(Z_{1})$ \\
$T_{ 7}$ & $H(Z_{1},X_{1,2})$ \\
$T_{ 8}$ & $H(Z_{1},W_{1})$ \\
$T_{ 9}$ & $H(Z_{1},Z_{2},X_{1,2})$ \\
$T_{10}$ & $H(Z_{1},Z_{2},W_{1})$ \\\hline
\end{tabular}
\caption{Terms needed to prove Proposition \ref{prop:NK32}\label{table:NK32}.}
\end{table}

\begin{table}
\centering
\begin{tabular}{|cccccccccc||c||c|}
\hline
$T_{ 1}$  &$T_{ 2}$  &$T_{ 3}$  &$T_{ 4}$  &$T_{ 5}$  &$T_{ 6}$  &$T_{ 7}$  &$T_{ 8}$  &$T_{ 9}$  &$T_{10}$ & & \\
\hline\hline
          &$  2$     &$ -2$     &          &          &          &          &          &          &          &$2(R-H(X_{1,2}))\geq0$ &(1)  \\
          &          &          &$ -1$     &          &          &          &$  2$     &          &$ -1$   &$I(Z_1;Z_2|W_1)\geq0$&(2) \\
          &          &$  2$     &          &          &$  2$     &$ -2$     &          &          &        &$2I(X_{1,2};Z_1)\geq0$ &(3)  \\
          &          &          &          &$ -1$     &          &          &          &$  2$     &$ -1$   &$I(X_{1,2};X_{1,3}|Z_1,Z_2,W_1)\geq0$&(4) \\
          &          &          &          &          &          &$  2$     &$ -2$     &$ -2$     &$  2$  &$2I(X_{1,2};Z_2|Z_1,W_1)\geq0$ &(5)  \\
$ -1$     &          &          &$  1$     &          &          &          &          &          &          &$H(W_1)-F \geq0$&(6) \\
$ -3$     &          &          &          &$  1$     &          &          &          &          &          &$H(W_1,W_2,W_3)-3F\geq0$ &(7) \\
\hline\hline
$ -4$     &   2      &          &          &           &  $2$      &          &          &          &        & $2R+ 2H(Z_1)-4F\geq 0$ & \\
\hline
\end{tabular}
\caption{Proof by Tabulation of Proposition \ref{prop:NK32}, with terms defined in Table \ref{table:NK32}.\label{table:NK32_Proof}}
\end{table}

The proof of the Proposition \ref{prop:NK32} is given in the Table \ref{table:NK32}-\ref{table:NK32_Proof}, and that of the Proposition \ref{prop:NK42} is given in the Table \ref{table:NK42}-\ref{table:NK42_Proof}. Each row in Table  \ref{table:NK32_Proof} and Table \ref{table:NK42_Proof}, except the last rows, are simple and known information inequalities, up to the symmetry defined in Section \ref{sec:symmetry}. The last rows in Table  \ref{table:NK32_Proof} and Table \ref{table:NK42_Proof} are the sum of all previous rows, which are the sought-after inequalities and they are simply the consequences of the known inequalities summed together. When represented in this form, the correctness of the proof is immediate, since the columns representing quantities not present in the final bound cancel out each other when being summed together. The rows in Table \ref{table:NK32_Proof} are labeled and it has more details in order to illustrate the meaning and usage of the tabulation proof in the example we provide next.

As mentioned previously, each row in Table \ref{table:NK32_Proof} is an information inequality, which involves multiple joint entropies but can also be represented in a mutual information form. For example row (2) is read as 
\begin{align}
2T_8-T_4-T_{10}\geq 0,
\end{align}
and in the last but one column of Table \ref{table:NK32_Proof}, an information inequality is given which is an equivalent  representation as a mutual information quantity
\begin{align}
I(Z_1;Z_2|W_1)\geq 0,
\end{align}
which can be seen by simply expanding the mutual information as
\begin{align}
I(Z_1;Z_2|W_1)&=H(Z_1,W_1)+H(Z_2,W_1)-H(W_1)-H(Z_1,Z_2,W_1)\nonumber\\
&=2H(Z_1,W_1)-H(W_1)-H(Z_1,Z_2,W_1)\nonumber\\
&=2T_8-T_4-T_{10}.
\end{align}
Directly summing up these information inequalities and cancel out redundant terms will result in the bound $2R+ 2H(Z_1)-4F\geq 0$, which clearly can be used to write $2R+ 2M-4F\geq 0$.

Using these proof tables, one can write down different versions of proofs, and one such example is provided next based on Table  \ref{table:NK32}-\ref{table:NK32_Proof} for Proposition \ref{prop:NK32} by invoking the inequalities in Table \ref{table:NK32_Proof} one by one. 
\begin{align}
2M+2R&\stackrel{(1)}{\geq} 2H(X_{1,2})+2H(Z_1)\stackrel{(3)}{\geq} 2H(Z_1,X_{1,2})\nonumber\\
&\stackrel{(5)}{\geq} 2H(Z_1,X_{1,2})-2I(X_{1,2};Z_2|Z_1,W_1)\nonumber\\
&=2H(Z_1,X_{1,2},W_1)-2I(X_{1,2};Z_2|Z_1,W_1)\nonumber\\
&\stackrel{(c)}{=}2H(Z_1,Z_2,W_1,X_{1,2})+2H(Z_1,W_1)-2H(Z_1,Z_2,W_1)\nonumber\\
&\stackrel{(2)}{\geq} 2H(Z_1,Z_2,W_1,X_{1,2})+2H(Z_1,W_1)-2H(Z_1,Z_2,W_1)-I(Z_1;Z_2|W_1)\nonumber\\
&\stackrel{}{=}2H(Z_1,Z_2,W_1,X_{1,2})-H(Z_1,Z_2,W_1)+H(W_1)\nonumber\\
&\stackrel{(4)}\geq 2H(Z_1,Z_2,W_1,X_{1,2})-H(Z_1,Z_2,W_1)+H(W_1)-I(X_{1,2};X_{1,3}|Z_1,Z_2,W_1)\nonumber\\
&\stackrel{}=H(W_1)+H(W_1,W_2,W_3)\nonumber\\
&\stackrel{(6,7)}\geq 4F,
\end{align}
where the inequalities match precisely the rows in Table \ref{table:NK32_Proof}, and the equality labeled (c) indicates the decoding requirement is used. In this version of the proof, we applied the inequalities in the order of (1)-(3)-(5)-(2)-(4)-(6,7), but this  is by no means critical, as any order will yield a valid proof. One can similarly produce many different versions of proofs for Proposition \ref{prop:NK42} based on Table \ref{table:NK42}-\ref{table:NK42_Proof}.

\begin{table}
\centering
\begin{tabular}{|c|c|}
\hline
$T_{ 1}$ & $F$ \\
$T_{ 2}$ & $R$ \\
$T_{ 3}$ & $H(X_{1,2})$ \\
$T_{ 4}$ & $H(W_{1})$ \\
$T_{ 5}$ & $H(W_{1},X_{1,2})$ \\
$T_{ 6}$ & $H(W_{1},X_{1,3},X_{2,1})$ \\
$T_{ 7}$ & $H(W_{1},W_{2})$ \\
$T_{ 8}$ & $H(W_{1},W_{3},X_{1,2})$ \\
$T_{ 9}$ & $H(W_{1},W_{2},W_{3},W_{4})$ \\
$T_{10}$ & $H(Z_{1})$ \\
$T_{11}$ & $H(Z_{1},X_{1,2})$ \\
$T_{12}$ & $H(Z_{1},X_{1,3},X_{2,1})$ \\
$T_{13}$ & $H(Z_{1},X_{1,4},X_{2,3})$ \\
$T_{14}$ & $H(Z_{1},W_{1})$ \\
$T_{15}$ & $H(Z_{1},W_{3},X_{1,2})$ \\
$T_{16}$ & $H(Z_{1},W_{2},X_{1,2})$ \\
$T_{17}$ & $H(Z_{1},W_{1},W_{2})$ \\
$T_{18}$ & $H(Z_{1},W_{2},W_{3},X_{1,2})$ \\
$T_{19}$ & $H(Z_{1},Z_{2},X_{1,2})$ \\
$T_{20}$ & $H(Z_{1},Z_{2},X_{1,3},X_{2,1})$ \\
$T_{21}$ & $H(Z_{1},Z_{2},W_{3},X_{1,2})$ \\
$T_{22}$ & $H(Z_{1},Z_{2},W_{1},W_{2})$ 
\\\hline
\end{tabular}
\caption{Terms needed to prove Proposition \ref{prop:NK42}\label{table:NK42}.}
\end{table}

\begin{table}
\setlength{\tabcolsep}{3pt}
\centering
\begin{tabular}{|cccccccccccccccccccccc|}
\hline
$T_{ 1}$  &$T_{ 2}$  &$T_{ 3}$  &$T_{ 4}$  &$T_{ 5}$  &$T_{ 6}$  &$T_{ 7}$  &$T_{ 8}$  &$T_{ 9}$  &$T_{10}$  &$T_{11}$  &$T_{12}$  &$T_{13}$  &$T_{14}$  &$T_{15}$  &$T_{16}$  &$T_{17}$  &$T_{18}$  &$T_{19}$  &$T_{20}$  &$T_{21}$  &$T_{22}$  \\\hline\hline
          &          &          &          &          &          &          &          &          &          &          &          &          &          &          &          &          &          &$  1$     &          &          &$ -1$     \\
          &$  8$     &$ -8$     &          &          &          &          &          &          &          &          &          &          &          &          &          &          &          &          &          &          &           \\
          &          &$  8$     &          &          &          &          &          &          &$  8$     &$ -8$     &          &          &          &          &          &          &          &          &          &          &           \\
          &          &          &          &          &          &          &          &          &$ -2$     &$  4$     &          &$ -2$     &          &          &          &          &          &          &          &          &           \\
          &          &          &          &          &          &          &          &$ -1$     &          &          &          &          &          &          &          &          &          &$ -1$     &$  2$     &          &           \\
          &          &          &          &          &$ -2$     &          &          &          &          &          &$  4$     &          &          &          &          &          &          &          &$ -2$     &          &           \\
          &          &          &$ -2$     &$  2$     &          &          &          &          &          &          &          &          &$  2$     &          &$ -2$     &          &          &          &          &          &           \\
          &          &          &          &          &          &$ -1$     &$  1$     &          &          &          &          &          &          &          &          &$  1$     &$ -1$     &          &          &          &           \\
          &          &          &          &          &          &          &          &          &          &$  2$     &$ -2$     &          &$ -2$     &          &$  2$     &          &          &          &          &          &           \\
          &          &          &          &          &          &          &          &          &          &          &          &          &          &$  1$     &          &$ -1$     &          &          &          &$ -1$     &$  1$     \\
          &          &          &          &$ -2$     &$  2$     &          &          &          &          &$  2$     &$ -2$     &          &          &          &          &          &          &          &          &          &           \\
          &          &          &          &          &          &          &$ -1$     &          &          &          &          &          &          &$  1$     &          &          &$  1$     &          &          &$ -1$     &           \\
          &          &          &          &          &          &          &          &$ -2$     &          &          &          &$  2$     &          &$ -2$     &          &          &          &          &          &$  2$     &           \\
$ -2$     &          &          &$  2$     &          &          &          &          &          &          &          &          &          &          &          &          &          &          &          &          &          &           \\
$ -2$     &          &          &          &          &          &$  1$     &          &          &          &          &          &          &          &          &          &          &          &          &          &          &           \\
$-12$     &          &          &          &          &          &          &          &$  3$     &          &          &          &          &          &          &          &          &          &          &          &          &           \\
\hline\hline
$-16$     &   $8$  &          &          &          &          &          &          &          &     $6$       &            &    &          &          &          &          &          &          &          &          &          &           \\
\hline
\end{tabular}
\caption{Proof by Tabulation of Proposition \ref{prop:NK42}, with terms defined in Table \ref{table:NK42}.\label{table:NK42_Proof}}
\end{table}

\section{Proofs of Lemma \ref{lemma:peeling} and Theorem \ref{theorem:NK_N_2}}
\label{appendix:theoremN2}

\begin{IEEEproof}[Proofs of Lemma \ref{lemma:peeling}]
We first write the following chain of inequalities
\begin{align}
(N-n)H(Z_1,W_{[1:n]},X_{n,n+1})&=(N-n)\left[H(Z_1,W_{[1:n]})+H(X_{n,n+1}|Z_1,W_{[1:n]})\right]\nonumber\\
&\stackrel{(a)}{=}(N-n)H(Z_1,W_{[1:n]})+\sum_{i=n+1}^{N}H(X_{n,i}|Z_1,W_{[1:n]})\nonumber\\
&\geq (N-n)H(Z_1,W_{[1:n]})+H(X_{n,[n+1:N]}|Z_1,W_{[1:n]})\nonumber\\
&=(N-n-1)H(Z_1,W_{[1:n]})+H(X_{n,[n+1:N]},Z_1,W_{[1:n]}), \label{eqn:stop1}
\end{align}
where $(a)$ is because of the file-index-symmetry. 
Next notice that by the user-index-symmetry
\begin{align}
H(Z_1,W_{[1:n]})=H(Z_2,W_{[1:n]}),
\end{align}
which implies that
\begin{align}
H(Z_1,W_{[1:n]})+H(X_{n,[n+1:N]},Z_1,W_{[1:n]})&\geq H(Z_2,W_{[1:n]})+H(X_{n,[n+1:N]},W_{[1:n]})\nonumber\\
&\stackrel{(b)}{\geq} H(W_{[1:n]})+H(X_{n,[n+1:N]},Z_2,W_{[1:n]})\nonumber\\
&\stackrel{(c)}{=} H(W_{[1:n]})+H(X_{n,[n+1:N]},Z_2,W_{[1:N]})\nonumber\\
&= H(W_{[1:n]})+H(W_{[1:N]})=N+n,\label{eqn:stop2}
\end{align}
where $(b)$ is by the sub-modularity of the entropy function, and $(c)$ is because of (\ref{eqn:reconstruction}). 
Now substituting (\ref{eqn:stop2}) into (\ref{eqn:stop1}) gives  (\ref{eqn:lemma}), which completes the proof.
\end{IEEEproof}

We are now ready to prove Theorem \ref{theorem:NK_N_2}.
\begin{IEEEproof}[Proof of Theorem \ref{theorem:NK_N_2}]
For $N\geq 3$, it can be verified that the three corner points of the given tradeoff region are 
\begin{align}
(0,2),\quad(\frac{N}{2},\frac{1}{2}),\quad(N,0),
\end{align}
which are achievable using the codes given in \cite{MaddahAliNiesen:14}. The outer bound $M+NR\geq N$ can also be obtained as one of the cut-set outer bounds in \cite{MaddahAliNiesen:14}, and it only remains to show that the inequality $3M+NR\geq 2N$ is true. For this purpose, we claim that for any integer $n\in\{1,2,\ldots,N-2\}$
\begin{align}
&3M+NR\geq 3\sum_{j=1}^{n}\left[\frac{N+j}{N-j} \prod_{i=1}^{j-1} \frac{N-(i+2)}{N-i}\right]
+3\prod_{j=1}^n \frac{N-(j+2)}{N-j}H(Z_1,W_{[1:n]})\nonumber\\
&\qquad\qquad\qquad+[N-(n+2)]\prod_{j=1}^{n-1} \frac{N-(j+2)}{N-j}H(X_{1,2}), \label{eqn:claim}
\end{align}
which we prove next by induction. 

First notice that
\begin{align}
&3M+NR\geq 3H(Z_1)+NH(X_{1,2})\nonumber\\
&\geq 3H(Z_1,X_{1,2})+(N-3)H(X_{1,2})\nonumber\\
&\stackrel{ (\ref{eqn:reconstruction})}{=}3H(Z_1,W_1,X_{1,2})+(N-3)H(X_{1,2})\nonumber\\
&\stackrel{(d)}\geq \frac{3(N-3)}{N-1}H(Z_1,W_1)+\frac{3(N+1)}{N-1}+(N-3)H(X_{1,2}), 
\end{align}
where we wrote $(\ref{eqn:reconstruction})$ to mean by Eqn. (\ref{eqn:reconstruction}), and $(d)$ is by Lemma \ref{lemma:peeling} with $n=1$. 
This is precisely the claim when $n=1$, when we take the convention $\prod_{k}^n( \cdot)=1$ when $n<k$ in (\ref{eqn:claim}).

Assume the claim is true for $n=n^*$, and we next prove it is true for $n=n^*+1$. Notice that the second and third terms in (\ref{eqn:claim}) has a common factor 
\begin{align}
\frac{N-(n^*+2)}{N-n^*}\prod_{j=1}^{n^*-1} \frac{N-(j+2)}{N-j}=\prod_{j=1}^{n^*} \frac{N-(j+2)}{N-j},
\label{eqn:ind1}
\end{align}
using which to normalize the last two terms gives 
\begin{align}
&3H(Z_1,W_{[1:n^*]})+(N-n^*)H(X_{1,2})\nonumber\\
&\stackrel{(e)}{=}3[H(Z_1,W_{[1:n^*]})+H(X_{n^*+1,n^*+2})]\nonumber\\
&\qquad\qquad\qquad\qquad+(N-n^*-3)H(X_{1,2})\nonumber\\
&\geq3[H(Z_1,W_{[1:n^*]},X_{n^*+1,n^*+2})]\nonumber\\
&\qquad\qquad\qquad\qquad+(N-n^*-3)H(X_{1,2})\nonumber\\
&\stackrel{(\ref{eqn:reconstruction})}{=}3[H(Z_1,W_{[1:n^*+1]},X_{n^*+1,n^*+2})]\nonumber\\
&\qquad\qquad\qquad\qquad+(N-n^*-3)H(X_{1,2})\nonumber\\
&\stackrel{(f)}{\geq} 3\frac{(N-n^*-3)}{N-n^*-1}H(Z_1,W_{[1:n^*+1]})+3\frac{N+n^*+1}{N-n^*-1}\nonumber\\
&\qquad\qquad\qquad\qquad+(N-n^*-3)H(X_{1,2}),\label{eqn:ind2}
\end{align}
where $(e)$ is by the file-index-symmetry, and $(f)$ is by Lemma \ref{lemma:peeling}. 
Substituting (\ref{eqn:ind1}) and (\ref{eqn:ind2}) into (\ref{eqn:claim}) for the case $n=n^*$ gives exactly (\ref{eqn:claim}) for the case $n=n^*+1$, which completes the proof for (\ref{eqn:claim}).

It remains to show that (\ref{eqn:claim}) implies the bound $3M+NR\geq 2N$. For this purpose, notice that when $n=N-2$, the last two terms in (\ref{eqn:claim}) reduce to zero, and thus we only need to show that 
\begin{align}
Q(N)\triangleq 3\sum_{j=1}^{N-2}\left[\frac{N+j}{N-j} \prod_{i=1}^{j-1} \frac{N-(i+2)}{N-i}\right]=2N.
\end{align}
For each summand, we have
\begin{align}
\frac{N+j}{N-j} \prod_{i=1}^{j-1} \frac{N-(i+2)}{N-i}
&=\frac{N+j}{N-j}\left[\frac{N-3}{N-1}\frac{N-4}{N-2}\frac{N-5}{N-3}\ldots\frac{N-j-1}{N-j+1}\right]\nonumber\\
&=\frac{(N-j-1)(N+j)}{(N-1)(N-2)}.
\end{align}
Thus we have
\begin{align*}
Q(N) = \frac{3}{(N-1)(N-2)}\sum_{j=1}^{N-2}(N-j-1)(N+j)=2N,
\end{align*}
where we have used the well-known formula for the sum of integer squares. The proof is thus complete.
\end{IEEEproof}

\section{Proof of Proposition \ref{prop:NK23}}

We first consider the achievability, for which only the achievability of the following extremal points needs to be shown because of the polytope structure of the region:
\begin{align}
(M,R)\in\left\{(0,2),\left(\frac{1}{3},\frac{4}{3}\right),\left(\frac{4}{3},\frac{1}{3}\right),(2,0)\right\}.
\end{align}
Achieving the rate pairs $(0,2)$ and $(2,0)$ is trivial. The scheme in \cite{MaddahAliNiesen:14} can achieve the rate pair $\left(\frac{4}{3},\frac{1}{3}\right)$. The rate pair  $\left(\frac{1}{3},\frac{4}{3}\right)$ can be achieved by a scheme given in \cite{chen2016fundamental}, which is a generalization of a special scheme given in \cite{MaddahAliNiesen:14}. To prove the converse, we note first that the cut-set-based approach can provide all bounds in (\ref{eqn:bounds2_3}) except 
\begin{align}
3M+3R\geq 5, 
\end{align}
which is a new inequality. As mentioned earlier, this inequality is a special case of Theorem \ref{theorem:2K} and there is no need to prove it separately.

\section{Proof of Proposition \ref{prop:NK24}}
\label{appendix:Prop4_9}

The inequality $14M+11R\geq 20$ is proved using Table \ref{table:NK24_1}-\ref{table:NK24_1_Proof}, and the inequality $9M+8R\geq 14$ is proved using Table \ref{table:NK24_2}- \ref{table:NK24_2_Proof}. 

\begin{table}
\centering
\begin{tabular}{|c|c|}
\hline
$T_{ 1}$ & $F$ \\
$T_{ 2}$ & $R$ \\
$T_{ 3}$ & $H(X_{1,1,1,2})$ \\
$T_{ 4}$ & $H(X_{1,1,2,2})$ \\
$T_{ 5}$ & $H(W_{1})$ \\
$T_{ 6}$ & $H(W_{1},X_{1,1,1,2})$ \\
$T_{ 7}$ & $H(W_{1},X_{1,1,2,2})$ \\
$T_{ 8}$ & $H(W_{1},W_{2})$ \\
$T_{ 9}$ & $H(Z_{1})$ \\
$T_{10}$ & $H(Z_{1},X_{1,1,1,2})$ \\
$T_{11}$ & $H(Z_{1},X_{1,1,2,2})$ \\
$T_{12}$ & $H(Z_{4},X_{1,1,1,2})$ \\
$T_{13}$ & $H(Z_{1},W_{1})$ \\
$T_{14}$ & $H(Z_{1},Z_{2},X_{1,1,1,2})$ \\
$T_{15}$ & $H(Z_{1},Z_{2},X_{1,1,2,2})$ \\
$T_{16}$ & $H(Z_{1},Z_{2},W_{1})$ \\
$T_{17}$ & $H(Z_{1},Z_{2},Z_{3},X_{1,1,1,2})$ \\
$T_{18}$ & $H(Z_{1},Z_{2},Z_{3},W_{1})$ 
\\\hline
\end{tabular}
\caption{Terms needed to prove Proposition \ref{prop:NK24}, inequality $14M+11R\geq 20$. \label{table:NK24_1}}
\end{table}

\begin{table}
\setlength{\tabcolsep}{3pt}
\centering
\begin{tabular}{|cccccccccccccccccc|}
\hline
$T_{ 1}$  &$T_{ 2}$  &$T_{ 3}$  &$T_{ 4}$  &$T_{ 5}$  &$T_{ 6}$  &$T_{ 7}$  &$T_{ 8}$  &$T_{ 9}$  &$T_{10}$  &$T_{11}$  &$T_{12}$  &$T_{13}$  &$T_{14}$  &$T_{15}$  &$T_{16}$  &$T_{17}$  &$T_{18}$  \\
\hline\hline
          &          &          &          &          &          &          &          &          &          &          &          &          &          &          &          &$  2$     &$ -2$     \\
          &$  8$     &$ -8$     &          &          &          &          &          &          &          &          &          &          &          &          &          &          &           \\
          &$  3$     &          &$ -3$     &          &          &          &          &          &          &          &          &          &          &          &          &          &           \\
          &          &$  6$     &          &          &          &          &          &$  6$     &$ -6$     &          &          &          &          &          &          &          &           \\
          &          &$  4$     &          &          &          &          &          &$  4$     &          &          &$ -4$     &          &          &          &          &          &           \\
          &          &          &$  4$     &          &          &          &          &$  4$     &          &$ -4$     &          &          &          &          &          &          &           \\
          &          &          &          &          &$ -4$     &          &          &          &$  8$     &          &          &          &$ -4$     &          &          &          &           \\
          &          &          &          &          &          &          &          &          &$ -2$     &          &          &          &$  4$     &          &          &$ -2$     &           \\
          &          &          &$ -1$     &          &          &$  2$     &$ -1$     &          &          &          &          &          &          &          &          &          &           \\
          &          &          &          &          &          &$ -2$     &          &          &          &$  4$     &          &          &          &$ -2$     &          &          &           \\
          &          &          &          &$ -2$     &$  2$     &          &$ -2$     &          &          &          &          &$  2$     &          &          &          &          &           \\
          &          &          &          &          &          &          &$ -2$     &          &          &          &$  2$     &$ -2$     &          &          &$  2$     &          &           \\
          &          &          &          &          &          &          &$ -2$     &          &          &          &          &          &          &$  2$     &$ -2$     &          &$  2$     \\
          &          &$ -2$     &          &          &$  2$     &          &$ -2$     &          &          &          &$  2$     &          &          &          &          &          &           \\
$ -2$     &          &          &          &$  2$     &          &          &          &          &          &          &          &          &          &          &          &          &           \\
$-18$     &          &          &          &          &          &          &$  9$     &          &          &          &          &          &          &          &          &          &           \\
\hline\hline
$-20$     &   $11$  &          &          &          &          &          &          &   $14$   &          &            &    &          &          &          &          &          &             \\
\hline
\end{tabular}
\caption{Tabulation proof of Proposition \ref{prop:NK24} inequality $14M+11R\geq 20$, with terms defined in Table \ref{table:NK24_1}.\label{table:NK24_1_Proof}}
\end{table}

\begin{table}
\centering
\begin{tabular}{|c|c|}
\hline
$T_{ 1}$ & $F$ \\
$T_{ 2}$ & $R$ \\
$T_{ 3}$ & $H(X_{1,1,1,2})$ \\
$T_{ 4}$ & $H(X_{1,1,2,2})$ \\
$T_{ 5}$ & $H(W_{1})$ \\
$T_{ 6}$ & $H(W_{1},X_{1,1,1,2})$ \\
$T_{ 7}$ & $H(W_{1},X_{1,1,2,2})$ \\
$T_{ 8}$ & $H(W_{1},X_{1,2,1,1},X_{1,1,2,2})$ \\
$T_{ 9}$ & $H(W_{1},W_{2})$ \\
$T_{10}$ & $H(Z_{1})$ \\
$T_{11}$ & $H(Z_{1},X_{1,1,1,2})$ \\
$T_{12}$ & $H(Z_{1},X_{1,1,2,2})$ \\
$T_{13}$ & $H(Z_{1},X_{1,2,1,1},X_{1,1,2,2})$ \\
$T_{14}$ & $H(Z_{1},W_{1})$ \\
$T_{15}$ & $H(Z_{1},Z_{2},X_{1,1,1,2})$ \\
$T_{16}$ & $H(Z_{1},Z_{2},X_{1,1,2,2})$ \\
$T_{17}$ & $H(Z_{1},Z_{2},W_{1})$ \\
$T_{18}$ & $H(Z_{1},Z_{2},Z_{3},X_{1,1,1,2})$ \\
$T_{19}$ & $H(Z_{1},Z_{2},Z_{3},W_{1})$ 
\\\hline
\end{tabular}
\caption{Terms needed to prove Proposition \ref{prop:NK24}, inequality $9M+8R\geq 14$. \label{table:NK24_2}}
\end{table}

\begin{table}
\setlength{\tabcolsep}{3pt}
\centering
\begin{tabular}{|ccccccccccccccccccc|}
\hline
$T_{ 1}$  &$T_{ 2}$  &$T_{ 3}$  &$T_{ 4}$  &$T_{ 5}$  &$T_{ 6}$  &$T_{ 7}$  &$T_{ 8}$  &$T_{ 9}$  &$T_{10}$  &$T_{11}$  &$T_{12}$  &$T_{13}$  &$T_{14}$  &$T_{15}$  &$T_{16}$  &$T_{17}$  &$T_{18}$  &$T_{19}$  \\
\hline\hline
          &          &          &          &          &          &          &          &          &          &          &          &          &          &          &          &          &$  1$     &$ -1$     \\
          &$  4$     &$ -4$     &          &          &          &          &          &          &          &          &          &          &          &          &          &          &          &           \\
          &$  4$     &          &$ -4$     &          &          &          &          &          &          &          &          &          &          &          &          &          &          &           \\
          &          &          &          &          &          &          &$ -1$     &          &          &          &          &$  1$     &          &          &          &          &          &           \\
          &          &$  4$     &          &          &          &          &          &          &$  4$     &$ -4$     &          &          &          &          &          &          &          &           \\
          &          &          &$  5$     &          &          &          &          &          &$  5$     &          &$ -5$     &          &          &          &          &          &          &           \\
          &          &          &          &          &$ -2$     &          &          &          &          &$  4$     &          &          &          &$ -2$     &          &          &          &           \\
          &          &          &          &          &          &          &          &          &          &$ -1$     &          &          &          &$  2$     &          &          &$ -1$     &           \\
          &          &          &$ -1$     &          &          &$  2$     &          &$ -1$     &          &          &          &          &          &          &          &          &          &           \\
          &          &          &          &          &          &$ -1$     &          &          &          &          &$  2$     &          &          &          &$ -1$     &          &          &           \\
          &          &          &          &$ -2$     &$  2$     &          &          &$ -2$     &          &          &          &          &$  2$     &          &          &          &          &           \\
          &          &          &          &          &          &          &          &$ -1$     &          &          &$  1$     &          &$ -1$     &          &          &$  1$     &          &           \\
          &          &          &          &          &          &          &          &          &          &$  1$     &$  1$     &$ -1$     &$ -1$     &          &          &          &          &           \\
          &          &          &          &          &          &          &          &$ -1$     &          &          &          &          &          &          &$  1$     &$ -1$     &          &$  1$     \\
          &          &          &          &          &          &$ -1$     &$  1$     &$ -1$     &          &          &$  1$     &          &          &          &          &          &          &           \\
$ -2$     &          &          &          &$  2$     &          &          &          &          &          &          &          &          &          &          &          &          &          &           \\
$-12$     &          &          &          &          &          &          &          &$  6$     &          &          &          &          &          &          &          &          &          &           \\
\hline\hline
$-14$     &   $8$  &          &          &          &          &          &          &       &  $9$         &            &    &          &          &          &          &          &             &\\
\hline
\end{tabular}
\caption{Tabulation proof of Proposition \ref{prop:NK24} inequality $9M+8R\geq 14$, with terms defined in Table \ref{table:NK24_2}.\label{table:NK24_2_Proof}}
\end{table}

\section{Proof of Lemma  \ref{lemma:2K}}
\label{appendix:lemma2K}

\begin{IEEEproof}[Proof of Lemma \ref{lemma:2K}]
We prove this lemma by induction. First consider the case when $k=K-1$, for which we write
\begin{align}
&2H(Z_1,W_1,X_{\rightarrow [2:K-1]})\nonumber\\
&\stackrel{(a)}{=}H(Z_1,W_1,X_{\rightarrow [2:K-1]})+H(Z_1,W_1,X_{\rightarrow [2:K-2]},X_{\rightarrow K})\nonumber\\
&=H(X_{\rightarrow K-1}|Z_1,W_1,X_{\rightarrow [2:K-2]})+H(X_{\rightarrow K}|Z_1,W_1,X_{\rightarrow [2:K-2]})+2H(Z_1,W_1,X_{\rightarrow [2:K-2]})\nonumber\\
&\geq H(Z_1,W_1,X_{\rightarrow [2:K]})+H(Z_1,W_1,X_{\rightarrow [2:K-2]}), \label{eqn:Kminus1_start}
\end{align}
where $(a)$ is by file-index symmetry.
The first quantity can be lower bounded as
\begin{align}
H(Z_1,W_1,X_{\rightarrow [2:K]})\geq H(W_1,X_{\rightarrow [2:K]}),
\end{align}
which leads to a bound on the following sum
\begin{align}
&H(Z_1,W_1,X_{\rightarrow [2:K]})+H(Z_1,W_1,X_{\rightarrow [2:K-1]})\nonumber\\
&\geq H(W_1,X_{\rightarrow [2:K]})+H(Z_1,W_1,X_{\rightarrow [2:K-1]})\nonumber\\
&\geq H(X_{\rightarrow K}|W_1,X_{\rightarrow [2:K-1]})+H(Z_1|W_1,X_{\rightarrow [2:K-1]})+2H(W_1,X_{\rightarrow [2:K-1]})\nonumber\\
&\stackrel{(b)}{=}H(X_{\rightarrow K}|W_1,X_{\rightarrow [2:K-1]})+H(Z_K|W_1,X_{\rightarrow [2:K-1]})+2H(W_1,X_{\rightarrow [2:K-1]})\nonumber\\
&\geq H(Z_K,X_{\rightarrow K}|W_1,X_{\rightarrow [2:K-1]})+2H(W_1,X_{\rightarrow [2:K-1]})\nonumber\\
&\stackrel{(c)}{=}H(Z_K,X_{\rightarrow K},W_2|W_1,X_{\rightarrow [2:K-1]})+2H(W_1,X_{\rightarrow [2:K-1]})\nonumber\\
&\stackrel{(d)}{=}H(W_1,W_2)+H(W_1,X_{\rightarrow [2:K-1]}), \label{eqn:Kminus1}
\end{align}
where $(b)$ is by the user index symmetry, and $(c)$ is because $Z_K$ and $X_{1,1,\ldots,2}$ can be used to produce $W_2$, and $(d)$ is because all other variables are deterministic functions of $(W_1,W_2)$.
Adding $H(Z_1,W_1,X_{\rightarrow [2:K-1]})$ on both sides of (\ref{eqn:Kminus1_start}), and then apply (\ref{eqn:Kminus1}) leads to
\begin{align}
3H(Z_1,W_1,X_{\rightarrow [2:K-1]})&\geq H(Z_1,W_1,X_{\rightarrow [2:K-2]})+H(W_1,X_{\rightarrow [2:K-1]})+H(W_1,W_2)\nonumber\\
&\stackrel{(e)}{=}H(Z_{K-1},W_1,X_{\rightarrow [2:K-2]})+H(W_1,X_{\rightarrow [2:K-1]})+H(W_1,W_2)\nonumber\\
&\stackrel{(f)}{\geq} H(Z_{K-1},W_1,X_{\rightarrow [2:K-1]})+H(W_1,X_{\rightarrow [2:K-2]})+H(W_1,W_2)\nonumber\\
&= H(Z_{K-1},W_1,X_{\rightarrow [2:K-1]},W_2)+H(W_1,X_{\rightarrow [2:K-2]})+H(W_1,W_2)\nonumber\\
&=H(W_1,X_{\rightarrow [2:K-2]})+2H(W_1,W_2),
\end{align}
which $(e)$ follows from the user-index symmetry, and $(f)$ by the sub-modularity of the entropy function. This is precisely (\ref{eqn:lemma2K}) for $k=K-1$.

Now suppose (\ref{eqn:lemma2K}) holds for $k=k^*+1$, we next prove it is true for $k=k^*$ for $K\geq 4$, since when $K=3$ there is nothing to prove beyond $k=K-1=2$. Using a similar decomposition as in  (\ref{eqn:Kminus1_start}), we can write 
\begin{align}
2H(Z_1,W_1,X_{\rightarrow [2:k^*]})\geq H(Z_1,W_1,X_{\rightarrow [2:k^*+1]})+H(Z_1,W_1,X_{\rightarrow [2:k^*-1]})
\end{align}
Next we apply the supposition for $k=k^*+1$ on the first term of the right hand side, which gives
\begin{align}
2H(Z_1,W_1,X_{\rightarrow [2:k^*]})\geq& \frac{ [{(K-k^*-1)(K-k^*)}-2]H(Z_1,W_1,X_{\rightarrow [2:k^*]})}{(K-k^*)(K-k^*+1)}+\frac{2H(W_1,X_{\rightarrow [2:k^*]})}{(K-k^*)(K-k^*+1)}\nonumber\\
&+\frac{2(K-k^*)H(W_1,W_2)}{(K-k^*)(K-k^*+1)}+H(Z_1,W_1,X_{\rightarrow [2:k^*-1]})\label{eqn:putintKminus1}
\end{align}
Notice that the coefficient in front of $H(W_1,X_{\rightarrow [2:k^*]})$ is always less than one for $K\geq 4$ and $k^*\in\{2,3,\ldots,K-1\}$, and we can thus bound the following sum
\begin{align}
&\frac{2H(W_1,X_{\rightarrow [2:k^*]})}{(K-k^*)(K-k^*+1)}+H(Z_1,W_1,X_{\rightarrow [2:k^*-1]})\nonumber\\
&=\frac{2[H(W_1,X_{\rightarrow [2:k^*]})+H(Z_1,W_1,X_{\rightarrow [2:k^*-1]})]}{(K-k^*)(K-k^*+1)}+\frac{(K-k^*)(K-k^*+1)-2}{(K-k^*)(K-k^*+1)}H(Z_1,W_1,X_{\rightarrow [2:k^*-1]})\nonumber\\
&\stackrel{(g)}{\geq} \frac{2[H(W_1,W_2)+H(W_1,X_{\rightarrow [2:k^*-1]})]}{(K-k^*)(K-k^*+1)}+\frac{(K-k^*)(K-k^*+1)-2}{(K-k^*)(K-k^*+1)}H(Z_1,W_1,X_{\rightarrow [2:k^*-1]}), \label{eqn:2Klast}
\end{align}
where $(g)$ follows the same line of argument as in (\ref{eqn:Kminus1}). Substituting (\ref{eqn:2Klast}) into (\ref{eqn:putintKminus1}) and canceling out the common terms of $H(Z_1,W_1,X_{\rightarrow [2:k^*]})$ on both sides now give  (\ref{eqn:lemma2K}) for $k=k^*$. The proof is thus complete.
\end{IEEEproof}

\section{Proof for the Converse of Proposition \ref{theorem:result2_4}}
\label{appendix6_1}

The inequalities $8M+6R\geq 11$, $3M+3R\geq 5$, and $5M+6R\geq 9$ in (\ref{eqn:bounds2_4_type3_1}) can be proved using Table \ref{table:theorem:result2_4_1}-\ref{table:theorem:result2_4_1proof}, Table \ref{table:theorem:result2_4_2}-\ref{table:theorem:result2_4_2proof}, and Table \ref{table:theorem:result2_4_3}-\ref{table:theorem:result2_4_3proof}, respectively. The inequality $3M+3R\geq 5$ in (\ref{eqn:bounds2_4_type2_2}) is proved using Table \ref{table:theorem:result2_4_4}-\ref{table:theorem:result2_4_4proof}. All other bounds in Proposition \ref{theorem:result2_4} follow from the cut-set bound.

\begin{table}
\centering
\begin{tabular}{|c|c|}
\hline
$T_{ 1}$ & $F$ \\
$T_{ 2}$ & $R$ \\
$T_{ 3}$ & $H(X_{1,1,1,2})$ \\
$T_{ 4}$ & $H(W_{1})$ \\
$T_{ 5}$ & $H(W_{1},X_{1,1,1,2})$ \\
$T_{ 6}$ & $H(W_{1},W_{2})$ \\
$T_{ 7}$ & $H(Z_{1})$ \\
$T_{ 8}$ & $H(Z_{1},X_{1,1,1,2})$ \\
$T_{ 9}$ & $H(Z_{1},W_{1})$ \\
$T_{10}$ & $H(Z_{1},Z_{2},X_{1,1,1,2})$ \\
$T_{11}$ & $H(Z_{1},Z_{2},W_{1})$ \\
$T_{12}$ & $H(Z_{1},Z_{2},Z_{3},X_{1,1,1,2})$ \\
$T_{13}$ & $H(Z_{1},Z_{2},Z_{3},W_{1})$
\\\hline
\end{tabular}
\caption{Terms needed to prove Proposition \ref{theorem:result2_4}, inequality $8M+6R\geq 11$. \label{table:theorem:result2_4_1}}
\end{table}

\begin{table}
\setlength{\tabcolsep}{3pt}
\centering
\begin{tabular}{|ccccc ccccc ccc|}
\hline
$T_{ 1}$  &$T_{ 2}$  &$T_{ 3}$  &$T_{ 4}$  &$T_{ 5}$  &$T_{ 6}$  &$T_{ 7}$  &$T_{ 8}$  &$T_{ 9}$  &$T_{10}$  &$T_{11}$  &$T_{12}$  &$T_{13}$  \\
\hline\hline
          &          &          &          &          &          &          &          &          &          &          &$  1$     &$ -1$     \\
          &$  6$     &$ -6$     &          &          &          &          &          &          &          &          &          &           \\
          &          &          &$  2$     &          &          &$  2$     &          &$ -2$     &          &          &          &           \\
          &          &$  6$     &          &          &          &$  6$     &$ -6$     &          &          &          &          &           \\
          &          &          &          &$ -3$     &          &          &$  6$     &          &$ -3$     &          &          &           \\
          &          &          &          &          &          &          &$ -1$     &          &$  2$     &          &$ -1$     &           \\
          &          &          &$ -3$     &$  3$     &$ -3$     &          &          &$  3$     &          &          &          &           \\
          &          &          &          &          &$ -1$     &          &$  1$     &$ -1$     &          &$  1$     &          &           \\
          &          &          &          &          &$ -1$     &          &          &          &$  1$     &$ -1$     &          &$  1$     \\
$ -1$     &          &          &$  1$     &          &          &          &          &          &          &          &          &           \\
$-10$     &          &          &          &          &$  5$     &          &          &          &          &          &          &           \\
\hline\hline
$-11$     &   $6$  &          &          &          &          &    $8$    &          &       &            &            &    &          \\
\hline
\end{tabular}
\caption{Tabulation proof of Proposition \ref{theorem:result2_4} inequality $8M+6R\geq 11$, with terms defined in Table \ref{table:theorem:result2_4_1}.\label{table:theorem:result2_4_1proof}}
\end{table}

\begin{table}
\centering
\begin{tabular}{|c|c|}
\hline
$T_{ 1}$ & $F$ \\
$T_{ 2}$ & $R$ \\
$T_{ 3}$ & $H(X_{1,1,1,2})$ \\
$T_{ 4}$ & $H(W_{1})$ \\
$T_{ 5}$ & $H(W_{1},X_{1,1,1,2})$ \\
$T_{ 6}$ & $H(W_{1},X_{1,1,1,2},X_{1,1,2,1})$ \\
$T_{ 7}$ & $H(W_{1},W_{2})$ \\
$T_{ 8}$ & $H(Z_{1})$ \\
$T_{ 9}$ & $H(Z_{1},X_{1,1,1,2})$ \\
$T_{10}$ & $H(Z_{1},X_{1,1,1,2},X_{1,1,2,1})$ \\
$T_{11}$ & $H(Z_{1},W_{1})$ 
\\\hline
\end{tabular}
\caption{Terms needed to prove Proposition \ref{theorem:result2_4}, inequality $3M+3R\geq 5$ in (\ref{eqn:bounds2_4_type3_1}). \label{table:theorem:result2_4_2}}
\end{table}

\begin{table}
\setlength{\tabcolsep}{3pt}
\centering
\begin{tabular}{|ccccc ccccc c|}
\hline
$T_{ 1}$  &$T_{ 2}$  &$T_{ 3}$  &$T_{ 4}$  &$T_{ 5}$  &$T_{ 6}$  &$T_{ 7}$  &$T_{ 8}$  &$T_{ 9}$  &$T_{10}$  &$T_{11}$  \\
\hline\hline
          &$  3$     &$ -3$     &          &          &          &          &          &          &          &           \\
          &          &          &          &          &$ -1$     &          &          &          &$  1$     &           \\
          &          &$  3$     &          &          &          &          &$  3$     &$ -3$     &          &           \\
          &          &          &          &          &          &          &          &$  2$     &$ -1$     &$ -1$     \\
          &          &          &$ -1$     &$  1$     &          &$ -1$     &          &          &          &$  1$     \\
          &          &          &          &$ -1$     &$  1$     &$ -1$     &          &$  1$     &          &           \\
$ -1$     &          &          &$  1$     &          &          &          &          &          &          &           \\
$ -4$     &          &          &          &          &          &$  2$     &          &          &          &           \\
\hline\hline
$-5$     &   $3$  &          &          &          &          &       &      $3$    &       &            &             \\
\hline
\end{tabular}
\caption{Tabulation proof of Proposition \ref{theorem:result2_4} inequality $3M+3R\geq 5$ in (\ref{eqn:bounds2_4_type3_1}), with terms defined in Table \ref{table:theorem:result2_4_2}.\label{table:theorem:result2_4_2proof}}
\end{table}

\begin{table}
\centering
\begin{tabular}{|c|c|}
\hline
$T_{ 1}$ & $F$ \\
$T_{ 2}$ & $R$ \\
$T_{ 3}$ & $H(X_{1,1,1,2})$ \\
$T_{ 4}$ & $H(W_{1})$ \\
$T_{ 5}$ & $H(W_{1},X_{1,1,1,2})$ \\
$T_{ 6}$ & $H(W_{1},X_{1,1,1,2},X_{1,1,2,1})$ \\
$T_{ 7}$ & $H(W_{1},X_{1,1,1,2},X_{1,1,2,1},X_{1,2,1,1})$ \\
$T_{ 8}$ & $H(W_{1},W_{2})$ \\
$T_{ 9}$ & $H(Z_{1})$ \\
$T_{10}$ & $H(Z_{1},X_{1,1,1,2})$ \\
$T_{11}$ & $H(Z_{1},X_{1,1,1,2},X_{1,1,2,1})$ \\
$T_{12}$ & $H(Z_{1},X_{1,1,1,2},X_{1,1,2,1},X_{1,2,1,1})$ \\
$T_{13}$ & $H(Z_{1},W_{1})$ 
\\\hline
\end{tabular}
\caption{Terms needed to prove Proposition \ref{theorem:result2_4}, inequality $5M+6R\geq 9$. \label{table:theorem:result2_4_3}}
\end{table}

\begin{table}
\setlength{\tabcolsep}{3pt}
\centering
\begin{tabular}{|ccccc ccccc ccc|}
\hline
$T_{ 1}$  &$T_{ 2}$  &$T_{ 3}$  &$T_{ 4}$  &$T_{ 5}$  &$T_{ 6}$  &$T_{ 7}$  &$T_{ 8}$  &$T_{ 9}$  &$T_{10}$  &$T_{11}$  &$T_{12}$  &$T_{13}$  \\
\hline\hline
          &$  6$     &$ -6$     &          &          &          &          &          &          &          &          &          &           \\
          &          &          &          &          &          &$ -1$     &          &          &          &          &$  1$     &           \\
          &          &          &          &          &          &          &$ -1$     &$ -1$     &          &          &          &$  2$     \\
          &          &$  6$     &          &          &          &          &          &$  6$     &$ -6$     &          &          &           \\
          &          &          &          &          &          &          &          &          &$  6$     &$ -3$     &          &$ -3$     \\
          &          &          &          &          &          &          &          &          &$ -1$     &$  2$     &$ -1$     &           \\
          &          &          &$ -1$     &$  1$     &          &          &$ -1$     &          &          &          &          &$  1$    \\
          &          &          &          &$ -1$     &$  1$     &          &$ -1$     &          &$  1$     &          &          &           \\
          &          &          &          &          &$ -1$     &$  1$     &$ -1$     &          &          &$  1$     &          &           \\
$ -1$     &          &          &$  1$     &          &          &          &          &          &          &          &          &           \\
$ -8$     &          &          &          &          &          &          &$  4$     &          &          &          &          &           \\
\hline\hline
$-9$     &   $6$  &          &          &          &          &       &         &   $5$    &            &          &           &   \\
\hline
\end{tabular}
\caption{Tabulation proof of Proposition \ref{theorem:result2_4} inequality $5M+6R\geq 9$, with terms defined in Table \ref{table:theorem:result2_4_3}.\label{table:theorem:result2_4_3proof}}
\end{table}

\begin{table}
\centering
\begin{tabular}{|c|c|}
\hline
$T_{ 1}$ & $F$ \\
$T_{ 2}$ & $R$ \\
$T_{ 3}$ & $H(X_{1,1,2,2})$ \\
$T_{ 4}$ & $H(W_{1})$ \\
$T_{ 5}$ & $H(W_{1},X_{1,1,2,2})$ \\
$T_{ 6}$ & $H(W_{1},X_{1,1,2,2},X_{1,2,1,2})$ \\
$T_{ 7}$ & $H(W_{1},W_{2})$ \\
$T_{ 8}$ & $H(Z_{1})$ \\
$T_{ 9}$ & $H(Z_{1},X_{1,1,2,2})$ \\
$T_{10}$ & $H(Z_{1},X_{1,1,2,2},X_{1,2,1,2})$ \\
$T_{11}$ & $H(Z_{1},W_{1})$
\\\hline
\end{tabular}
\caption{Terms needed to prove Proposition \ref{theorem:result2_4}, inequality $3M+3R\geq 5$ in (\ref{eqn:bounds2_4_type2_2}). \label{table:theorem:result2_4_4}}
\end{table}

\begin{table}
\setlength{\tabcolsep}{3pt}
\centering
\begin{tabular}{|ccccc ccccc c|}
\hline
$T_{ 1}$  &$T_{ 2}$  &$T_{ 3}$  &$T_{ 4}$  &$T_{ 5}$  &$T_{ 6}$  &$T_{ 7}$  &$T_{ 8}$  &$T_{ 9}$  &$T_{10}$  &$T_{11}$  \\
\hline\hline
          &$  3$     &$ -3$     &          &          &          &          &          &          &          &           \\
          &          &          &          &          &$ -1$     &          &          &          &$  1$     &           \\
          &          &$  3$     &          &          &          &          &$  3$     &$ -3$     &          &           \\
          &          &          &          &          &          &          &          &$  2$     &$ -1$     &$ -1$     \\
          &          &          &$ -1$     &$  1$     &          &$ -1$     &          &          &          &$  1$     \\
          &          &          &          &$ -1$     &$  1$     &$ -1$     &          &$  1$     &          &           \\
$ -1$     &          &          &$  1$     &          &          &          &          &          &          &           \\
$ -4$     &          &          &          &          &          &$  2$     &          &          &          &           \\
\hline\hline
$-5$     &   $3$  &          &          &          &          &       &    $3$  &       &            &          \\
\hline
\end{tabular}
\caption{Tabulation proof of Proposition \ref{theorem:result2_4} inequality $3M+3R\geq 5$ in (\ref{eqn:bounds2_4_type2_2}), with terms defined in Table \ref{table:theorem:result2_4_4}.\label{table:theorem:result2_4_4proof}}
\end{table}

\section{Proof for the Forward of Proposition \ref{theorem:result2_4}}
\label{appendix:theorem:result2_4}

Note that the optimal tradeoff for the single demand type $(3,1)$ system has the following corner points 
\begin{align*}
(M,R)=(0,2),\left(\frac{1}{4},\frac{3}{2}\right), \left(\frac{1}{2},\frac{7}{6}\right), \left(1,\frac{2}{3}\right), \left(\frac{3}{2},\frac{1}{4}\right), (2,0).
\end{align*}
The corner points  $\left(1,\frac{2}{3}\right)$ and $\left(\frac{3}{2},\frac{1}{4}\right)$ are achievable using the Maddah-Ali-Niesen scheme \cite{MaddahAliNiesen:14}. The point $\left(\frac{1}{4},\frac{3}{2}\right)$ is achievable by the code given in \cite{chen2016fundamental} or \cite{tian2018caching}. The only remaining corner point of interest is thus  $\left(\frac{1}{2},\frac{7}{6}\right)$, in the binary field. This can be achieved by the following strategy in Table \ref{tab:0001}, where the first file has 6 symbols $(A_1,A_2,\ldots,A_6)$ and the second file $(B_1,B_2,\ldots,B_6)$.
\begin{table}[h]
\begin{center}
\begin{tabular}{|c || c | c | c |}
\hline
User 1 &$A_1+B_1$  & $A_2+B_2$ & $A_3+B_3$ \\\hline
User 2 &$A_1+B_1$ & $A_4+B_4$ & $A_5+B_5$ \\\hline
User 3 &$A_2+B_2$ & $A_4+B_4$ & $A_6+B_6$ \\\hline
User 4 &$A_3+B_3$ & $A_5+B_5$ & $A_6+B_6$\\
\hline
\end{tabular}
\caption{Code for the tradeoff point $\left(\frac{1}{2},\frac{7}{6}\right)$ for demand type $(3,1)$  when $(N,K)=(2,4)$.\label{tab:0001}}
\end{center}
\end{table}
By the symmetry, we only need to consider the demand when the first three users request $A$ and the last user request $B$. The server can send the following symbols in this case
\begin{align*}
A_3,A_5,A_6,B_1,B_2,B_4,A_1+A_2+A_4.
\end{align*}

Let us consider now the single demand type $(2,2)$ system, for which the corner points on the optimal tradeoff are:
\begin{align*}
(M,R)=(0,2),\left(\frac{1}{3},\frac{4}{3}\right), \left(\frac{4}{3},\frac{1}{3}\right), (2,0).
\end{align*}
Let us denote the first file as $(A_1,A_2,A_3)$, and the second file as $(B_1,B_2,B_3)$, which are in the binary field. To achieve the corner point $\left(\frac{1}{3},\frac{4}{3}\right)$, we use the caching code in Table \ref{tab:0011_1}.
\begin{table}[h]
\begin{center}
\begin{tabular}{|c || c |}
\hline
User 1 &$A_1+B_1$   \\\hline
User 2 &$A_2+B_2$ \\\hline
User 3 &$A_3+B_3$ \\\hline
User 4 &$A_1+A_2+A_3+B_1+B_2+B_3$\\
\hline
\end{tabular}
\caption{Code for the tradeoff point $\left(\frac{1}{3},\frac{4}{3}\right)$ for demand type $(2,2)$ when $(N,K)=(2,4)$.\label{tab:0011_1}}
\end{center}
\end{table}
Again due to the symmetry, we only need to consider  the case when the first two users request $A$, and the other two request $B$. For this case, the server can send
\begin{align*}
B_1,B_2,A_3,A_1+A_2+A_3.
\end{align*}
For the other corner point $\left(\frac{4}{3},\frac{1}{3}\right)$ the following placement in Table \ref{tab:0011_2} can be used.
\begin{table}[h]
\begin{center}
\begin{tabular}{|c || c | c| c| c|}
\hline
User 1 &$A_1$& $A_2$ & $B_1$ & $B_2$ \\\hline
User 2 &$A_2$& $A_3$ & $B_2$ & $B_3$ \\\hline
User 3 &$A_1$& $A_3$ & $B_1$ & $B_3$ \\\hline
User 4 &$A_1+A_2$& $A_2+A_3$ & $B_1+B_2$ & $B_2+B_3$\\
\hline
\end{tabular}
\caption{Code for the tradeoff point  $\left(\frac{4}{3},\frac{1}{3}\right)$ for demand type $(2,2)$  when $(N,K)=(2,4)$.\label{tab:0011_2}}
\end{center}
\end{table}
Again for the case when the first two users request $A$, and the other two request $B$, the server can send
\begin{align*}
A_1-A_3+B_2.
\end{align*}

\begin{table}
\centering
\begin{tabular}{|c|c|}
\hline
$T_{ 1}$ & $F$ \\
$T_{ 2}$ & $R$ \\
$T_{ 3}$ & $H(X_{1,1,2})$ \\
$T_{ 4}$ & $H(W_{1})$ \\
$T_{ 5}$ & $H(W_{2},X_{1,1,2})$ \\
$T_{ 6}$ & $H(W_{1},W_{2},W_{3})$ \\
$T_{ 7}$ & $H(Z_{1})$ \\
$T_{ 8}$ & $H(Z_{3},X_{1,1,2})$ \\
$T_{ 9}$ & $H(Z_{1},W_{1})$ \\
$T_{10}$ & $H(Z_{1},W_{2},X_{1,1,2})$ \\
$T_{11}$ & $H(Z_{1},Z_{3},X_{1,1,2})$ \\
$T_{12}$ & $H(Z_{1},Z_{2},W_{1})$ 
\\\hline
\end{tabular}
\caption{Terms needed to prove Proposition \ref{theorem:firstresult3_3A}, inequality $M+R\geq 2$ in (\ref{eqn:3_3bounds2_1_0}). \label{table:theorem:result3_3_1}}
\end{table}

\begin{table}
\setlength{\tabcolsep}{2pt}
\centering
\begin{tabular}{|ccccc ccccc cc|}
\hline
$T_{ 1}$  &$T_{ 2}$  &$T_{ 3}$  &$T_{ 4}$  &$T_{ 5}$  &$T_{ 6}$  &$T_{ 7}$  &$T_{ 8}$  &$T_{ 9}$  &$T_{10}$  &$T_{11}$  &$T_{12}$  \\
\hline\hline
          &$  2$     &$ -2$     &          &          &          &          &          &          &          &          &           \\
          &          &$  2$     &          &          &          &$  2$     &$ -2$     &          &          &          &           \\
          &          &          &          &          &$ -1$     &          &          &          &          &$  2$     &$ -1$     \\
          &          &          &$ -1$     &$  1$     &          &          &          &$  1$     &$ -1$     &          &           \\
          &          &          &          &          &          &          &$  1$     &$ -1$     &          &$ -1$     &$  1$     \\
          &          &          &          &$ -1$     &          &          &$  1$     &          &$  1$     &$ -1$     &           \\
$ -1$     &          &          &$  1$     &          &          &          &          &          &          &          &           \\
$ -3$     &          &          &          &          &$  1$     &          &          &          &          &          &           \\
\hline\hline
$-4$     &   $2$  &          &          &          &          &     $2$  &             &       &            &         &         \\
\hline
\end{tabular}
\caption{Tabulation proof of Proposition \ref{theorem:firstresult3_3A} inequality $M+R\geq 2$ in (\ref{eqn:3_3bounds2_1_0}), with terms defined in Table \ref{table:theorem:result3_3_1}.\label{table:theorem:result3_3_1proof}}
\end{table}

\section{Proof of Proposition  \ref{theorem:firstresult3_3A}}
\label{appendix33}
The inequalities $M+R\geq 2$ and $2M+3R\geq 5$  in (\ref{eqn:3_3bounds2_1_0}) are proved in Table \ref{table:theorem:result3_3_1}-\ref{table:theorem:result3_3_1proof}, and Table \ref{table:theorem:result3_3_2}-\ref{table:theorem:result3_3_2proof}, respectively. The inequalities $6M+3R\geq 8$, $M+R\geq 2$, $12M+18R\geq 29$, and $3M+6R\geq 8$ in (\ref{eqn:3_3bounds1_1_1}) are proved in Table \ref{table:theorem:result3_3_3}-\ref{table:theorem:result3_3_3proof}, Table \ref{table:theorem:result3_3_4}- \ref{table:theorem:result3_3_4proof}, Table \ref{table:theorem:result3_3_5}-\ref{table:theorem:result3_3_5proof}, and Table \ref{table:theorem:result3_3_6}- \ref{table:theorem:result3_3_6proof}, respectively. All other bounds in Proposition \ref{theorem:firstresult3_3A} can be deduced from the cut-set bound thus do not need a proof.

\begin{table}
\centering
\begin{tabular}{|c|c|}
\hline
$T_{ 1}$ & $F$ \\
$T_{ 2}$ & $R$ \\
$T_{ 3}$ & $H(X_{1,2,2})$ \\
$T_{ 4}$ & $H(X_{2,3,3},X_{2,1,2})$ \\
$T_{ 5}$ & $H(W_{1})$ \\
$T_{ 6}$ & $H(W_{1},X_{1,2,2})$ \\
$T_{ 7}$ & $H(W_{2},X_{1,2,2})$ \\
$T_{ 8}$ & $H(W_{3},X_{1,3,3},X_{2,3,3})$ \\
$T_{ 9}$ & $H(W_{1},X_{1,3,3},X_{2,1,1})$ \\
$T_{10}$ & $H(W_{1},W_{2})$ \\
$T_{11}$ & $H(W_{2},W_{3},X_{1,2,2})$ \\
$T_{12}$ & $H(W_{1},W_{2},W_{3})$ \\
$T_{13}$ & $H(Z_{1})$ \\
$T_{14}$ & $H(Z_{2},X_{1,2,2})$ \\
$T_{15}$ & $H(Z_{1},X_{1,2,2})$ \\
$T_{16}$ & $H(Z_{2},X_{1,3,3},X_{2,3,3})$ \\
$T_{17}$ & $H(Z_{1},X_{1,3,3},X_{2,1,1})$ \\
$T_{18}$ & $H(Z_{2},X_{2,3,3},X_{2,1,2})$ \\
$T_{19}$ & $H(Z_{1},X_{2,3,3},X_{2,1,2})$ \\
$T_{20}$ & $H(Z_{1},X_{2,3,3},X_{3,1,1},X_{2,1,2})$ \\
$T_{21}$ & $H(Z_{1},W_{1})$ \\
$T_{22}$ & $H(Z_{1},W_{2},X_{1,2,2})$ \\
$T_{23}$ & $H(Z_{1},W_{3},X_{1,2,2})$ \\
$T_{24}$ & $H(Z_{2},W_{1},X_{1,2,2})$ \\
$T_{25}$ & $H(Z_{2},W_{3},X_{1,2,2})$ \\
$T_{26}$ & $H(Z_{1},W_{3},X_{2,3,3},X_{2,1,2})$ \\
$T_{27}$ & $H(Z_{1},W_{1},X_{2,3,3},X_{2,1,2})$ \\
$T_{28}$ & $H(Z_{1},W_{1},W_{2})$ \\
$T_{29}$ & $H(Z_{2},Z_{3},X_{1,2,2})$ \\
$T_{30}$ & $H(Z_{1},Z_{2},X_{1,2,2})$ \\
$T_{31}$ & $H(Z_{2},Z_{3},X_{1,3,3},X_{2,3,3})$ \\
$T_{32}$ & $H(Z_{1},Z_{3},X_{2,3,3},X_{2,1,2})$ \\
$T_{33}$ & $H(Z_{1},Z_{2},W_{1})$ \\
$T_{34}$ & $H(Z_{2},Z_{3},W_{3},X_{1,2,2})$
\\\hline
\end{tabular}
\caption{Terms needed to prove Proposition \ref{theorem:firstresult3_3A}, inequality $2M+3R\geq 5$ in (\ref{eqn:3_3bounds2_1_0}). \label{table:theorem:result3_3_2}}
\end{table}

\begin{landscape}
\begin{table}
\setlength{\tabcolsep}{2pt}
\centering
\begin{tabular}{|ccccc ccccc ccccc ccccc ccccc ccccc cccc|}
\hline
$T_{ 1}$  &$T_{ 2}$  &$T_{ 3}$  &$T_{ 4}$  &$T_{ 5}$  &$T_{ 6}$  &$T_{ 7}$  &$T_{ 8}$  &$T_{ 9}$  &$T_{10}$  &$T_{11}$  &$T_{12}$  &$T_{13}$  &$T_{14}$  &$T_{15}$  &$T_{16}$  &$T_{17}$  &$T_{18}$  &$T_{19}$  &$T_{20}$  &$T_{21}$  &$T_{22}$  &$T_{23}$  &$T_{24}$  &$T_{25}$  &$T_{26}$  &$T_{27}$  &$T_{28}$  &$T_{29}$  &$T_{30}$  &$T_{31}$  &$T_{32}$  &$T_{33}$  &$T_{34}$  \\
\hline\hline
          &          &          &          &          &          &          &          &          &          &          &          &          &          &          &          &          &          &          &          &          &          &          &          &          &          &          &          &$  2$     &          &          &          &$ -2$     &           \\
          &$ 27$     &$-27$     &          &          &          &          &          &          &          &          &          &          &          &          &          &          &          &          &          &          &          &          &          &          &          &          &          &          &          &          &          &          &           \\
          &          &$  8$     &$ -4$     &          &          &          &          &          &          &          &          &          &          &          &          &          &          &          &          &          &          &          &          &          &          &          &          &          &          &          &          &          &           \\
          &          &          &          &          &          &          &          &          &          &          &          &          &          &          &          &          &          &          &          &          &          &          &          &          &          &          &          &$ -3$     &          &$  3$     &          &          &           \\
          &          &          &          &          &          &          &          &          &          &          &          &$ -2$     &          &          &          &          &          &          &          &$  4$     &          &          &          &          &          &          &$ -2$     &          &          &          &          &          &           \\
          &          &$  8$     &          &          &          &          &          &          &          &          &          &$  8$     &          &$ -8$     &          &          &          &          &          &          &          &          &          &          &          &          &          &          &          &          &          &          &           \\
          &          &$ 15$     &          &          &          &          &          &          &          &          &          &$ 15$     &$-15$     &          &          &          &          &          &          &          &          &          &          &          &          &          &          &          &          &          &          &          &           \\
          &          &          &          &          &          &          &          &          &          &          &          &          &$ 12$     &          &$ -6$     &          &          &          &          &$ -6$     &          &          &          &          &          &          &          &          &          &          &          &          &           \\
          &          &          &          &          &          &          &          &          &          &$ -3$     &          &          &          &          &          &          &          &          &          &          &          &          &          &$  6$     &          &          &          &          &          &          &          &          &$ -3$     \\
          &          &          &          &          &          &          &$ -3$     &          &          &          &          &          &          &          &$  6$     &          &          &          &          &          &          &          &          &          &          &          &          &          &          &$ -3$     &          &          &           \\
          &          &          &          &$ -2$     &          &$  2$     &          &          &          &          &          &          &          &          &          &          &          &          &          &$  2$     &$ -2$     &          &          &          &          &          &          &          &          &          &          &          &           \\
          &          &          &          &$ -1$     &$  1$     &$  1$     &          &$ -1$     &          &          &          &          &          &          &          &          &          &          &          &          &          &          &          &          &          &          &          &          &          &          &          &          &           \\
          &          &          &          &          &          &          &          &          &$ -3$     &$  3$     &$ -3$     &          &          &          &          &          &          &          &          &          &          &          &          &          &          &          &$  3$     &          &          &          &          &          &           \\
          &          &          &          &          &          &          &          &          &          &          &          &$ -3$     &$  3$     &          &          &          &          &          &          &$  3$     &          &          &          &$ -3$     &          &          &          &          &          &          &          &          &           \\
          &          &          &          &          &          &          &          &          &          &          &          &          &          &$  2$     &          &          &          &          &          &$ -2$     &          &          &          &          &          &          &          &          &$ -2$     &          &          &$  2$     &           \\
          &          &          &          &          &          &          &          &          &          &          &          &          &$  1$     &$  1$     &          &          &          &$ -1$     &          &$ -1$     &          &          &          &          &          &          &          &          &          &          &          &          &           \\
          &          &          &          &          &          &          &          &          &          &          &          &          &          &          &          &          &          &          &          &          &          &$  1$     &$  1$     &          &          &$ -1$     &$ -1$     &          &          &          &          &          &           \\
          &          &$ -4$     &$  4$     &          &          &          &          &          &          &          &          &          &          &$  4$     &          &          &$ -4$     &          &          &          &          &          &          &          &          &          &          &          &          &          &          &          &           \\
          &          &          &          &          &$ -1$     &          &          &$  1$     &          &          &          &          &          &$  1$     &          &$ -1$     &          &          &          &          &          &          &          &          &          &          &          &          &          &          &          &          &           \\
          &          &          &          &          &          &$ -3$     &$  3$     &          &          &          &$ -3$     &          &          &          &          &          &          &          &          &          &$  3$     &          &          &          &          &          &          &          &          &          &          &          &           \\
          &          &          &          &          &          &          &          &          &          &          &$ -1$     &          &          &          &          &          &          &          &          &          &$ -1$     &          &          &          &$  1$     &          &          &          &$  1$     &          &          &          &           \\
          &          &          &          &          &          &          &          &          &          &          &          &          &          &          &          &$  1$     &$  1$     &          &$ -1$     &          &          &$ -1$     &          &          &          &          &          &          &          &          &          &          &           \\
          &          &          &          &          &          &          &          &          &          &          &          &          &$ -1$     &          &          &          &          &$  1$     &          &          &          &          &          &          &          &          &          &$  1$     &          &          &$ -1$     &          &           \\
          &          &          &          &          &          &          &          &          &          &          &$ -1$     &          &          &          &          &          &          &          &          &          &          &          &$ -1$     &          &          &$  1$     &          &          &$  1$     &          &          &          &           \\
          &          &          &          &          &          &          &          &          &          &          &$ -3$     &          &          &          &          &          &$  3$     &          &          &          &          &          &          &$ -3$     &          &          &          &          &          &          &          &          &$  3$     \\
          &          &          &          &          &          &          &          &          &          &          &$ -1$     &          &          &          &          &          &          &          &$  1$     &          &          &          &          &          &$ -1$     &          &          &          &          &          &$  1$     &          &           \\
$ -3$     &          &          &          &$  3$     &          &          &          &          &          &          &          &          &          &          &          &          &          &          &          &          &          &          &          &          &          &          &          &          &          &          &          &          &           \\
$ -6$     &          &          &          &          &          &          &          &          &$  3$     &          &          &          &          &          &          &          &          &          &          &          &          &          &          &          &          &          &          &          &          &          &          &          &           \\
$-36$     &          &          &          &          &          &          &          &          &          &          &$ 12$     &          &          &          &          &          &          &          &          &          &          &          &          &          &          &          &          &          &          &          &          &          &           \\
\hline\hline
$-45$     &   $27$  &          &          &          &          &       &     &       &            &         & & $18$ & &&&&&&&&&&&&&&&&&&&&\\
\hline
\end{tabular}
\caption{Tabulation proof of Proposition \ref{theorem:firstresult3_3A} inequality $2M+3R\geq 5$ in (\ref{eqn:3_3bounds2_1_0}), with terms defined in Table \ref{table:theorem:result3_3_2}.\label{table:theorem:result3_3_2proof}}
\end{table}
\end{landscape}

\begin{table}
\centering
\begin{tabular}{|c|c|}
\hline
$T_{ 1}$ & $F$ \\
$T_{ 2}$ & $R$ \\
$T_{ 3}$ & $H(X_{1,2,3,)})$ \\
$T_{ 4}$ & $H(W_{1},W_{2})$ \\
$T_{ 5}$ & $H(W_{1},W_{2},X_{1,2,3})$ \\
$T_{ 6}$ & $H(W_{1},W_{2},W_{3})$ \\
$T_{ 7}$ & $H(Z_{1})$ \\
$T_{ 8}$ & $H(Z_{1},X_{1,2,3,)})$ \\
$T_{ 9}$ & $H(Z_{1},W_{2},X_{1,2,3})$ \\
$T_{10}$ & $H(Z_{1},W_{1},W_{2})$ \\
$T_{11}$ & $H(Z_{1},Z_{2},X_{1,2,3})$ \\
$T_{12}$ & $H(Z_{1},Z_{2},W_{1},W_{2})$ 
\\\hline
\end{tabular}
\caption{Terms needed to prove Proposition \ref{theorem:firstresult3_3A}, inequality $6M+3R\geq 8$ in (\ref{eqn:3_3bounds1_1_1}). \label{table:theorem:result3_3_3}}
\end{table}

\begin{table}
\setlength{\tabcolsep}{3pt}
\centering
\begin{tabular}{|ccccc ccccc cc|}
\hline
$T_{ 1}$  &$T_{ 2}$  &$T_{ 3}$  &$T_{ 4}$  &$T_{ 5}$  &$T_{ 6}$  &$T_{ 7}$  &$T_{ 8}$  &$T_{ 9}$  &$T_{10}$  &$T_{11}$  &$T_{12}$  \\
\hline\hline
          &          &          &          &          &          &          &          &          &          &$  1$     &$ -1$     \\
          &$  3$     &$ -3$     &          &          &          &          &          &          &          &          &           \\
          &          &          &          &$ -1$     &          &          &          &$  1$     &          &          &           \\
          &          &          &          &          &          &          &          &$ -2$     &          &$  2$     &           \\
          &          &$  6$     &          &          &          &$  6$     &$ -6$     &          &          &          &           \\
          &          &$ -3$     &          &          &          &          &$  6$     &          &          &$ -3$     &           \\
          &          &          &$ -1$     &$  1$     &$ -1$     &          &          &          &$  1$     &          &           \\
          &          &          &          &          &$ -1$     &          &          &$  1$     &$ -1$     &          &$  1$     \\
$ -2$     &          &          &$  1$     &          &          &          &          &          &          &          &           \\
$ -6$     &          &          &          &          &$  2$     &          &          &          &          &          &           \\
\hline\hline
$-8$     &   $3$  &          &          &          &          &    $6$&     &       &            &         & \\
\hline
\end{tabular}
\caption{Tabulation proof of Proposition \ref{theorem:firstresult3_3A} inequality $6M+3R\geq 8$ in (\ref{eqn:3_3bounds1_1_1}), with terms defined in Table \ref{table:theorem:result3_3_3}.\label{table:theorem:result3_3_3proof}}
\end{table}

\begin{table}
\centering
\begin{tabular}{|c|c|}
\hline
$T_{ 1}$ & $F$ \\
$T_{ 2}$ & $R$ \\
$T_{ 3}$ & $H(X_{1,2,3})$ \\
$T_{ 4}$ & $H(W_{1})$ \\
$T_{ 5}$ & $H(W_{1},X_{1,2,3})$ \\
$T_{ 6}$ & $H(W_{1},X_{1,2,3},X_{1,3,2})$ \\
$T_{ 7}$ & $H(W_{1},W_{2},W_{3})$ \\
$T_{ 8}$ & $H(Z_{1})$ \\
$T_{ 9}$ & $H(Z_{1},X_{1,2,3})$ \\
$T_{10}$ & $H(Z_{1},X_{1,2,3},X_{1,3,2})$ \\
$T_{11}$ & $H(Z_{1},W_{1})$ \\
$T_{12}$ & $H(Z_{1},W_{2},X_{1,2,3})$ 
\\\hline
\end{tabular}
\caption{Terms needed to prove Proposition \ref{theorem:firstresult3_3A}, inequality $M+R\geq 2$ in (\ref{eqn:3_3bounds1_1_1}). \label{table:theorem:result3_3_4}}
\end{table}

\begin{table}
\setlength{\tabcolsep}{3pt}
\centering
\begin{tabular}{|ccccc ccccc cc|}
\hline
$T_{ 1}$  &$T_{ 2}$  &$T_{ 3}$  &$T_{ 4}$  &$T_{ 5}$  &$T_{ 6}$  &$T_{ 7}$  &$T_{ 8}$  &$T_{ 9}$  &$T_{10}$  &$T_{11}$  &$T_{12}$  \\
\hline\hline
          &$  2$     &$ -2$     &          &          &          &          &          &          &          &          &           \\
          &          &          &          &          &$ -1$     &          &          &          &$  1$     &          &           \\
          &          &$  2$     &          &          &          &          &$  2$     &$ -2$     &          &          &           \\
          &          &          &          &          &          &          &          &$  2$     &$ -1$     &$ -1$     &           \\
          &          &          &$ -1$     &$  1$     &          &          &          &          &          &$  1$     &$ -1$     \\
          &          &          &          &$ -1$     &$  1$     &$ -1$     &          &          &          &          &$  1$     \\
$ -1$     &          &          &$  1$     &          &          &          &          &          &          &          &           \\
$ -3$     &          &          &          &          &           &  $ 1$     &          &          &          &          &           \\
\hline\hline
$-8$      &    $3$  &          &          &          &           &    $6$   &          &             &            &         &         \\
\hline
\end{tabular}
\caption{Tabulation proof of Proposition \ref{theorem:firstresult3_3A} inequality $M+R\geq 2$ in (\ref{eqn:3_3bounds1_1_1}), with terms defined in Table \ref{table:theorem:result3_3_4}.\label{table:theorem:result3_3_4proof}}
\end{table}

\begin{table}
\centering
\begin{tabular}{|c|c|}
\hline
$T_{ 1}$ & $F$ \\
$T_{ 2}$ & $R$ \\
$T_{ 3}$ & $H(X_{1,2,3})$ \\
$T_{ 4}$ & $H(X_{1,2,3},X_{1,3,2})$ \\
$T_{ 5}$ & $H(X_{1,2,3},X_{1,3,2},X_{2,1,3})$ \\
$T_{ 6}$ & $H(W_{1})$ \\
$T_{ 7}$ & $H(W_{1},X_{1,2,3})$ \\
$T_{ 8}$ & $H(W_{2},X_{1,2,3},X_{1,3,2})$ \\
$T_{ 9}$ & $H(W_{1},W_{2})$ \\
$T_{10}$ & $H(W_{1},W_{2},X_{1,2,3})$ \\
$T_{11}$ & $H(W_{1},W_{2},X_{1,3,2},X_{2,1,3})$ \\
$T_{12}$ & $H(W_{2},W_{3},X_{1,2,3},X_{1,3,2})$ \\
$T_{13}$ & $H(W_{1},W_{2},X_{1,2,3},X_{1,3,2},X_{2,1,3})$ \\
$T_{14}$ & $H(W_{1},W_{2},W_{3})$ \\
$T_{15}$ & $H(Z_{1})$ \\
$T_{16}$ & $H(Z_{1},X_{1,2,3})$ \\
$T_{17}$ & $H(Z_{1},X_{1,3,2},X_{2,1,3})$ \\
$T_{18}$ & $H(Z_{2},X_{1,2,3},X_{1,3,2})$ \\
$T_{19}$ & $H(Z_{1},X_{1,2,3},X_{1,3,2})$ \\
$T_{20}$ & $H(Z_{1},X_{1,2,3},X_{1,3,2},X_{2,1,3})$ \\
$T_{21}$ & $H(Z_{1},W_{1})$ \\
$T_{22}$ & $H(Z_{1},W_{2},X_{1,2,3})$ \\
$T_{23}$ & $H(Z_{1},W_{2},X_{1,2,3},X_{1,3,2})$ \\
$T_{24}$ & $H(Z_{1},W_{1},W_{2})$ 
\\\hline
\end{tabular}
\caption{Terms needed to prove Proposition \ref{theorem:firstresult3_3A}, inequality $12M+18R\geq 29$ in (\ref{eqn:3_3bounds1_1_1}). \label{table:theorem:result3_3_5}}
\end{table}

\begin{table}
\setlength{\tabcolsep}{2pt}
\centering
\begin{tabular}{|ccccc ccccc ccccc ccccc cccc|}
\hline
$T_{ 1}$  &$T_{ 2}$  &$T_{ 3}$  &$T_{ 4}$  &$T_{ 5}$  &$T_{ 6}$  &$T_{ 7}$  &$T_{ 8}$  &$T_{ 9}$  &$T_{10}$  &$T_{11}$  &$T_{12}$  &$T_{13}$  &$T_{14}$  &$T_{15}$  &$T_{16}$  &$T_{17}$  &$T_{18}$  &$T_{19}$  &$T_{20}$  &$T_{21}$  &$T_{22}$  &$T_{23}$  &$T_{24}$  \\
\hline\hline
          &          &$  2$     &$ -1$     &          &          &          &          &          &          &          &          &          &          &          &          &          &          &          &          &          &          &          &           \\
          &$ 18$     &$-18$     &          &          &          &          &          &          &          &          &          &          &          &          &          &          &          &          &          &          &          &          &           \\
          &          &          &          &          &          &          &          &          &          &$ -1$     &          &          &          &          &          &$  1$     &          &          &          &          &          &          &           \\
          &          &          &          &          &          &          &          &          &          &          &          &$ -3$     &          &          &          &          &          &          &$  3$     &          &          &          &           \\
          &          &$ 17$     &          &          &          &          &          &          &          &          &          &          &          &$ 17$     &$-17$     &          &          &          &          &          &          &          &           \\
          &          &          &          &          &          &          &          &          &          &          &          &          &          &$ -2$     &$  4$     &          &$ -2$     &          &          &          &          &          &           \\
          &          &          &          &          &          &          &          &          &          &          &          &          &          &$ -3$     &$  6$     &$ -3$     &          &          &          &          &          &          &           \\
          &          &          &          &          &          &          &          &          &          &          &          &          &          &          &$  2$     &          &          &$ -1$     &          &$ -1$     &          &          &           \\
          &          &          &          &          &          &          &          &          &          &          &          &          &          &          &          &          &          &          &          &          &$  6$     &$ -3$     &$ -3$     \\
          &          &$ -1$     &$  2$     &$ -1$     &          &          &          &          &          &          &          &          &          &          &          &          &          &          &          &          &          &          &           \\
          &          &          &          &          &$ -3$     &$  3$     &          &          &          &          &          &          &          &          &          &          &          &          &          &$  3$     &$ -3$     &          &           \\
          &          &          &          &          &          &          &          &$ -1$     &$  1$     &          &          &          &$ -1$     &          &          &          &          &          &          &          &          &          &$  1$     \\
          &          &          &          &          &          &          &          &          &          &          &          &          &          &          &$  2$     &          &          &          &          &$ -2$     &$ -2$     &          &$  2$     \\
          &          &          &          &          &          &$ -3$     &$  3$     &          &          &          &          &          &          &          &$  3$     &          &$ -3$     &          &          &          &          &          &           \\
          &          &          &          &          &          &          &          &          &$ -1$     &$  1$     &          &          &$ -1$     &          &          &          &          &          &          &          &$  1$     &          &           \\
          &          &          &          &          &          &          &          &          &          &          &          &          &          &          &          &$  2$     &$  2$     &          &$ -2$     &          &$ -2$     &          &           \\
          &          &          &$ -1$     &$  1$     &          &          &          &          &          &          &          &          &          &          &          &          &          &$  1$     &$ -1$     &          &          &          &           \\
          &          &          &          &          &          &          &$ -3$     &          &          &          &$  3$     &          &$ -3$     &          &          &          &          &          &          &          &          &$  3$             & \\
          &          &          &          &          &          &          &          &          &          &          &$ -3$     &$  3$     &$ -3$     &          &          &          &$  3$     &          &          &          &          &          &           \\
$ -3$     &          &          &          &          &$  3$     &          &          &          &          &          &          &          &          &          &          &          &          &          &          &          &          &          &           \\
$ -2$     &          &          &          &          &          &          &          &$  1$     &          &          &          &          &          &          &          &          &          &          &          &          &          &          &           \\
$-24$     &          &          &          &          &          &          &          &          &          &          &          &          &$  8$     &          &          &          &          &          &          &          &          &          &           \\
\hline\hline
$-29$      &    $18$  &          &          &          &           &       &          &             &            &         &         &                &            &   $12$      &          &          &           & &&&&& \\
\hline
\end{tabular}
\caption{Tabulation proof of Proposition \ref{theorem:firstresult3_3A} inequality $12M+18R\geq 29$ in (\ref{eqn:3_3bounds1_1_1}), with terms defined in Table \ref{table:theorem:result3_3_5}.\label{table:theorem:result3_3_5proof}}
\end{table}

\begin{table}
\centering
\begin{tabular}{|c|c|}
\hline
$T_{ 1}$ & $F$ \\
$T_{ 2}$ & $R$ \\
$T_{ 3}$ & $H(X_{1,2,3})$ \\
$T_{ 4}$ & $H(W_{1},W_{2})$ \\
$T_{ 5}$ & $H(W_{1},W_{2},X_{1,2,3})$ \\
$T_{ 6}$ & $H(W_{2},W_{3},X_{1,2,3},X_{1,3,2,)})$ \\
$T_{ 7}$ & $H(W_{1},W_{2},X_{1,2,3},X_{1,3,2},X_{2,1,3})$ \\
$T_{ 8}$ & $H(W_{1},W_{2},W_{3})$ \\
$T_{ 9}$ & $H(Z_{1})$ \\
$T_{10}$ & $H(Z_{1},X_{1,2,3})$ \\
$T_{11}$ & $H(Z_{2},X_{1,2,3},X_{1,3,2})$ \\
$T_{12}$ & $H(Z_{1},X_{1,3,2},X_{2,1,3})$ \\
$T_{13}$ & $H(Z_{1},X_{1,2,3},X_{1,3,2},X_{2,1,3})$ \\
$T_{14}$ & $H(Z_{1},W_{2},X_{1,2,3})$ \\
$T_{15}$ & $H(Z_{1},W_{1},W_{2})$ 
\\\hline
\end{tabular}
\caption{Terms needed to prove Proposition \ref{theorem:firstresult3_3A}, inequality $3M+6R\geq 8$ in (\ref{eqn:3_3bounds1_1_1}). \label{table:theorem:result3_3_6}}
\end{table}

\begin{table}
\setlength{\tabcolsep}{2pt}
\centering
\begin{tabular}{|ccccc ccccc ccccc|}
\hline
$T_{ 1}$  &$T_{ 2}$  &$T_{ 3}$  &$T_{ 4}$  &$T_{ 5}$  &$T_{ 6}$  &$T_{ 7}$  &$T_{ 8}$  &$T_{ 9}$  &$T_{10}$  &$T_{11}$  &$T_{12}$  &$T_{13}$  &$T_{14}$  &$T_{15}$  \\
\hline\hline
          &          &          &          &          &          &          &          &          &          &          &          &          &$  1$     &$ -1$     \\
          &$  6$     &$ -6$     &          &          &          &          &          &          &          &          &          &          &          &           \\
          &          &          &          &$ -1$     &$  1$     &          &          &          &          &          &          &          &          &           \\
          &          &          &          &          &          &$ -1$     &          &          &          &          &          &$  1$     &          &           \\
          &          &$  6$     &          &          &          &          &          &$  6$     &$ -6$     &          &          &          &          &           \\
          &          &          &          &          &          &          &          &$ -2$     &$  4$     &$ -2$     &          &          &          &           \\
          &          &          &          &          &          &          &          &$ -1$     &$  2$     &          &$ -1$     &          &          &           \\
          &          &          &$ -1$     &$  1$     &          &          &$ -1$     &          &          &          &          &          &          &$  1$     \\
          &          &          &          &          &          &          &          &          &          &$  1$     &$  1$     &$ -1$     &$ -1$     &           \\
          &          &          &          &          &$ -1$     &$  1$     &$ -1$     &          &          &$  1$     &          &          &          &           \\
$ -2$     &          &          &$  1$     &          &          &          &          &          &          &          &          &          &          &           \\
$ -6$     &          &          &          &          &          &          &$  2$     &          &          &          &          &          &          &           \\
\hline\hline
$-8$      &    $6$  &         &          &          &           &       &          &   $3$          &            &         &         &                & &            \\
\hline
\end{tabular}
\caption{Tabulation proof of Proposition \ref{theorem:firstresult3_3A} inequality $3M+6R\geq 8$ in (\ref{eqn:3_3bounds1_1_1}), with terms defined in Table \ref{table:theorem:result3_3_6}.\label{table:theorem:result3_3_6proof}}
\end{table}

\end{appendices}

\section*{Acknowledgment}

The author wishes to thank Dr. Urs Niesen and Dr. Vaneet Aggarwal for early discussions which partly motivated this work. He also wishes to thank Dr. Jun Chen for several discussions as well as the insightful comments on an early draft. Additionally the author wishes to thank the authors of \cite{ghasemi2017improved} for making the source code to compute several outer bounds available online, which was conveniently used in the comparisons of several existing bounds in this work.

\bibliographystyle{IEEEtran}

\begin{thebibliography}{10}
\providecommand{\url}[1]{#1}
\csname url@samestyle\endcsname
\providecommand{\newblock}{\relax}
\providecommand{\bibinfo}[2]{#2}
\providecommand{\BIBentrySTDinterwordspacing}{\spaceskip=0pt\relax}
\providecommand{\BIBentryALTinterwordstretchfactor}{4}
\providecommand{\BIBentryALTinterwordspacing}{\spaceskip=\fontdimen2\font plus
\BIBentryALTinterwordstretchfactor\fontdimen3\font minus
  \fontdimen4\font\relax}
\providecommand{\BIBforeignlanguage}[2]{{%
\expandafter\ifx\csname l@#1\endcsname\relax
\typeout{** WARNING: IEEEtran.bst: No hyphenation pattern has been}%
\typeout{** loaded for the language `#1'. Using the pattern for}%
\typeout{** the default language instead.}%
\else
\language=\csname l@#1\endcsname
\fi
#2}}
\providecommand{\BIBdecl}{\relax}
\BIBdecl

\bibitem{Yeung:97}
R.~W. Yeung, ``A framework for linear information inequalities,'' \emph{IEEE
  Trans. on Information Theory}, vol.~43, no.~6, pp. 1924--1934, Nov. 1997.

\bibitem{Tian:JSAC13}
C.~Tian, ``Characterizing the rate region of the (4, 3, 3) exact-repair
  regenerating codes,'' \emph{IEEE Journal on Selected Areas in
  Communications}, vol.~32, no.~5, pp. 967--975, May 2014.

\bibitem{TianLiu:15}
C.~Tian and T.~Liu, ``Multilevel diversity coding with regeneration,''
  \emph{IEEE Trans. on Information Theory}, vol.~62, no.~9, pp. 4833--4847,
  Sep. 2016.

\bibitem{Tian:15-2}
C.~Tian, ``A note on the rate region of exact-repair regenerating codes,''
  \emph{arXiv:1503.00011}, Mar. 2015.

\bibitem{li2017multilevel}
C.~Li, S.~Weber, and J.~M. Walsh, ``Multilevel diversity coding systems: Rate
  regions, codes, computation, \& forbidden minors,'' \emph{IEEE Trans. on
  Information Theory}, vol.~63, no.~1, pp. 230--251, 2017.

\bibitem{MaddahAliNiesen:14}
M.~A. Maddah-Ali and U.~Niesen, ``Fundamental limits of caching,'' \emph{IEEE
  Trans. on Information Theory}, vol.~60, no.~5, pp. 2856--2867, May 2014.

\bibitem{MaddahAliNiesen:14Networking}
------, ``Decentralized coded caching attains order-optimal memory-rate
  tradeoff,'' \emph{IEEE/ACM Trans. on Networking}, vol.~23, no.~4, pp.
  1029--1040, Aug. 2015.

\bibitem{niesen2017coded}
U.~Niesen and M.~A. Maddah-Ali, ``Coded caching with nonuniform demands,''
  \emph{IEEE Trans. on Information Theory}, vol.~63, no.~2, pp. 1146--1158,
  Feb. 2017.

\bibitem{pedarsani2016online}
R.~Pedarsani, M.~A. Maddah-Ali, and U.~Niesen, ``Online coded caching,''
  \emph{IEEE/ACM Trans. on Networking}, vol.~24, no.~2, pp. 836--845, Mar.
  2016.

\bibitem{karamchandani2016hierarchical}
N.~Karamchandani, U.~Niesen, M.~A. Maddah-Ali, and S.~N. Diggavi,
  ``Hierarchical coded caching,'' \emph{IEEE Trans. on Information Theory},
  vol.~62, no.~6, pp. 3212--3229, 2016.

\bibitem{ji2017order}
M.~Ji, A.~M. Tulino, J.~Llorca, and G.~Caire, ``Order-optimal rate of caching
  and coded multicasting with random demands,'' \emph{IEEE Trans. on
  Information Theory}, vol.~63, no.~6, pp. 3923--3949, Jun. 2017.

\bibitem{ghasemi2017improved}
H.~Ghasemi and A.~Ramamoorthy, ``Improved lower bounds for coded caching,''
  \emph{IEEE Trans. on Information Theory}, vol.~63, no.~7, pp. 4388--4413,
  2017.

\bibitem{sengupta2017improved}
A.~Sengupta and R.~Tandon, ``Improved approximation of storage-rate tradeoff
  for caching with multiple demands,'' \emph{IEEE Trans. on Communications},
  vol.~65, no.~5, pp. 1940--1955, May 2017.

\bibitem{ajaykrishnan2015critical}
N.~Ajaykrishnan, N.~S. Prem, V.~M. Prabhakaran, and R.~Vaze, ``Critical
  database size for effective caching,'' in \emph{Proc. 2015 Twenty First
  National Conference on Communications (NCC)}, 2015, pp. 1--6.

\bibitem{chen2016fundamental}
Z.~Chen, P.~Fan, and K.~B. Letaief, ``Fundamental limits of caching: improved
  bounds for users with small buffers,'' \emph{IET Communications}, vol.~10,
  no.~17, pp. 2315--2318, Nov. 2016.

\bibitem{Sahraei:15}
S.~Sahraei and M.~Gastpar, ``$k$ users caching two files: An improved
  achievable rate,'' \emph{arXiv:1512.06682}, Dec. 2015.

\bibitem{Amiri:17}
M.~M. Amiri and D.~Gunduz, ``Fundamental limits of caching: Improved delivery
  rate-cache capacity trade-off,'' \emph{IEEE Trans. on Communications},
  vol.~65, no.~2, pp. 806--815, Feb. 2017.

\bibitem{Wan:16}
K.~Wan, D.~Tuninetti, and P.~Piantanida, ``On caching with more users than
  files,'' \emph{arXiv:1601.06383}, Jan. 2016.

\bibitem{yu2018exact}
Q.~Yu, M.~A. Maddah-Ali, and A.~S. Avestimehr, ``The exact rate-memory tradeoff
  for caching with uncoded prefetching,'' \emph{IEEE Trans. on Information
  Theory}, vol.~64, no.~2, pp. 1281--1296, 2018.

\bibitem{tian2018caching}
C.~Tian and J.~Chen, ``Caching and delivery via interference elimination,''
  \emph{IEEE Trans. on Information Theory}, vol.~64, no.~3, pp. 1548--1560,
  2018.

\bibitem{gomez2018fundamental}
J.~G{\'o}mez-Vilardeb{\'o}, ``Fundamental limits of caching: Improved bounds
  with coded prefetching,'' \emph{IEEE Trans. Communications}, to appear, 2018.

\bibitem{CoverThomas}
T.~M. Cover and J.~A. Thomas, \emph{Elements of Information Theory},
  1st~ed.\hskip 1em plus 0.5em minus 0.4em\relax New York: Wiley, 1991.

\bibitem{Tian:16symmetry}
C.~Tian, ``Symmetry, demand types and outer bounds in caching systems,'' in
  \emph{2016 IEEE International Symposium on Information Theory (ISIT)}, 2016,
  pp. 825--829.

\bibitem{yu2017characterizing}
Q.~Yu, M.~A. Maddah-Ali, and A.~S. Avestimehr, ``Characterizing the rate-memory
  tradeoff in cache networks within a factor of 2,'' \emph{arXiv:1702.04563},
  Feb. 2017.

\bibitem{wang2017improved}
C.-Y. Wang, S.~S. Bidokhti, and M.~Wigger, ``Improved converses and gap-results
  for coded caching,'' \emph{arXiv:1702.04834}, Feb. 2017.

\bibitem{Yeung:book}
R.~Yeung, \emph{A First Course in Information Theory}.\hskip 1em plus 0.5em
  minus 0.4em\relax New York: Kluwer Academic Publishers, 2002.

\bibitem{Lassez:92}
C.~Lassez and J.-L. Lassez, ``Quantifier elimination for conjunctions of linear
  constraints via a convex hull algorithm,'' in \emph{Symbolic and numerical
  computation for artificial intelligence}, B.~R. Donald, D.~Kapur, and J.~L.
  Mundy, Eds.\hskip 1em plus 0.5em minus 0.4em\relax San Diego, CA: Academic
  Press, 1992, ch.~4, pp. 103--1199.

\bibitem{Apte:15}
J.~Apte and J.~M. Walsh, ``Exploiting symmetry in computing polyhedral bounds
  on network coding rate regions,'' in \emph{Proc. of International Symposium
  on Network Coding (NetCod) 2015}, Sydney, Australia, Jun. 2015, pp. 76--80.

\bibitem{Ho:14}
S.-W. Ho, C.~W. Tan, and R.~W. Yeung, ``Proving and disproving information
  inequalities,'' in \emph{Proc. of 2014 IEEE International Symposium on
  Information Theory (ISIT)}, Honolulu, HI, Jun. 2014.

\bibitem{ZhangTian:17TCOM}
K.~Zhang and C.~Tian, ``On the symmetry reduction of information
  inequalities,'' \emph{IEEE Transactions on Communications}, pp. 2396--2408, 6
  2018.

\bibitem{harary1953graph}
F.~Harary and R.~Z. Norman, \emph{Graph theory as a mathematical model in
  social science}.\hskip 1em plus 0.5em minus 0.4em\relax University of
  Michigan Press, 1953.

\bibitem{Andrews:book}
G.~E. Andrews, \emph{The Theory of Partitions}.\hskip 1em plus 0.5em minus
  0.4em\relax Cambridge University Press, 1976.

\bibitem{Dougherty:05}
R.~Dougherty, C.~Freiling, and K.~Zeger, ``Insufficiency of linear coding in
  network information flow,'' \emph{IEEE Trans. on Information Theory},
  vol.~51, no.~8, pp. 2745--2759, Aug. 2005.

\bibitem{Zhang:97}
Z.~Zhang and R.~W. Yeung, ``A non-shannon-type conditional inequality of
  information quantities,'' \emph{IEEE Trans. on Information Theory}, vol.~43,
  no.~6, pp. 1982--1986, Nov. 1997.

\bibitem{Zhang:98}
------, ``On characterization of entropy function via information
  inequalities,'' \emph{IEEE Trans. on Information Theory}, vol.~44, no.~4, pp.
  1440--1452, Jul. 1998.

\bibitem{Yeung:99}
R.~W. Yeung and Z.~Zhang, ``On symmetrical multilevel diversity coding,''
  \emph{IEEE Trans. on Information Theory}, vol.~45, no.~2, pp. 609--621, Mar.
  1999.

\bibitem{Tian:11}
C.~Tian, ``Latent capacity region: A case study on symmetric broadcast with
  common messages,'' \emph{IEEE Trans. on Information Theory}, vol.~57, no.~6,
  pp. 3273--3285, Jun. 2011.

\bibitem{tian2017uncoded}
K.~Zhang and C.~Tian, ``Fundamental limits of coded caching: From uncoded
  prefetching to coded prefetching,'' \emph{IEEE Journal on Selected Areas in
  Communications}, to appear, 2018.

\end{thebibliography}
 \newcommand{\noop}[1]{}

\end{document}